\documentclass[journal,english]{IEEEtran}

\usepackage{epsfig}
\usepackage{times}
\usepackage{float}
\usepackage{afterpage}
\usepackage{amsmath}
\usepackage{amstext,cite}
\usepackage{amssymb,bm}
\usepackage{latexsym}
\usepackage{color}
\usepackage{graphicx}
\usepackage{amsmath}
\usepackage{amsthm}
\usepackage{graphicx}
\usepackage[center]{caption}
\usepackage{pstricks}
\usepackage{caption}
\usepackage{subcaption}
\usepackage{booktabs}
\usepackage{multicol}
\usepackage{lipsum}
\usepackage[T1]{fontenc}
\usepackage{hyperref}
\usepackage{aecompl}

\usepackage{soul}
\usepackage{cancel}

\allowdisplaybreaks

\long\def\comment#1{}

\newcommand{\Ac}{{\mathcal A}}

\newcommand{\Hc}{{\mathcal H}}

\newcommand{\Jc}{{\mathcal J}}
\newcommand{\Kc}{{\mathcal K}}

\newcommand{\Sc}{{\mathcal S}}

\newcommand{\Uc}{{\mathcal U}}
\newcommand{\Wc}{{\mathcal W}}
\newcommand{\Vc}{{\mathcal V}}

\newcommand{\Yc}{{\mathcal Y}}

\newcommand{\qsf}{{\mathsf q}}

\newcommand{\usf}{{\mathsf u}}

\newcommand{\Bsf}{{\mathsf B}}

\newcommand{\Ksf}{{\mathsf K}}

\newcommand{\Msf}{{\mathsf M}}
\newcommand{\Nsf}{{\mathsf N}}

\newcommand{\Rsf}{{\mathsf R}}

\newcommand{\Tsf}{{\mathsf T}}

\newtheorem{thm}{Theorem}
\newtheorem{cor}{Corollary}
\newtheorem{lem}{Lemma}
\newtheorem{prop}{Proposition}
\newtheorem{rem}{Remark}
\newtheorem{example}{Example}

\providecommand{\definitionname}{Definition}

\newcommand{\mgt}{\magenta}

\newcommand*{\rom}[1]{\expandafter\@slowromancap\romannumeral #1@}
\usepackage{changes}
\definechangesauthor[color=blue,name={Mingyue Ji}]{MJ}

\begin{document}

\title{Fundamental Limits of Decentralized Data Shuffling}

\author{
Kai~Wan,~\IEEEmembership{Member,~IEEE,} 
Daniela Tuninetti,~\IEEEmembership{Senior~Member,~IEEE,}
Mingyue~Ji,~\IEEEmembership{Member,~IEEE,}
~Giuseppe Caire,~\IEEEmembership{Fellow,~IEEE,}
~and Pablo Piantanida,~\IEEEmembership{Senior~Member,~IEEE}

\thanks{A short version of this paper was presented the 56th Annual Allerton Conference (2018) on Communication, Control, and Computing in Monticello, USA. }
\thanks{%
K.~Wan and G.~Caire are with the Electrical Engineering and Computer
Science Department, Technische Universit\"at Berlin, 10587 Berlin, Germany
(e-mail:  kai.wan@tu-berlin.de; caire@tu-berlin.de). The work of K.~Wan and G.~Caire was partially funded by the European Research Council  under the ERC Advanced Grant N. 789190, CARENET.}
\thanks{%
D.~Tuninetti is with the Electrical and Computer Engineering Department, University of Illinois at Chicago, Chicago, IL 60607, USA (e-mail: danielat@uic.edu). The work of   D.~Tuninetti was supported in parts by NSF 1527059 and NSF 1910309.}
\thanks{%
M.~Ji is with the Electrical and Computer Engineering Department, University of Utah, Salt Lake City, UT 84112, USA (e-mail: mingyue.ji@utah.edu). The work of M.~Ji was supported by NSF 1817154 and NSF 1824558.}
\thanks{%
P.~Piantanida is with CentraleSup{\'e}lec--French National Center for Scientific Research (CNRS)--Universit{\'e} Paris-Sud, 
91192 Gif-sur-Yvette, France, and with Montreal Institute for Learning Algorithms (MILA) at Universit{\'e} de Montr{\'e}al, 
QC H3T 1N8, Canada (e-mail: pablo.piantanida@centralesupelec.fr). The work of  P.~Piantanida was supported by the European Commission's Marie Sklodowska-Curie Actions (MSCA), through the Marie Sklodowska-Curie IF (H2020-MSCAIF-2017-EF-797805-STRUDEL).}
}

\maketitle

\begin{abstract}
Data shuffling of training data among different computing nodes (workers) has been identified as a core element to improve the statistical performance of modern large-scale machine learning algorithms. Data shuffling is often considered as one of the most significant bottlenecks in such systems due to the heavy communication load. Under a master-worker architecture (where a master has access to the entire dataset and only communication between the master and the workers is allowed) coding has been recently proved to considerably reduce the communication load.
This work considers a different communication paradigm referred to as {\it decentralized data shuffling}, where workers are allowed to communicate with one another via a shared link. The decentralized data shuffling problem has two phases: workers communicate with each other during the {\it data shuffling} phase, and then workers update their stored content during the {\it storage} phase.
 The main challenge is to derive novel converse bounds and achievable schemes for decentralized data shuffling by   considering the asymmetry of the workers' storages (i.e., workers are constrained to store different files in their storages based on the problem setting), in order to characterize the fundamental limits of this problem.

For the case of uncoded storage (i.e., each worker directly stores a subset of bits of the dataset), this paper proposes converse 
and achievable bounds (based on distributed interference alignment and distributed clique-covering strategies)   that are   within a factor of $3/2$ of one another.
The proposed schemes are also exactly optimal under the constraint of uncoded storage   for either large storage size or at most four workers in the system. 
%
\end{abstract}



\section{Introduction}
\label{sec:intro}

 \IEEEPARstart{R}{ecent} years have witnessed the emergence of big data and machine learning with wide applications in both business and consumer worlds. To cope with such a large size/dimension of data and the complexity of machine learning algorithms, it is increasingly popular to use distributed computing platforms such as Amazon Web Services Cloud,  Google Cloud, and Microsoft Azure services, where large-scale distributed machine learning algorithms can be implemented. The approach of data shuffling has been identified as one of the core elements to improve the statistical performance of modern large-scale machine learning algorithms~\cite{ChungUber2017,randomreshuffling2015}. In particular, data shuffling consists of re-shuffling the training data among all computing nodes (workers) once every few iterations, according to some given learning algorithms. However, due to the huge communication cost, data shuffling  may become one of the main system bottlenecks. 

To tackle this communication bottleneck problem, under a master-worker setup where the master  has access to the entire dataset, coded data shuffling has been recently proposed to significantly reduce the communication load between master and workers~\cite{speedup2018Lee}. 
However, when the whole dataset is stored across the workers, data shuffling can be implemented in a distributed fashion by allowing direct communication between the workers\footnote{In practice, workers communicate with each other as described in~\cite{ChungUber2017}.}. In this way, the communication bottleneck between a master and the workers can be considerably alleviated. 
This can be advantageous if the transmission capacity among workers is much higher than that between the master and workers, and the communication load between this two setups are similar. 

In this work, we consider such a {\it decentralized data shuffling} framework, where workers, connected by the same communication bus (common shared link), are allowed to communicate\footnote{  Notice that putting all nodes on the same bus (typical terminology in Compute Science) is very common and practically relevant since this is what happens for example with Ethernet, or with the  Peripheral Component Interconnect Express (PCI Express) bus inside a multi-core computer, where all cores share a common bus for intercommunication. The access of such bus is regulated by some collision avoidance protocol such as Carrier Sense Multiple Access (CSMA)~\cite{tobagiCSMA} or Token ring~\cite{tokenring}, such that once one node talks at a time, and all other listen. Therefore, this architecture is relevant in practice.  }. 
Although a master node may be present for the initial data distribution and/or for collecting the results of the training phase in a machine learning application, it is not involved in the data shuffling process which is entirely managed by the worker nodes in a distributed manner. In the following, we will review the literature of coded data shuffling (which we shall refer to as centralized data shuffling) and introduce the decentralized data shuffling framework studied in this paper. 

\subsection{Centralized Data Shuffling}
The coded data shuffling problem was originally proposed in~\cite{speedup2018Lee} in a {\it master-worker centralized model}  as illustrated in Fig.~\ref{fig:numerical 0a}.
In this setup, a master, with the access to the whole dataset containing $\Nsf$ data units, 
is connected to $\Ksf=\Nsf/\qsf$ workers, where $\qsf:=\Nsf/\Ksf$ is a positive integer.  
Each shuffling epoch is divided into {\it data shuffling} and {\it storage update} phases. 
In the data shuffling phase, a subset of the data units is assigned to each worker and each worker must recover these data units from the broadcasted packets of the master and its own stored content from the previous epoch. 
In the  storage update phase, each worker must store the newly assigned data units and, in addition, some information about other data units that can be retrieved from the  storage content and master transmission in the current epoch.  
Such additional information should be strategically designed in order to help the coded delivery of the required data units in the following epochs.
Each worker can store up to 
$\Msf$ data units in its local storage. If each worker directly copies some bits of the data units in its storage, the storage update phase is said to be {\it uncoded}. On the other hand, if the workers store functions (e.g., linear combinations) of the data objects' bits, the storage update is said to be {\it coded}.
The goal is, for a given $(\Msf,\Nsf,\qsf)$, to find the best two-phase strategy that minimizes the communication load during the data shuffling phase regardless of the shuffle.

The scheme proposed in~\cite{speedup2018Lee} uses a random uncoded storage (to fill users' extra memories independently when $\Msf > \qsf$) and a coded multicast transmission from the master to the workers, and yields a gain of a factor of $O(\Ksf)$ in terms of communication load with respect to the naive scheme for which the master simply broadcasts the missing, but required data bits to the workers.

The centralized coded data shuffling scheme with coordinated (i.e., deterministic) uncoded storage update phase was originally proposed in~\cite{informationAttia2016,worstAttia2016}, in order to minimize  the 
worst-case communication load $\Rsf$ among all
the possible shuffles, i.e., $\Rsf$ is smallest possible such that any shuffle can be realized.  
The proposed schemes in~\cite{informationAttia2016,worstAttia2016} are optimal under the constraint of uncoded storage for the cases where there is no extra storage for each worker (i.e., $\Msf=\qsf $) or there are less than or equal to three workers in the systems. Inspired by the achievable and converse bounds for the single-bottleneck-link caching problem in~\cite{dvbt2fundamental,ontheoptimality,exactrateuncoded}, the authors  in~\cite{neartoptimalAttia2018} then proposed a general coded data shuffling scheme, which was shown to be order optimal  to within a factor of $2$ under the constraint of uncoded storage. Also in~\cite{neartoptimalAttia2018}, the authors improved the performance of the general coded shuffling scheme by introducing an aligned coded delivery,
which was shown to be optimal under the constraint of uncoded storage for either $\Msf=\qsf$ or $\Msf\geq (\Ksf-2) \qsf$.
 
Recently, inspired by the 
improved data shuffling scheme in~\cite{neartoptimalAttia2018}, the authors in~\cite{fundamentalshuffling2018} proposed a linear coding scheme based on interference alignment, which achieves the optimal worst-case communication load under the constraint of uncoded storage for all system parameters. In addition, under the constraint of uncoded storage, the proposed coded data shuffling scheme in~\cite{fundamentalshuffling2018} was shown to be optimal for any shuffles (not just for the worst-case shuffles) when $\qsf=1$. 

\subsection{Decentralized Data Shuffling}

\begin{figure}
    \centering
    \begin{subfigure}[t]{0.5\textwidth}
        \centering
        \includegraphics[scale=0.3]{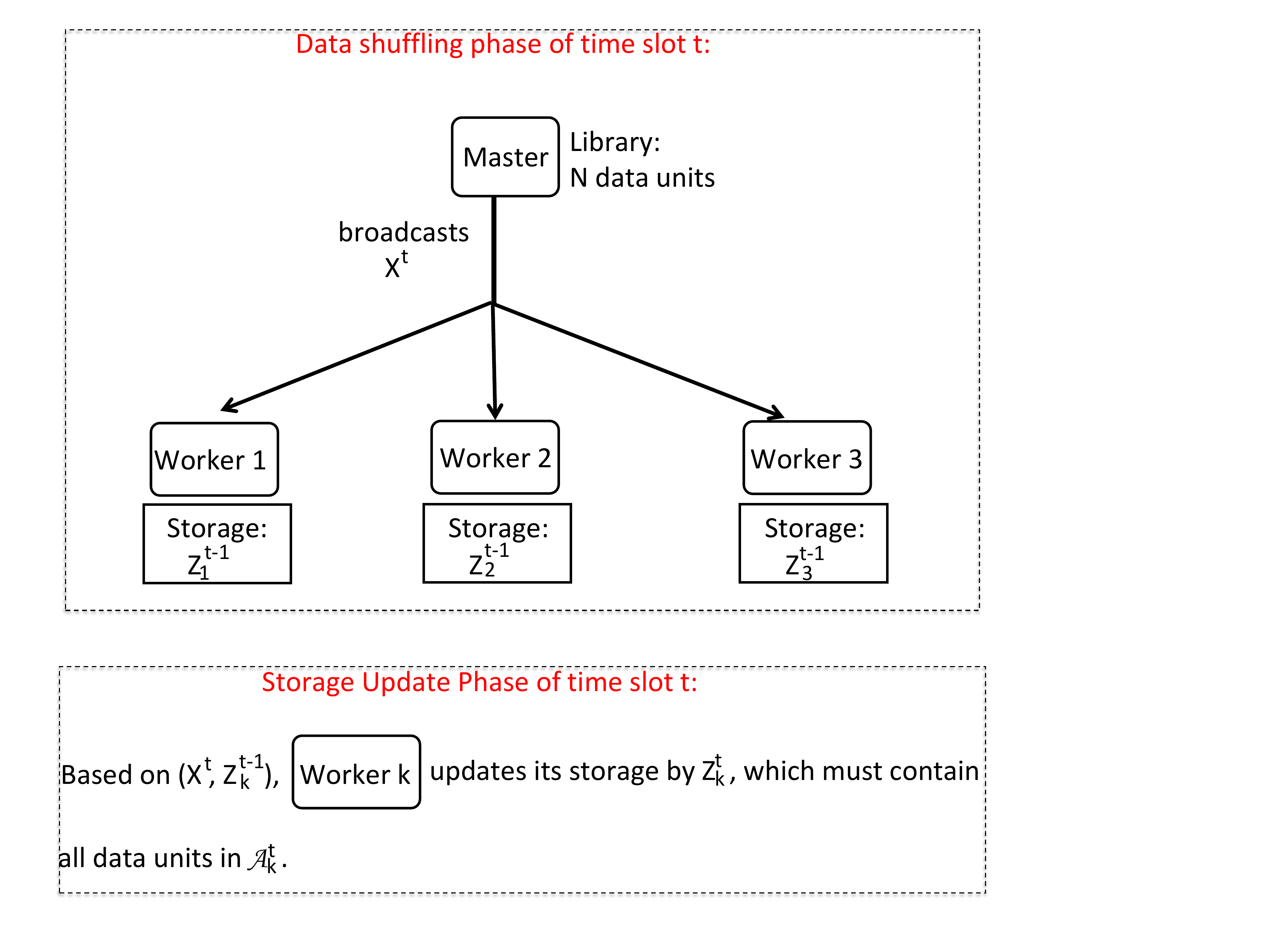}
        \caption{\small Centralized data shuffling.}
        \label{fig:numerical 0a}
    \end{subfigure}%
    \\ 
    \begin{subfigure}[t]{0.5\textwidth}
        \centering
        \includegraphics[scale=0.3]{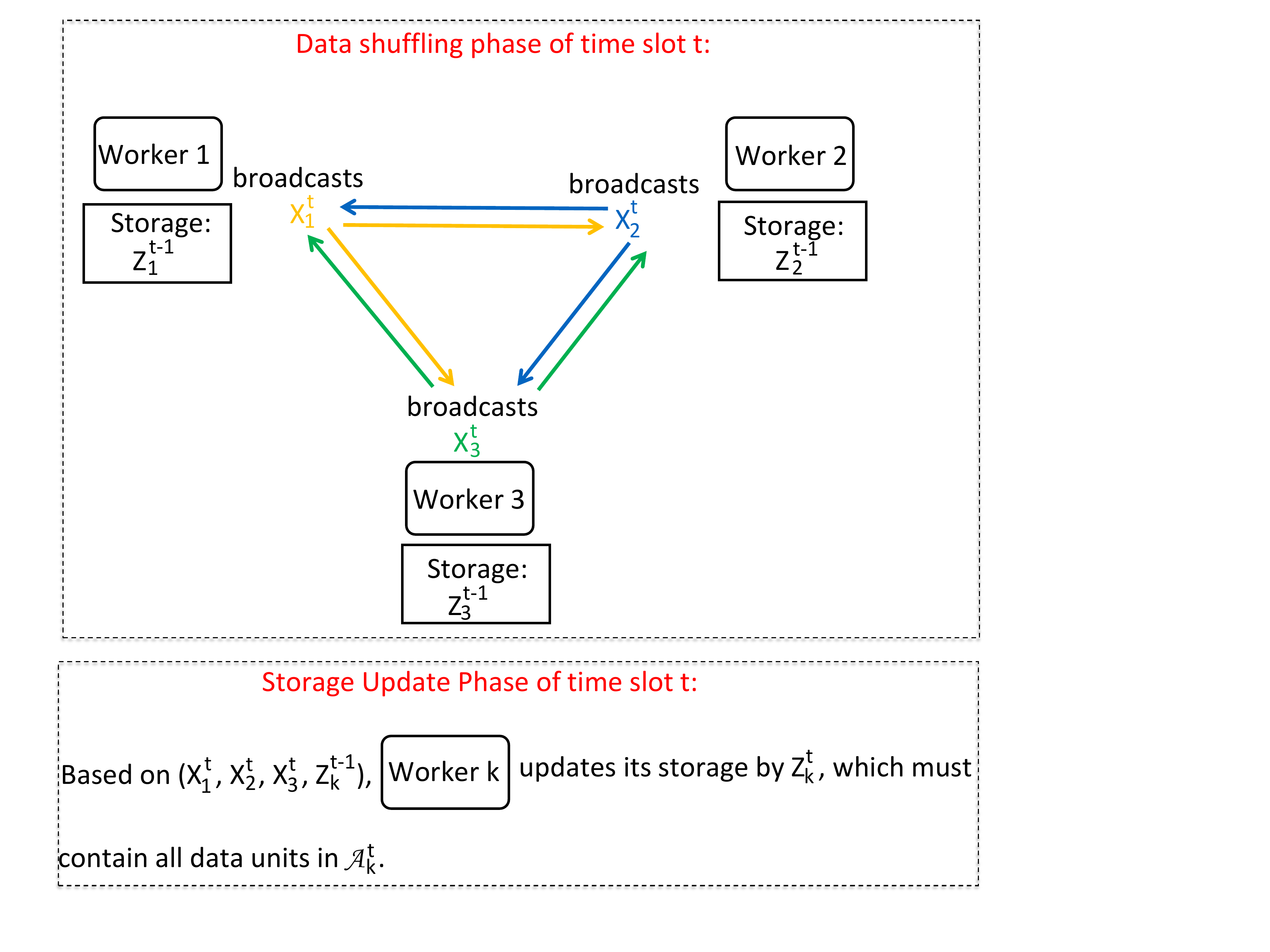}
        \caption{\small Decentralized data shuffling.}
        \label{fig:numerical 0b}
    \end{subfigure}
    \caption{\small The system models of the $3$-worker centralized and decentralized data shuffling problems in time slot $t$.  The data units in $\Ac^t_k$ are assigned to worker $k$, where $k\in\{1,2,3\}$ at time $t$.}
    \label{fig:numerical 0}
\end{figure}

An important limitation of the centralized framework is the assumption that workers can only receive packets from the master.
Since the entire dataset is stored in a decentralized fashion across the  workers at each epoch of the distributed learning algorithm, the master may not be needed in the data shuffling phase if workers can communicate with each other (e.g., \cite{ChungUber2017}). 
In addition, the communication among workers can be much more efficient compared to the communication from the master node to the workers~\cite{ChungUber2017},  e.g.,  the connection between the master node and workers is via a single-ported interface, where only one message can be passed for a given time/frequency slot.
In this paper, we propose the {\it decentralized data shuffling} problem as illustrated in Fig.~\ref{fig:numerical 0b}, where only communications among workers are allowed during the shuffling phase. This means that in the data shuffling phase, each worker broadcasts well designed coded packets (i.e., representations of the data) based on its stored content in the previous epoch. Workers take turn in transmitting, and transmissions are received error-free by all other workers through the common communication bus.
The objective is to design  the data shuffling and storage update phases in order to minimize the total communication load across all the workers in the worst-case shuffling scenario.

\paragraph*{Importance of decentralized data shuffling in practice}
In order to make the decentralized topology work in practice, we need to firstly guarantee that all the data units are already stored across the nodes so that the communication among computing nodes is sufficient. This condition is automatically satisfied from the definition of the decentralized data shuffling problem. 
Although the decentralized coded data shuffling incurs a larger load compared to its centralized counterpart, in practice, we may prefer the decentralized coded shuffling framework. This is due to the fact that the transmission delay/latency of the data transmission in real distributed computing system may depend on other system properties besides the total communication load, and the decentralized topology may achieve a better transmission delay/latency. This could be due to that 1) the connection between the master node and the worker clusters is normally via a single-ported interference, where only one message can be transmitted per time/frequency slot \cite{ChungUber2017}; 2) computing nodes are normally connected (e.g., via grid, or ring topologies) and the link bandwidth is generally much faster, in addition, computing nodes can transmit in parallel.  
 
\subsection{Relation to  Device-to-device (D2D) Caching  and   Distributed Computing}
The coded decentralized data shuffling problem considered in this paper is related to the {\it coded device-to-device (D2D) caching problem}~\cite{d2dcaching} and the {\it coded distributed computing problem}~\cite{distributedcomputing} -- see also Remark~\ref{rem: decentralized data shuffling vs D2D caching} next. 

The coded caching problem was originally proposed in~\cite{dvbt2fundamental} for a shared-link  broadcast model. The authors in~\cite{d2dcaching} extended the coded caching model to the case of D2D networks under the so-called protocol model. By choosing the communication radius of the protocol model such that each node can broadcast messages to all other nodes in the network,  the delivery phase of D2D coded caching is resemblant (as far as the topology of communication between the nodes is concerned) to the shuffling phase of our decentralized data shuffling problem.

 Recently, the scheme for coded D2D caching in~\cite{d2dcaching} has been extended to the coded distributed computing problem~\cite{distributedcomputing}, which consists of two stages named {\it Map} and {\it Reduce}.
In the Map stage, workers compute a fraction of intermediate computation values using local input data according to the designed Map functions. 
In the Reduce stage, according to the designed Reduce functions, workers exchange among each other a set of well designed (coded) intermediate computation values,  in order to compute the final output results. The coded distributed computing problem can be seen as a coded D2D caching problem under the constraint of uncoded and symmetric cache placement, where the symmetry  means that  each worker uses the same cache function for each file.
A converse bound 
was proposed in~\cite{distributedcomputing} to show that the proposed coded distributed computing scheme is   optimal in terms of communication load.  
This coded distributed computing framework was extended to   cases such as computing only necessary intermediate values~\cite{alternative2017,combinatoricsCDC2018}, 
 reducing file partitions and number of output functions~\cite{combinatoricsCDC2018,leveraging2018},  and considering random network topologies~\cite{CDCrandomconnect2018},  random connection graphs~\cite{CDCgraph2018,topologygraphCDC2019},
 stragglers~\cite{straggleswireless2017},  storage cost~\cite{yan2018distributedcom}, and heterogeneous computing power, function assignment and storage space~\cite{cascaded2019,CDChetero2019}.


Compared to coded D2D caching and coded distributed computing, the decentralized data shuffling problem differs as follows. 
On the one hand, an {\it asymmetric} constraint on the stored contents for the workers is present (because each worker must store all bits of each assigned data unit in the previous epoch, which breaks the symmetry of the stored contents across data units of the other settings).
On the other hand, each worker also needs to {\it dynamically update its storage} based on the received packets and its own stored content in the previous epoch. Therefore the decentralized data shuffling problem over multiple data assignment epochs is indeed a dynamic system where the evolution across the epochs of the node stored content plays a key role, while in the other problems reviewed above the cache content is static and determined at a single initial placement setup phase.


\subsection{Relation to  Centralized, Distributed,  and  Embedded  Index Codings}
\label{sub:index coding}
 In a {\it distributed index coding} problem~\cite{distribuedindexcoding,liu2018distributedIC}, there are multiple senders  connected to several receivers, where each sender or receiver can access to a subset of messages in the library. Each receiver demands one message and according to the users' demands and side informations, the senders cooperatively broadcast packets to all users to satisfy the users' demands.
The difference in a {\it centralized index coding} problem~\cite{birk1998informedsource} compared to the distributed one is that only one sender exists and this sender can access the whole library.  Very recently, the authors in~\cite{porter2019embeddedindex} considered a special case of distributed index coding, referred to as {\it embedded index coding}, where  each node  acts as both a sender  and a receiver in the system. 
It was shown in~\cite{porter2019embeddedindex}  that a linear 
code for this embedded index coding problem  can be obtained from a linear index code for the centralized version of the problem by   doubling the communication load. 

The centralized and decentralized data shuffling phases with uncoded storage are special cases of centralized and embedded index coding problems, respectively.
By using the construction in~\cite{porter2019embeddedindex} we could thus design a code for the decentralized data shuffling problem by using the optimal (linear) code  for the centralized case~\cite{fundamentalshuffling2018}; this would give a  decentralized data shuffling scheme with a load  twice that of~\cite{fundamentalshuffling2018}. 
It will be clarified later (in Remark~\ref{rem:comparison to direct extension}) that the proposed decentralized data shuffling schemes are strictly better than the those derived with the construction in~\cite{porter2019embeddedindex}. This is so because the construction in~\cite{porter2019embeddedindex} is general, while our design is for the specific topology considered.



\subsection{Contributions}
In this paper, we study the decentralized data shuffling problem, 
for which we propose converse and achievable bounds as follows.
\begin{enumerate}

\item
{\it Novel converse bound under the constraint of uncoded storage.}
Inspired by the induction method in~\cite[Thm.1]{distributedcomputing} for the distributed computing problem, we derive a converse bound under the constraint of uncoded storage. Different from the converse bound for the distributed computing problem, in our proof we propose  a novel approach  to account for the additional constraint on the ``asymmetric'' stored content. 

\item
{\it Scheme~A: General scheme for any $\Msf$.} 
By extending the general centralized data shuffling scheme from~\cite{neartoptimalAttia2018} to our decentralized model, we propose a general decentralized data shuffling scheme, where the analysis holds for any system parameters.  

\item
{\it Scheme~B: Improved scheme for  $\Msf\geq (\Ksf-2) \qsf$.}  
It can be seen later that  Scheme~A does not fully leverage the workers' stored content. With the storage update phase inspired by the converse bound and also used in the improved centralized data shuffling scheme in~\cite{neartoptimalAttia2018}, we propose a two-step scheme for decentralized data shuffling to improve on Scheme~A. In the first step we generate multicast messages as in~\cite{dvbt2fundamental}, and in the second step we encode these multicast messages by a  linear code based on distributed interference alignment (see Remark~\ref{rem:dia e scheme B}). 

By comparing our proposed converse bound and Scheme~B, we prove that Scheme~B is exactly optimal under the constraint of uncoded storage for $\Msf\geq (\Ksf-2) \qsf$. 
Based on this result, we can also characterize the exact optimality under the constraint of uncoded storage  when the number of workers satisfies $\Ksf\leq 4$.

\item
{\it Scheme~C: Improved scheme for $\Msf=2\qsf$.}
The delivery schemes proposed in~\cite{dvbt2fundamental,d2dcaching,neartoptimalAttia2018} for coded caching with a shared-link, D2D caching, and centralized data shuffling, all belong to the class of {\it clique-covering} method from a graph theoretic viewpoint. 
 By a   non-trivial extension from  a {\it distributed clique-covering approach}  for  the two-sender distributed index coding problems~\cite{thapa2017chandra}   to our decentralized data shuffling problem for the case $\Msf=2\qsf$, we propose a novel decentralized data shuffling scheme. The resulting scheme outperforms the previous two schemes  for this specific storage size.  


\item
{\it Order optimality under the constraint of uncoded storage.}
By combing the three proposed schemes and comparing with the proposed converse bound, we prove the order optimality of the combined scheme within a factor of  $3/2$  under the constraint of uncoded storage.
\end{enumerate}

\subsection{Paper Organization}
The rest of the paper is organized as follows.
The system model and problem formulation   for the decentralized data shuffling problem are given in Section~\ref{sec:model}.
Results from decentralized data shuffling related to our work are compiled in Section~\ref{sec:distributed vs shared-link data shuffling}.
Our main results are summarized in Section~\ref{sec:main}. 
The proof of the proposed converse bound can be found in Section~\ref{sec:proof of converse bound}, while the analysis of the proposed achievable schemes is in Section~\ref{sec:achievable}. Section~\ref{sec:conclusion} concludes the paper. 
The proofs of some auxiliary results can be   found in the Appendix.

\subsection{Notation Convention}
We use the following notation convention.
Calligraphic symbols denote sets, 
bold symbols denote vectors, 
and sans-serif symbols denote system parameters.
We use $|\cdot|$ to represent the cardinality of a set or the length of a vector;
$[a:b]:=\left\{ a,a+1,\ldots,b\right\}$ and $[n] := \{1,2,\ldots,n\}$; 
$\oplus$ represents bit-wise XOR; $\mathbb{N}$ denotes the set of all positive integers.

\section{System Model} 
\label{sec:model}
The $(\Ksf,\qsf,\Msf)$ decentralized data shuffling problem illustrated in Fig.~\ref{fig:numerical 0b} is defined as follows.
There are $\Ksf\in\mathbb{N}$ workers, each of which is charged to process and store $\qsf\in\mathbb{N}$ data units from a dataset of $\Nsf := \Ksf\qsf$ data units. Data units are denoted as $(F_{1},F_{2},\ldots,F_{\Nsf})$ and   each data unit is a binary vector containing $\Bsf$ i.i.d. bits. 
Each worker  has a local storage of   $\Msf\Bsf$ bits, where $\qsf \leq \Msf \leq \Ksf\qsf=\Nsf$. 
The workers are interconnected through a noiseless multicast network. 

The computation process occurs over $\Tsf$ time slots/epochs.
At the end of time slot $t-1, \ t\in[\Tsf],$ the content of the local storage of worker $k\in[\Ksf]$ is denoted by $Z^{t-1}_{k}$; the content of all storages is denoted by $Z^{t-1} := (Z^{t-1}_{1}, Z^{t-1}_{2}, \ldots,Z^{t-1}_{\Ksf})$.
At the beginning of time slot $t\in[\Tsf],$ the $\Nsf$ data units are partitioned into $\Ksf$ disjoint batches, each containing $\qsf$ data units. The data units indexed by $\Ac^{t}_{k}\subseteq[N]$ are assigned to worker  $k\in[\Ksf]$ who must store them in its local storage by the end of time slot $t\in[\Tsf]$. 
The dataset partition (i.e., data shuffle) in time slot $t\in[\Tsf]$ is denoted by $\Ac^{t} = (\Ac^{t}_{1},\Ac^{t}_{2},\ldots,\Ac^{t}_{\Ksf})$ and must satisfy
\begin{subequations}
\begin{align}
&
|\Ac^{t}_{k}|=\qsf, \ \forall k\in [\Ksf], 
\\&
 \Ac^{t}_{k_1} \cap \Ac^{t}_{k_2}=\emptyset,  \ \forall (k_1,k_2)\in [\Ksf]^2 : k_1\not= k_2, 
\\&
\cup_{k\in [\Ksf]} \Ac^{t}_{k} =[\Nsf]  \quad \text{(dataset partition)}.
\end{align}
\label{eq:dataset partition}
\end{subequations}
If $\qsf=1$, we let $\Ac^{t}_{k}=\{d^t_k\}$ for each $k\in[\Ksf]$.

We denote the worker who must store data unit $F_i$ at the end of time slot $t$ by $\usf^t_i$, where 
\begin{align}
\usf^t_i=k \textrm{ if and only if } i\in \Ac^{t}_{k}.
\label{eq:def usf^t_i}
\end{align}

The following two-phase scheme allows workers to store the requested data units. 

\paragraph*{Initialization} 
We first focus on the initial time slot $t=0$, where a master node broadcasts to all the workers.
Given partition $\Ac^{0}$, worker  $k\in[\Ksf]$ must store all the data units $F_{i}$ where $i\in \Ac^0_k$;
if there is excess storage, that is, if $\Msf>\qsf$, 
worker $k\in[\Ksf]$ can store in its local storage parts of the data units indexed by $[\Nsf]\setminus\Ac^{0}_{k}$. The storage function for worker $k\in[\Ksf]$ in time slot $t=0$ is denoted by $\psi^0_k$, where
\begin{subequations}
\begin{align}
   & Z^0_k:=\psi^0_k \Big(\Ac^0,(F_{i}:i\in \Nsf)\Big) \quad \text{(initial storage placement)} : 
\\ & H\Big(Z^{0}_{k}\Big)\leq \Msf\Bsf, \ \forall k\in[\Ksf] \quad \text{(initial storage size constraint)},
\\ & H\Big(\big(F_i:i\in \Ac^{0}_{k} \big)|Z^{0}_{k}\Big)=0  \quad \text{(initial storage content constraint)}.
\end{align}
\end{subequations}
Notice that the storage initialization and the storage update phase (which will be described later) are without knowledge of later shuffles.
In subsequent time slots $t\in [\Tsf]$, the master is not needed and the workers communicate with one another.

\paragraph*{Data Shuffling Phase} 
Given global knowledge of the stored content $Z^{t-1}$ at all workers, and of the data shuffle from $\Ac^{t-1}$ to $\Ac^{t}$ (indicated as $\Ac^{t-1} \to \Ac^{t}$) worker $k\in[\Ksf]$ broadcasts a message $X^{t}_{k}$ 
to all other workers, where $X^{t}_{k}$ is based only on the its local storage content $Z^{t-1}_{k}$, that is,
\begin{align}
H\Big(X^{t}_{k}| Z^{t-1}_{k}\Big)=0  \quad \text{(encoding)}.
\label{eq:encoding node k}
\end{align}
The collection of all sent messages is denoted by $X^{t} := (X^{t}_{1}, X^{t}_{2}, \ldots,X^{t}_{\Ksf})$.
Each worker  $k\in[\Ksf]$ must recover all data units indexed by $\Ac^{t}_{k}$ from the sent messages $X^{t}$ and its local storage content $Z^{t-1}_{k}$, that is,
\begin{align}
H\Big(\big(F_i:i\in \Ac^{t}_{k}\big)|Z^{t-1}_{k}, X^{t}\Big)=0 \quad \text{(decoding)}.
\label{eq:decoding node k}
\end{align}
The rate $\Ksf$-tuple $(\Rsf^{\Ac^{t-1} \to \Ac^{t}}_{1},\ldots,\Rsf^{\Ac^{t-1} \to \Ac^{t}}_{\Ksf})$ is said to be {\it feasible}
if there exist 
 delivery functions $\phi^{t}_{k} : X^{t}_{k} = \phi^{t}_{k}(Z^{t-1}_{k})$ for all $t\in[\Tsf]$ and $k\in[\Ksf]$
satisfying the constraints~\eqref{eq:encoding node k} and~\eqref{eq:decoding node k}, and such that 
\begin{align}
H\Big(X^{t}_{k}\Big) \leq \Bsf\Rsf^{\Ac^{t-1} \to \Ac^{t}}_{k} \quad \text{(load)}.
\label{eq:load node k}
\end{align}

\paragraph*{Storage Update Phase} 
After the data shuffling phase in time slot $t$, we have the storage update phase in time slot $t\in [\Tsf]$.
Each worker  $k\in[\Ksf]$ must update its local storage based on the sent messages $X^{t}$ and its local stored content $Z^{t-1}_{k}$, that is, 
\begin{align}
H\Big(Z^{t}_{k}|Z^{t-1}_{k}, X^{t}\Big)=0 \quad \text{(storage update)},
\label{eq:storage update node k}
\end{align}
by placing in it all the recovered data units, that is, 
\begin{align}
H\Big(\big(F_i:i\in \Ac^{t}_{k} \big)|Z^{t}_{k}\Big)=0, \quad \text{(stored content)}.
\label{eq:additional storage constraint node k}
\end{align}
Moreover, the local storage has limited size bounded by
\begin{align}
H\Big(Z^{t}_{k}\Big)\leq \Msf\Bsf, \ \forall k\in[\Ksf], \quad \text{(storage size)}.
\label{eq:cache size node k}
\end{align}
A storage update for worker $k\in[\Ksf]$ is said to be feasible 
if there exist 
functions $\psi^{t}_{k} : Z^{t}_{k} = \psi^{t}_{k}(\Ac_k^t,Z^{t-1}_k,X^t)$ for all $t\in[\Tsf]$ and $k\in[\Ksf]$
satisfying the constraints in~\eqref{eq:storage update node k}, \eqref{eq:additional storage constraint node k} and \eqref{eq:cache size node k}.

Note: if for any $k_1, k_2\in[\Ksf]$ and $t_1,t_2 \in [\Tsf]$ we have $\Psi^{t_1}_{k_1} \equiv \Psi^{t_2}_{k_2}$ (i.e., $\Psi^{t_1}_{k_1} $ is equivalent to $\Psi^{t_2}_{k_2}$), the storage  phase is called {\it structural invariant}.

\paragraph*{Objective} 
The objective is to minimize the {\it worst-case total communication load}, or just load for short in the following, among all possible consecutive data shuffles, that is we aim to characterized $\Rsf^{\star}$ defined as 
\begin{align}
\Rsf^{\star} := 
&\lim_{\Tsf\to \infty}
\min_{\substack{\psi^{t^{\prime}}_k,\phi^{t^{\prime}}_k :\\ t^{\prime}\in[\Tsf],k\in[\Ksf]} } \
\max_{(\Ac^0,\ldots,\Ac^{\Tsf})} \ 
\bigg\{\max_{t\in [  \Tsf ]}\sum_{k\in[\Ksf]}\Rsf^{\Ac^{t-1} \to \Ac^{t}}_{k} :   \nonumber\\ &   \textrm{the rate $\Ksf$-tuple and the storage are feasible}\bigg\}.
\label{eq:goal}
\end{align}

The minimum 
load under the constraint of uncoded storage  is denoted by $\Rsf^{\star}_{\mathrm{u}}$. 
In general, $\Rsf^{\star}_{\mathrm{u}} \geq \Rsf^{\star}$,  because the set of all general data shuffling schemes is a superset of all data shuffling schemes with uncoded storage.

\begin{rem}[Decentralized Data Shuffling vs D2D Caching]
\label{rem: decentralized data shuffling vs D2D caching}
The   D2D caching problem studied in~\cite{d2dcaching} differs from our setting as follows:
\begin{enumerate}
\item
in the decentralized data shuffling problem one has the constraint on the stored content in~\eqref{eq:additional storage constraint node k} that imposes that each worker stores the whole requested files, which is not present in the D2D caching problem; and
\item
in the D2D caching problem each worker fills its local cache by accessing the whole library of files, while in the decentralized data shuffling problem each worker updates its local storage based on the received packets in the current time slot and its stored content in the previous time slot as in~\eqref{eq:storage update node k}.
\end{enumerate}
Because of these differences, 
achievable and converse bounds for the decentralized data shuffling problem can not be obtained by trivial renaming of variables in the D2D caching problem.

\hfill$\square$
\end{rem}

\section{Relevant Results for Centralized Data Shuffling}
\label{sec:distributed vs shared-link data shuffling}

Data shuffling was originally proposed in~\cite{speedup2018Lee} for the centralized scenario, where communications only exists between the master and the workers, that is, the $\Ksf$  decentralized encoding conditions in~\eqref{eq:encoding node k} are replaced by $H(X^{t}| F_1,\ldots, F_{\Nsf})=0$ where $X^{t}$ is broadcasted by the master  to all the workers. We summarize next some key results from~\cite{neartoptimalAttia2018}, which will be used in the following sections. We shall use the subscripts ``$\text{u,cen,conv}$'' and ``$\text{u,cen,ach}$'' for converse (conv) and achievable (ach) bounds, respectively, for the centralized problem (cen) with uncoded storage (u). We have
\begin{enumerate}

\item
{\it Converse for centralized data shuffling:}
For a $(\Ksf,\qsf,\Msf)$ centralized data shuffling system, the worst-case communication load under the constraint of uncoded storage  is lower bounded by the lower convex envelope of the following storage-load pairs~\cite[Thm.2]{neartoptimalAttia2018}
\begin{align}
\left(
\frac{\Msf}{\qsf}=m, \ 
\frac{\Rsf}{\qsf}=\frac{\Ksf-m}{m}
\right)_\text{\rm u,cen,conv}, \
\forall m\in[\Ksf].
\label{eq:shared-link shuffling converse bound}
\end{align}

\item
{\it Achievability for centralized data shuffling:}
In~\cite{neartoptimalAttia2018} it was also shown that the lower convex envelope of the following storage-load pairs is achievable with  uncoded storage~\cite[Thm.1]{neartoptimalAttia2018}
\begin{align}
\left(
\frac{\Msf}{\qsf}=1+g\frac{\Ksf-1}{\Ksf}, \ 
\frac{\Rsf}{\qsf}=\frac{\Ksf-g}{g+1}
\right)_\text{\rm u,cen,ach}, \
\forall g\in[0:\Ksf].
\label{eq:shared-link shuffling achievable bound}
\end{align}
The achievable bound in~\eqref{eq:shared-link shuffling achievable bound} was shown to be within a factor $\frac{\Ksf}{\Ksf-1}\leq 2$ of the converse bound in~\eqref{eq:shared-link shuffling converse bound} under the constraint of uncoded storage~\cite[Thm.3]{neartoptimalAttia2018}.

\item
{\it Optimality for centralized data shuffling:}
It was shown in~\cite[Thm.4]{fundamentalshuffling2018} that the converse bound in~\eqref{eq:shared-link shuffling converse bound} can be achieved by a scheme that uses linear network coding and interference alignement/elimination. An optimality result similar to~\cite[Thm.4]{fundamentalshuffling2018} was shown in~\cite[Thm.4]{neartoptimalAttia2018}, but only for $m\in\{1,\Ksf-2,\Ksf-1\}$; note that $m=\Ksf$ is trivial.

\end{enumerate}

Although the scheme that achieves the load in~\eqref{eq:shared-link shuffling achievable bound} is not optimal in general, we shall next describe its inner workings as we will generalize it to the case of decentralized data shuffling.

\paragraph*{Structural Invariant Data Partitioning and Storage}
Fix $g\in[0:\Ksf]$ and divide each data unit into $\binom{\Ksf}{g}$ non-overlapping and equal-length sub-blocks of length $\Bsf/\binom{\Ksf}{g}$ bits.
  Let each data unit be
$F_i = (G_{i,\Wc} : \Wc\subseteq[\Ksf] : |\Wc|=g), \ \forall i\in[\Nsf]$.
The storage of worker $k\in[\Ksf]$ at the end of time slot $t$ is as follows,\footnote{\label{foot:GiW}  Notice that here each sub-block $G_{i,\Wc}$ is stored by workers $\{\usf^{t}_i\} \cup \Wc$. In addition, later in our proofs of the converse bound and  proposed achievable schemes for decentralized data shuffling, the notation $F_{i,\Wc}$ denotes the sub-block of $F_i$, which is stored by workers in $\Wc$.}
\begin{align}
&Z_k^{t} \nonumber\\
  &=  \Big(\negmedspace 
\underbrace{(G_{i,\Wc} : \forall \Wc, \forall i\in \Ac_k^{t})}_\text{required data units} \cup 
\underbrace{(G_{i,\Wc} : k \in\Wc, \forall i\in[\Nsf]\setminus \Ac_k^{t})}_\text{other data units}
 \negmedspace \Big) 
\label{eq:ravi scheme 1 storageI}
\\&=  \Big(\negmedspace 
\underbrace{(G_{i,\Wc} : k \not\in \Wc,  \forall i\in \Ac_k^{t})}_\text{variable part of the storage} \cup 
\underbrace{(G_{i,\Wc} : k \in \Wc, \forall i\in[\Nsf])}_\text{fixed part of the storage}
 \negmedspace \Big). 
\label{eq:ravi scheme 1 storageII}
\end{align}
Worker $k\in[\Ksf]$ stores all the $\binom{\Ksf}{g}$ sub-blocks of the required $\qsf$ data units indexed by $\Ac^{t}_{k}$, and also $\binom{\Ksf-1}{g-1}$ sub-blocks of each data unit indexed by $[\Nsf] \setminus \Ac^{t}_{k}$ (see~\eqref{eq:ravi scheme 1 storageI}), thus the required storage space is
\begin{align}
\Msf = \qsf+(\Nsf-\qsf)\frac{\binom{\Ksf-1}{g-1}}{\binom{\Ksf}{g}}
=\Big(1+g\frac{\Ksf-1}{\Ksf} \Big)\qsf.
\label{eq:ravi scheme 1 M}
\end{align}
It can be seen (see~\eqref{eq:ravi scheme 1 storageII} and also Table~\ref{tab:example Giw}) that the storage of worker $k\in[\Ksf]$ at time $t\in[\Tsf]$ is partitioned in two parts: (i) the ``fixed part'' contains all the sub-blocks of all data points that have the index $k$ in the second subscript; this part of the storage will not be changed over time; and
(ii) the ``variable part'' contains all the sub-blocks of all required data points at time $t$ that do not have the index $k$ in the second subscript; this part of the storage will be updated over time.

\begin{table*}
\protect\caption{ Example of   file partitioning and storage in~\eqref{eq:ravi scheme 1 storageII} at the end of time slot $t$ for the decentralized data shuffling problem with $(\Ksf,\qsf,\Msf)=(3,1,7/3)$ and $\Ac^{t} = (3,1,2)$ where $g=2$.}\label{tab:example Giw}
\centering{}
\begin{tabular}{|c|c|c|c|}
\hline 
Workers & Sub-blocks of $F_1$ & Sub-blocks of $F_2$ & Sub-blocks of $F_3$  \tabularnewline
\hline 
\hline 
Worker $1$ stores & $G_{1,\{1,2\}}$, $G_{1,\{1,3\}}$  & $G_{2,\{1,2\}}$, $G_{2,\{1,3\}}$  &$G_{3,\{1,2\}}$, $G_{3,\{1,3\}}$, $G_{3,\{2,3\}}$  \tabularnewline
\hline 
Worker $2$ stores & $G_{1,\{1,2\}}$, $G_{1,\{1,3\}}$, $G_{1,\{2,3\}}$  & $G_{2,\{1,2\}}$, $G_{2,\{2,3\}}$  &$G_{3,\{1,2\}}$, $G_{3,\{2,3\}}$  \tabularnewline
\hline 
Worker $3$ stores & $G_{1,\{1,3\}}$, $G_{1,\{2,3\}}$  & $G_{2,\{1,2\}}$, $G_{2,\{1,3\}}$, $G_{2,\{2,3\}}$  & $G_{3,\{1,3\}}$, $G_{3,\{2,3\}}$   \tabularnewline
\hline
\end{tabular}
\end{table*}

\paragraph*{Initialization (for the achievable bound in~\eqref{eq:shared-link shuffling achievable bound})}
The master directly transmits all data units. 
The storage is as in~\eqref{eq:ravi scheme 1 storageII} given  $\Ac^0$.

\paragraph*{Data Shuffling Phase of time slot $t\in [T]$ (for the achievable bound in~\eqref{eq:shared-link shuffling achievable bound})}After the end of storage update phase at time $t-1$, the new assignment $\Ac^{t}$ is revealed. 
For notation convenience, let 
\begin{align}
G^{\prime}_{k,\Wc} = \Big( G_{i,\Wc} :  i\in \Ac^{t}_{k}\setminus \Ac^{t-1}_{k} \Big), 
\label{eq:ravi scheme 1 storageIII}
\end{align}
for all $k\in[\Ksf]$ and all $\Wc\subseteq [\Ksf]$, where $|\Wc|=g$ and $k\notin \Wc$.
Note that in~\eqref{eq:ravi scheme 1 storageIII} we have $|G^{\prime}_{k,\Wc}|\leq \Bsf\frac{\qsf}{\binom{\Ksf}{g}}$, with equality (i.e., worst-case scenario) if and only if $\Ac^{t}_{k}\cap \Ac^{t-1}_{k}=\emptyset$. 
To allow the workers to recover their missing sub-blocks, the central server broadcasts $X^{t}$ defined as
\begin{align}
&X^{t} = (W^{t}_{\Jc} : \Jc\subseteq [\Ksf] : |\Jc|=g+1),
\label{eq:ravi scheme 1 X}
\\& \text{ where }
W^{t}_{\Jc} = \oplus_{k\in\Jc} G^{\prime}_{k,\Jc\setminus\{k\}},
\label{eq:ravi scheme 1 W}
\end{align}
where in the 
 multicast message $W^{t}_{\Jc}$ in~\eqref{eq:ravi scheme 1 W} the sub-blocks $G^{\prime}_{k,\Wc}$ involved in the sum are zero-padded to meet the length of the longest one.
Since worker  $k\in\Jc$ requests $G^{\prime}_{k,\Jc\setminus\{k\}}$ and has stored all the remaining sub-blocks in $W^{t}_{\Jc}$ defined in~\eqref{eq:ravi scheme 1 W}, it can recover $G^{\prime}_{k,\Jc\setminus\{k\}}$ from $W^{t}_{\Jc}$, and thus all its missing sub-blocks from $X^{t}$.

\paragraph*{Storage Update Phase of time slot $t\in [T]$  (for the achievable bound in~\eqref{eq:shared-link shuffling achievable bound})}
Worker $k\in[\Ksf]$ evicts from the (variable part of its) storage the sub-blocks
$(G_{i,\Wc} : k \not\in \Wc,  \forall i\in \Ac_k^{t-1} \setminus \Ac_k^{t})$
and replaces them with the sub-blocks
$(G_{i,\Wc} : k \not\in \Wc,  \forall i\in \Ac_k^{t} \setminus \Ac_k^{t-1})$.
This procedure maintains the structural invariant storage structure of the storage in~\eqref{eq:ravi scheme 1 storageII}.

\paragraph*{Performance Analysis (for the achievable bound in~\eqref{eq:shared-link shuffling achievable bound})}
The total worst-case communication load satisfies
\begin{align}
\Rsf
\leq \qsf\frac{\binom{\Ksf}{g+1}}{\binom{\Ksf}{g}}
=    \qsf\frac{\Ksf-g}{g+1},
\label{eq:ravi scheme 1 R}
\end{align}
with equality (i.e., worst-case scenario) if and only if $\Ac^{t}_{k}\cap \Ac^{t-1}_{k}=\emptyset$ for all $k\in[\Ksf]$.

\section{Main Results}
\label{sec:main}
In this section, we summarize our main results for the decentralized data shuffling problem.
We shall use the subscripts ``$\text{\rm u,dec,conv}$'' and ``$\text{\rm u,dec,ach}$'' for converse (conv) and achievable (ach) bounds, respectively, for the decentralized problem (dec) with uncoded storage (u).
We have:
\begin{enumerate}

\item
{\it Converse:} 
We start with a converse bound for the decentralized data shuffling problem under the constraint of uncoded storage.
\begin{thm}[Converse]
\label{thm:converse bound}
For a $(\Ksf,\qsf,\Msf)$ decentralized data shuffling system, the worst-case 
load under the constraint of uncoded storage  is lower bounded by the lower convex envelope of the following storage-load pairs
\begin{align}
\left(
\frac{\Msf}{\qsf}=m,
\frac{\Rsf}{\qsf}=\frac{\Ksf-m}{m} \frac{\Ksf}{\Ksf-1}
\right)_\text{\rm u,dec,conv}, \
\forall m\in[\Ksf].
\label{eq:distributed shuffling converse bound}
\end{align}
\end{thm}
Notice that   the proposed converse bound is a piecewise linear curve with the corner points in~\eqref{eq:distributed shuffling converse bound} and these corner points are  successively convex.

The proof of Theorem~\ref{thm:converse bound} can be found in Section~\ref{sec:proof of converse bound} and is inspired by the induction method proposed in~\cite[Thm.1]{distributedcomputing} for the distributed computing problem. 
However, there are two main differences in our proof compared to~\cite[Thm.1]{distributedcomputing}:
(i) we need to account for the additional constraint on the stored content in~\eqref{eq:additional storage constraint node k}, 
(ii) our storage update phase is by problem definition in~\eqref{eq:additional storage constraint node k} asymmetric across data units, while it is symmetric in the distributed computing problem.

\item
{\it Achievability:} 
We next extend the centralized data shuffling scheme in Section~\ref{sec:distributed vs shared-link data shuffling} 
to our decentralized setting.
\begin{thm}[Scheme~A]
\label{thm:Achievable scheme A}
For a $(\Ksf,\qsf,\Msf)$ decentralized data shuffling system, the worst-case 
load under the constraint of uncoded storage is upper bounded by the lower convex envelope of the following storage-load pairs
\begin{align}
\left(
\frac{\Msf}{\qsf}=1+g\frac{\Ksf-1}{\Ksf},
\frac{\Rsf}{\qsf}=\frac{\Ksf-g}{g} 
\right)_\text{\rm u,dec,ach}, \
\forall g\in  [\Ksf-1].
\label{eq:distributed shuffling achievable bound 1}
\\ \text{ and (smallest storage) }
\left(
\frac{\Msf}{\qsf}=1,
\frac{\Rsf}{\qsf}=\Ksf 
\right)_\text{\rm u,dec,ach}, \
\label{eq:distributed shuffling achievable bound 1 trivial M=q}
\\ \text{ and (largest storage) }
\left(
\frac{\Msf}{\qsf}=\Ksf,
\frac{\Rsf}{\qsf}=0 
\right)_\text{\rm u,dec,ach}. \
\label{eq:distributed shuffling achievable bound 1 trivial M=Kq}
\end{align}
\end{thm}
The proof is given in Section~\ref{sub:achievable a}.

\bigskip
A limitation of Scheme~A in Theorem~\ref{thm:Achievable scheme A} is that, in time slot $t\in[\Tsf]$ worker $k\in[\Ksf]$ does not fully leverage all its stored content. 
We overcome this limitation by developing Scheme~B described in Section~\ref{sub:achievable b}. 
\begin{thm}[Scheme~B]
\label{thm:Achievable scheme B}
For a $(\Ksf,\qsf,\Msf)$ decentralized data shuffling system, the worst-case 
load under the constraint of uncoded storage for $\Msf \geq (\Ksf-2)\qsf$ is upper bounded by the lower convex envelope of the following storage-load pairs
\begin{align}
&\left(
\frac{\Msf}{\qsf}= m,
\frac{\Rsf}{\qsf}= \frac{\Ksf-m}{m} \frac{\Ksf}{\Ksf-1} 
\right)_\text{\rm u,dec,ach}, \nonumber\\
&\forall m\in \{\Ksf-2,\Ksf-1,\Ksf\}.
\label{eq:distributed shuffling achievable bound 2}
\end{align}
\end{thm}

We note that Scheme~B is neither a direct extension of~\cite[Thm.4]{neartoptimalAttia2018} nor of~\cite[Thm.4]{fundamentalshuffling2018} from the centralized to the decentralized setting.
As it will become clear from the details in Section~\ref{sub:achievable b}, our scheme works  with a rather simple way to generate the multicast messages transmitted by the workers, and it applies to any shuffle, not just to the worst-case one. In Remark~\ref{rem:extension of achievable scheme B}, we also extend this scheme for the general storage size regime.

\bigskip
Scheme~B in Theorem~\ref{thm:Achievable scheme B} uses a distributed clique-covering method to generate multicast messages similar to what is done for D2D caching~\cite{dvbt2fundamental},  
where distributed clique cover is for the side information graph (more details in Section~\ref{subsec:Sub-block Division and Graphical Representation}). Each  multicast message corresponds to one distributed clique and includes one linear combination of all nodes in this clique. However, due to the asymmetry of the decentralized data shuffling problem (not present in D2D coded  caching), 
the lengths of most distributed cliques are small and thus 
the multicast messages based on cliques and sent by a worker in general include only a small number of messages (i.e., small multicast gain).
To overcome this limitation,    the key idea of Scheme~C for $\Msf/\qsf=2$ (described in Section~\ref{sub:achievable c}) is to augment some of the cliques and send them in $\Msf/\qsf=2$ linear combinations.
%
\begin{thm}[Scheme~C]
\label{thm:Achievable Scheme C}
For a $(\Ksf,\qsf,\Msf)$ decentralized data shuffling system, the worst-case 
load under the constraint of uncoded storage for $\Msf/\qsf=2$ is upper bounded by 
\begin{align}
\left(
\frac{\Msf}{\qsf}= 2,
\frac{\Rsf}{\qsf}= \frac{2 \Ksf (\Ksf-2)}{3(\Ksf-1)} 
\right)_\text{\rm u,dec,ach}.
\label{eq:distributed shuffling achievable bound 3}
\end{align}
\end{thm}

  It will be seen later that the proposed schemes only use binary codes, and only XOR operations are needed for the decoding procedure. 

Finally, we combine the proposed three schemes (by considering the one among Schemes~A, B or~C that attains the lowest load for each storage size).
\begin{cor}[Combined Scheme]
\label{cor:combined scheme}
For a $(\Ksf,\qsf,\Msf)$ decentralized data shuffling system,
the achieved storage-load tradeoff of the combined scheme is the lower convex envelope of the corner points  is as follows:
\begin{itemize}
\item  $\Msf=  \qsf$. With Scheme~A, the worst-case load is $\qsf \frac{\Ksf-m}{m}\frac{\Ksf}{\Ksf-1}$.
\item  $\Msf=2 \qsf$. With Scheme C, the worst-case load is $\qsf \frac{\Ksf-m}{m}\frac{\Ksf}{\Ksf-1} \frac{4}{3}$. 
\item $\Msf=\big(1+g\frac{\Ksf-1}{\Ksf} \big)\qsf$ where $g\in[2:\Ksf-3]$. With Scheme~A, the worst-case load is $\qsf \frac{\Ksf-g}{g}$.
\item $\Msf=m \qsf$ where $m\in[\Ksf-2:\Ksf]$. With Scheme B, the worst-case load is $\qsf \frac{\Ksf-m}{m}\frac{\Ksf}{\Ksf-1}$.
\end{itemize}

\end{cor}

\item
{\it Optimality:} 
By comparing our achievable and converse bounds, we have the following exact optimality results. 
\begin{thm}[Exact Optimality for $\Msf/\qsf \geq \Ksf-2$] 
\label{thm:exact optimality scheme 2}
For a $(\Ksf,\qsf,\Msf)$ decentralized data shuffling system, the optimal worst-case load under the constraint of uncoded storage  for $\Msf/\qsf\in [\Ksf-2,\Ksf]$ is given in Theorem~\ref{thm:converse bound} and is attained by Scheme~B in Theorem~\ref{thm:Achievable scheme B}. 
\end{thm} 
Note that the converse bound on the load for the case $\Msf/\qsf=1$ is trivially achieved by Scheme~A in Theorem~\ref{thm:Achievable scheme A}.

\bigskip
From Theorem~\ref{thm:exact optimality scheme 2} 
we can immediately conclude the following.
\begin{cor}[Exact Optimality for $\Ksf \leq 4$]
\label{cor:optimality until 4}
For a $(\Ksf,\qsf,\Msf)$ decentralized data shuffling system, the optimal worst-case load under the constraint of uncoded storage is given by Theorem~\ref{thm:converse bound} for $\Ksf \leq 4$.
\end{cor}

\bigskip
Finally, by combining the three proposed achievable schemes, we have the following order optimality result proved in Section~\ref{sub:optimality}. 
\begin{thm}[Order Optimality for $\Ksf > 4$]
\label{thm:order optimality}
For a $(\Ksf,\qsf,\Msf)$ decentralized data shuffling system under the constraint of uncoded storage, for the cases not covered by Theorem~\ref{thm:exact optimality scheme 2}, the combined scheme in Corollary~\ref{cor:combined scheme} achieves  the converse bound in Theorem~\ref{thm:converse bound}   within a factor of  $3/2$.  More precisely,  when $m \qsf \leq \Msf \leq (m+1)\qsf$,  the multiplicative gap 
between the achievable load in Corollary~\ref{cor:combined scheme} and the converse bound in Theorem~\ref{thm:converse bound}  is upper bounded by
\begin{itemize}
\item $4/3$, if  $m=1$;
\item   $1-\frac{1}{\Ksf}+\frac{1}{2}$, if $m=2$;
\item   $1-\frac{1}{\Ksf}+\frac{1}{m-1}$, if $m \in [3: \Ksf-3]$;
\item  $1$, if $m  \in \{ \Ksf-2,\Ksf-1\}$. 
\end{itemize}
\end{thm}

\item
Finally,
by directly comparing the minimum load for the centralized data shuffling system (the master-worker framework) in~\eqref{eq:shared-link shuffling achievable bound} with the load  achieved by the combined scheme in Corollary~\ref{cor:combined scheme}, we can quantify the communication cost of peer-to-peer operations  (i.e., the multiplicative gap on the minimum worst-case load under the constraint of uncoded storage between decentralized and centralized data shufflings), which   will be proved in Section~\ref{sub:optimality}.
\begin{cor}
\label{rem:cost}
For a $(\Ksf,\qsf,\Msf)$ decentralized data shuffling system under the constraint of uncoded storage,
the communication cost of peer-to-peer operations     is no more than a factor of  $2$.  More precisely, when $\Ksf\leq 4$,  this cost   is $\frac{\Ksf}{\Ksf-1}$; when $\Ksf \geq 5$ and $m \qsf \leq \Msf \leq (m+1)\qsf$, 
this   cost   is upper bounded by
\begin{itemize}
\item  $\frac{4 \Ksf}{3 (\Ksf-1)}$, if $m=1$;
\item   $1 +\frac{\Ksf}{2(\Ksf-1)}$, if $m=2$; 
\item  $1+\frac{\Ksf}{(m-1)(\Ksf-1)}$, if $m \in [3: \Ksf-3]$;  
\item $\frac{\Ksf}{\Ksf-1}$, if $m  \in \{ \Ksf-2,\Ksf-1\}$. 
\end{itemize}
\end{cor}

\end{enumerate}

\begin{rem}[Comparison  to the direct extension from~\cite{porter2019embeddedindex}]
\label{rem:comparison to direct extension}
As mentioned in Section~\ref{sub:index coding}, the result in~\cite{porter2019embeddedindex} guarantees that from the optimal (linear) centralized data shuffling scheme in~\cite{fundamentalshuffling2018} one can derive a linear scheme for the decentralized setting with     twice the number of transmissions (by the construction given in~\cite[Proof of Theorem 4]{porter2019embeddedindex}), that is, the following storage-load corner points can be achieved,
\begin{align}
\left(
\frac{\Msf}{\qsf}=m, \ 
\frac{\Rsf}{\qsf}=2\frac{\Ksf-m}{m}
\right), \
\forall m\in[\Ksf].
\label{eq:trivial extension}
\end{align}
The multiplicative gap between the  data shuffling scheme in~\eqref{eq:trivial extension} and the proposed converse bound in Theorem~\ref{thm:converse bound}, is $\frac{2(\Ksf-1)}{\Ksf}$, which is close to $2$ when $\Ksf$ is large. 
Our proposed combined scheme in Corollary~\ref{cor:combined scheme} does better:
for $\Ksf \leq 4$, it exactly matches the proposed converse bound in Theorem~\ref{thm:converse bound},
while for $\Ksf >4$ it is order optimal to within a factor of $3/2$.

In addition, the multiplicative gap $\frac{2(\Ksf-1)}{\Ksf}$ is independent of  the storage size $\Msf$. It is shown in Theorem~\ref{thm:order optimality} that    
the multiplicative gap between the combined scheme and the converse decreases towards $1$ when $\Msf$ increases.

Similar observation can be obtained for the communication cost of peer-to-peer operations. With the data shuffling scheme in~\eqref{eq:trivial extension}, we can only prove this cost upper is bounded by $2$, which is independent of $\Msf$ and $\Ksf$. With the combined scheme, it is shown in Corollary~\ref{rem:cost} that this cost decreases towards to $1$ when $\Msf$ and $\Ksf$ increase. 

\end{rem}

\begin{figure}
    \centering
    \begin{subfigure}[t]{0.5\textwidth}
        \centering
        \includegraphics[scale=0.6]{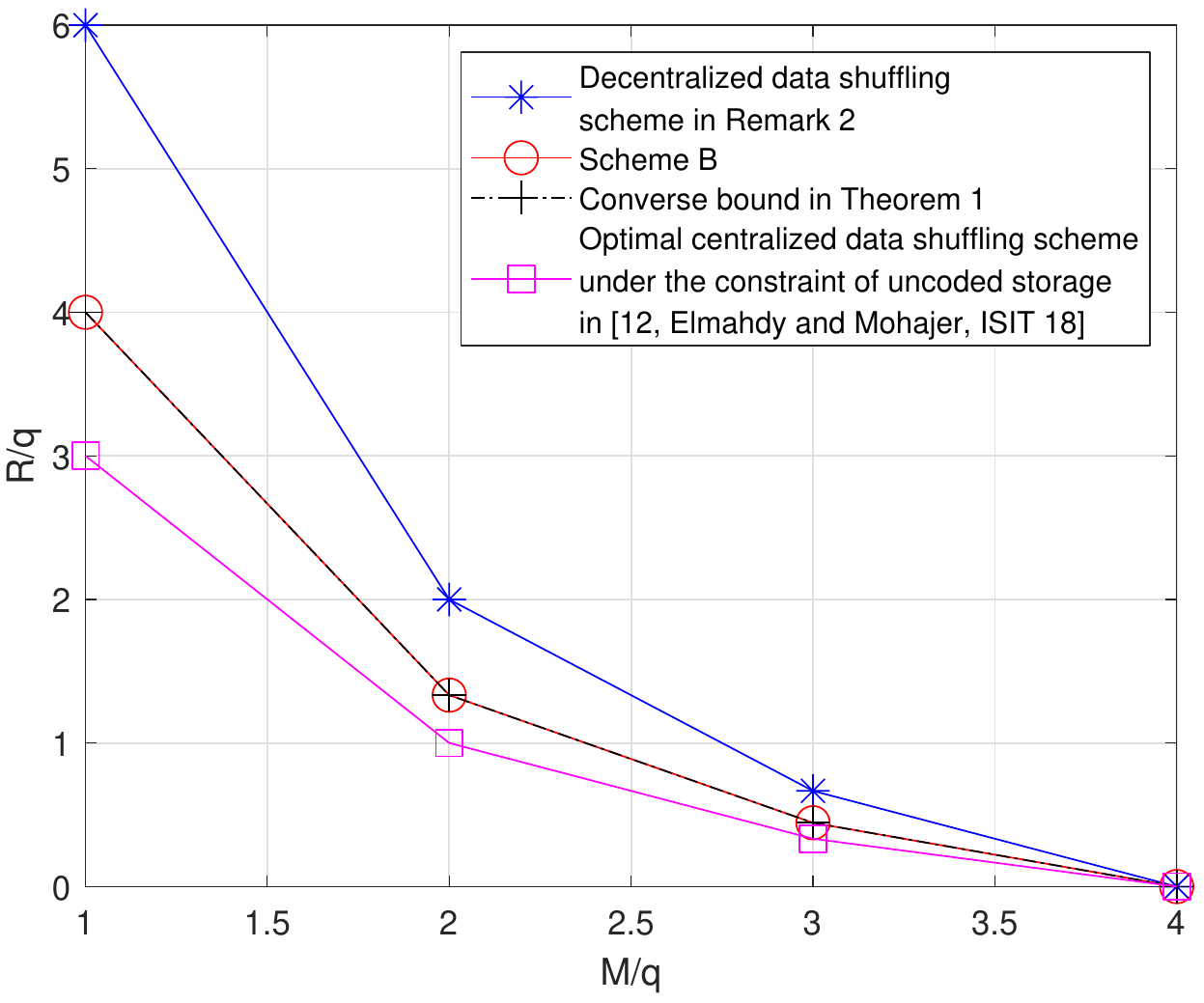}
        \caption{\small $\Ksf=4$.}
        \label{fig:numerical 1a}
    \end{subfigure}%
   \\
    \begin{subfigure}[t]{0.5\textwidth}
        \centering
        \includegraphics[scale=0.6]{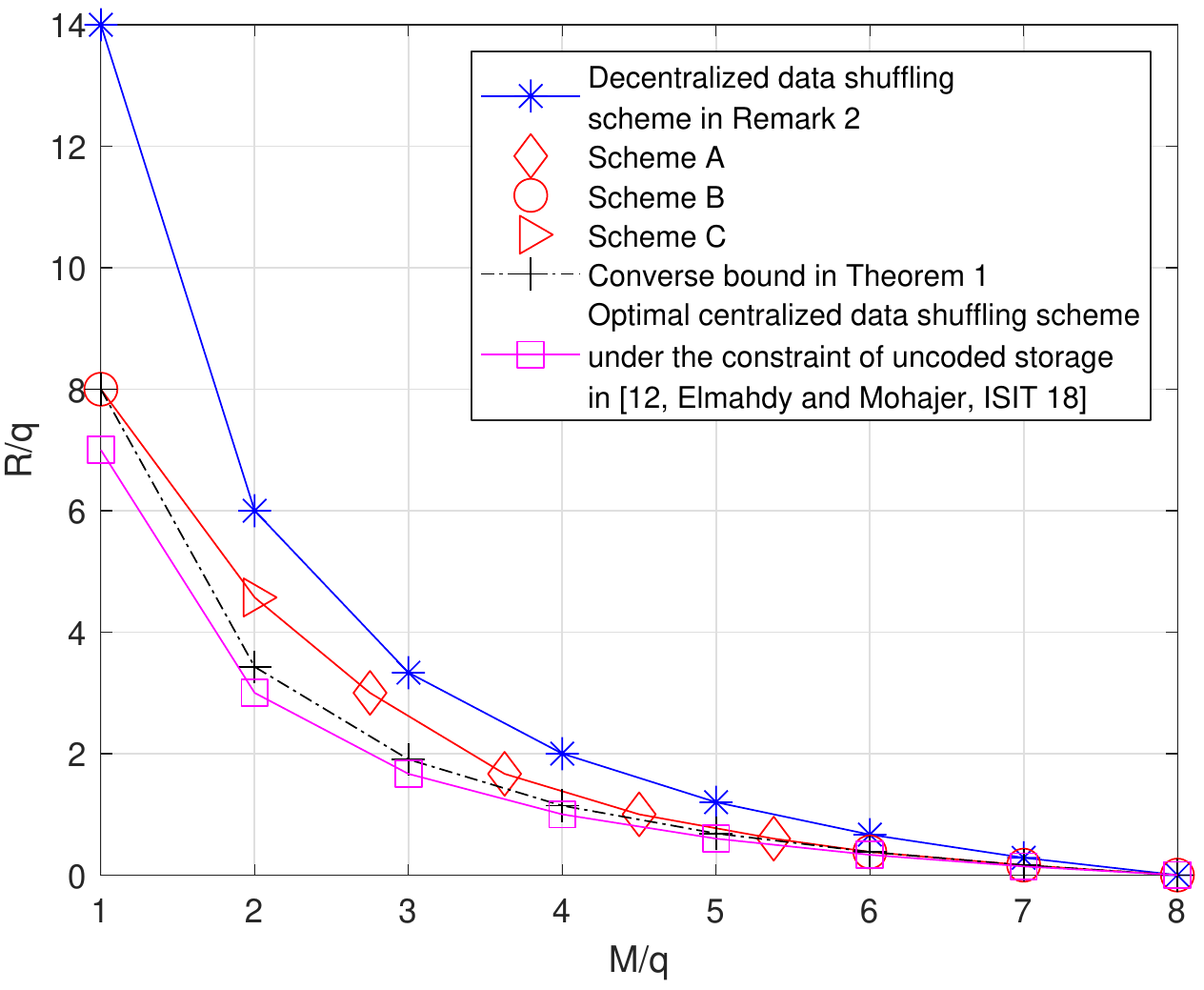}
        \caption{\small $\Ksf=8$.}
        \label{fig:numerical 1b}
    \end{subfigure}
    \caption{\small The storage-load tradeoff for the decentralized data shuffling problem.}
    \label{fig:numerical 1}
\end{figure}

We conclude this section by providing some numerical results. 
Fig.~\ref{fig:numerical 1} plots our converse bound and the best convex combination of the proposed achievable bounds on the worst-case load under the constraint of uncoded storage for the decentralized data shuffling systems with $\Ksf=4$ (Fig.~\ref{fig:numerical 1a}) and $\Ksf=8$ (Fig.~\ref{fig:numerical 1b}) workers. 
For comparison, we also  plot the achieved load by the decentralized data shuffling scheme in Remark~\ref{rem:comparison to direct extension}, and
 the optimal load for the corresponding centralized system in~\eqref{eq:shared-link shuffling converse bound} under the constraint of uncoded storage. 
For the case of $\Ksf=4$ workers, 
Theorem~\ref{thm:converse bound} is tight under the constraint of uncoded storage. 
For the case of $\Ksf=8$ workers, 
Scheme~B meets our converse bound when   $\Msf/\qsf\in [6,8]$, and also trivially when $\Msf/\qsf=1$.

\section{Proof of Theorem~\ref{thm:converse bound}: Converse Bound under the Constraint of Uncoded Storage}
\label{sec:proof of converse bound}
 We want to lower bound $\max_{\Ac^{t}} \sum_{k\in[\Ksf]}\Rsf^{\Ac^{t-1} \to \Ac^{t}}_k$ for a fixed $t\in [ \Tsf ]$ and a fixed $\Ac^{t-1}$. It can be also checked that this lower bound is also a lower bound on the 
 worst-case total communication load in~\eqref{eq:goal} among all $t\in [  \Tsf ]$ and all possible $(\Ac^0,\ldots,\Ac^{\Tsf})$.
Recall that the excess storage is said to be {\it uncoded} if each worker simply copies bits from the data units in its local storage.
When the storage update phase is uncoded, we can divide each data unit into sub-blocks depending on the set of workers who store them. 

\subsection{Sub-block Division of the Data Shuffling Phase under Uncoded Storage} 
\label{subsec:Sub-block Division and Graphical Representation}
Because of the data shuffling constraint in~\eqref{eq:dataset partition}, all the bits of all data units are stored by at least one worker at the end of any time slot. Recall that the worker who must store data unit $F_i$ at the end of time slot $t$ is denoted  by $\usf^t_i$ in~\eqref{eq:def usf^t_i}.
In the case of excess storage, some bits of some files may be stored by multiple workers. 
We denote by $F_{i,\Wc}$ the sub-block of bits of data unit $F_i$ exclusively  stored by workers in $\Wc$ where $i\in[\Nsf]$ and $\Wc \subseteq [\Ksf]$. 
By definition, at the end of step $t-1$, we have that $\usf^{t-1}_{i}$ must be in $\Wc$ for all sub-blocks $F_{i,\Wc}$ of data unit $F_i$; we also let $F_{i,\Wc} = \emptyset$ for all $\Wc\subseteq [\Ksf]$ if $\usf^{t-1}_{i}\not\in\Wc$.
Hence, at the end of step $t-1$, each data unit $F_i$ can be written as
\begin{align}
F_i = \{ F_{i,\Wc} : \Wc \subseteq [\Ksf], \usf_i^{t-1} \in \Wc  \},
\label{eq:part}
\end{align}
and the storage content as
\begin{align}
Z_k^{t-1}
  &= \{ F_{i,\Wc} : \Wc \subseteq [\Ksf], \{\usf_i^{t-1},k\}\subseteq\Wc , i\in [\Nsf]  \}
\notag
\\&= \underbrace{\{ F_{i} :  i\in \Ac^{t-1}_{k} \}}_{\text{required data units}}
\cup \underbrace{\{ F_{i,\Wc} : i\not\in\Ac^{t-1}_{k},\{\usf_i^{t-1},k\}\subseteq\Wc \}}_{\text{other data units}}.
\label{eq:store}
\end{align}
We note that the sub-blocks $F_{i,\Wc}$ have different   content at different times (as the partition in~\eqref{eq:part} is a function of $\Ac^{t-1}$ through $(\usf_1^{t-1},\ldots,\usf_\Nsf^{t-1})$); however, in order not to clutter the notation, we will not explicitly denote the dependance of $F_{i,\Wc}$ on time. Finally, please note that the definition of sub-block $F_{i,\Wc}$, as defined here for the converse bound, is not the same as $G_{i,\Wc}$ defined in Section~\ref{sec:achievable} for the achievable scheme (see Footnote~\ref{foot:GiW}).

\subsection{Proof of Theorem~\ref{thm:converse bound}}
We are interested in deriving an information theoretic lower bound on the worst-case communication load. We will first obtain a number of lower bounds on the load for some carefully chosen shuffles. Since the load of any shuffle is at most as large as the worst-case shuffle, the obtained lower bounds are valid lower bounds for the worst-case load as well. We will then average the obtained lower bounds.

In particular, the shuffles are chosen as follows.
Consider a permutation of $[\Ksf]$ denoted by $\mathbf{d}=(d_1,\ldots,d_{\Ksf})$ where $d_k\neq k$ for each $k\in [\Ksf]$ and consider the shuffle 
\begin{align}
\Ac^{t}_{k}=\Ac^{t-1}_{d_k}, \ \forall k\in [\Ksf].
\label{eq:something:Actk}
\end{align} 

We define   $X^{t}_{\Sc}$ as the messages sent by the workers in $\Sc$ during time slot $t$, that is,
\begin{align}
X^{t}_{\Sc}:=\Big\{ X^{t}_{k}:k\in \Sc \Big\}.
\label{eq:def X^{t+1}_{Sc}}
\end{align} 


From Lemma~\ref{lem:induction lemma} in the Appendix  with $\Sc=[\Ksf]$, which is the key novel contribution of our proof and that was inspired by the induction argument in~\cite{distributedcomputing}, we have
\begin{align}
 &\frac{\Rsf^{\star}_{\mathrm{u}}}{\Bsf} \geq H\Big(  X^{t}_{[\Ksf]} \Big) 
 \nonumber\\&
 \geq 
\sum_{m=1}^{\Ksf} \ \
\sum_{k\in  [\Ksf]} \ \ 
\sum_{i\in \Ac^{t}_{k}} \ \    
\sum_{\Wc\subseteq [\Ksf]\setminus \{k\}: \ \usf^{t-1}_{i}\in\Wc, \ |\Wc|=m}
\frac{|F_{i,\Wc}|}{m}.
\label{eq:K worker nodes lemma}
\end{align}

To briefly illustrate the main  ingredients on the derivation of~\eqref{eq:K worker nodes lemma}, we provide the following example.
\begin{example}[$\Ksf=\Nsf=3$]
\label{ex:converse}
\rm
We focus on the decentralized data shuffling problem with $\Ksf=\Nsf=3$. Without loss of generality, we assume 
\begin{align}
\Ac^{t-1}_{1}=\{3\}, \ \Ac^{t-1}_{2}=\{1\}, \ \Ac^{t-1}_{3}=\{2\}. \label{eq:example shuffle time t-1}
\end{align}
Based on $(\Ac^{t-1}_1,\Ac^{t-1}_2,\Ac^{t-1}_3)$, we can divide each data unit into sub-blocks as in~\eqref{eq:part}. More precisely, we have 
\begin{align*}
&F_1=\{F_{1,\{2\}}, F_{1,\{1,2\}},  F_{1,\{2,3\}}, F_{1,\{1,2,3\}}\};\\
&F_2=\{F_{2,\{3\}}, F_{2,\{1,3\}}, F_{2,\{2,3\}},  F_{2,\{1,2,3\}}\};\\
&F_3=\{F_{3,\{1\}}, F_{3,\{1,2\}}, F_{3,\{1,3\}},  F_{3,\{1,2,3\}}\}.
\end{align*}
At the end of time slot $t-1$, each worker $k\in [3]$ stores $F_{i,\Wc}$ if $k\in \Wc$. Hence, we have 
\begin{align*}
Z_1^{t-1}=&\{F_{1,\{1,2\}},   F_{1,\{1,2,3\}},  F_{2,\{1,3\}},   F_{2,\{1,2,3\}}, \nonumber\\ & F_{3,\{1\}}, F_{3,\{1,2\}}, F_{3,\{1,3\}},  F_{3,\{1,2,3\}} \};\\
Z_2^{t-1}=&\{F_{1,\{2\}}, F_{1,\{1,2\}},  F_{1,\{2,3\}}, F_{1,\{1,2,3\}},  F_{2,\{2,3\}}, \nonumber\\ &  F_{2,\{1,2,3\}},  F_{3,\{1,2\}},   F_{3,\{1,2,3\}} \};\\
Z_3^{t-1}=&\{F_{1,\{1,3\}},   F_{1,\{1,2,3\}}, F_{2,\{3\}}, F_{2,\{1,3\}}, F_{2,\{2,3\}}, \nonumber\\ & F_{2,\{1,2,3\}},   F_{3,\{1,3\}},  F_{3,\{1,2,3\}} \}.
\end{align*}

Now we consider a permutation of $[3]$ denoted by $\mathbf{d}=(d_1,d_2,d_3)$ where $d_k\neq k$ for each $k\in [3]$ and assume the considered permutation is $(2,3,1)$.
Based on $\mathbf{d}$,  from~\eqref{eq:something:Actk}, we consider the shuffle 
\begin{align}
\Ac^{t}_{1}=\Ac^{t-1}_{2}=\{1\}, \ \Ac^{t}_{2}=\Ac^{t-1}_{3}=\{2\}, \ \Ac^{t}_{3}=\Ac^{t-1}_{1}=\{3\}. \label{eq:example shuffle time t}
\end{align}

We first prove 
\begin{align}
H(X^t_{[2]}|Z^{t-1}_{3}, F_{3} )\geq |F_{1, \{2\}}|.\label{eq:12,3}
\end{align}
More precisely, by the decoding constraint in~\eqref{eq:decoding node k}, we have 
\begin{align}
H(F_{1,\{2\}}|Z^{t-1}_1,  X^t_{[3]})=0,
\end{align}
which implies 
\begin{align}
H(F_{1,\{2\}}|Z^{t-1}_1,  X^t_{[3]},Z^{t-1}_{3}, F_{3})=0.\label{eq:12,3 step 1}
\end{align}
Since $ F_{1, \{2\}}$ is not stored by workers $1$ and $3$, we have 
\begin{subequations}
\begin{align}
& |F_{1, \{2\}}|\leq H(F_{1,\{2\}},  X^t_{[2]}|Z^{t-1}_1,  X^t_3,Z^{t-1}_{3}, F_{3} ) \\
 &=H(  X^t_{[2]}|Z^{t-1}_1,  X^t_3,Z^{t-1}_{3}, F_{3} )+ \nonumber\\ &H(F_{1,\{2\}}|Z^{t-1}_1,  X^t_{[3]},Z^{t-1}_{3}, F_{3}) \\
 &=H(  X^t_{[2]}|Z^{t-1}_1,  X^t_3,Z^{t-1}_{3}, F_{3} ) \label{eq:12,3 step 2}\\
 &\leq H(   X^t_{[2]}| X^t_3, Z^{t-1}_{3}, F_{3} )\\
 &= H(   X^t_{[2]}|   Z^{t-1}_{3}, F_{3} ), \label{eq:12,3 step 3}
\end{align}
\end{subequations}
where~\eqref{eq:12,3 step 2} comes from~\eqref{eq:12,3 step 1} and~\eqref{eq:12,3 step 3} comes from   the fact that $ X^t_3$ is a function of $Z^{t-1}_{3}$.  Hence, we prove~\eqref{eq:12,3}.

Similarly, we can also prove
\begin{align}
&H( X^t_{\{1,3\}}|Z^{t-1}_{2}, F_{2} )\geq |F_{3, \{1\}}|.\label{eq:13,2}\\
&H( X^t_{\{2,3\}}|Z^{t-1}_{1}, F_{1} )\geq |F_{2, \{3\}}|.\label{eq:23,1}
\end{align}

In addition, we have 
\begin{subequations}
\begin{align}
&H(X^t_{[3]})=\frac{1}{3} \sum_{k\in [3]} \left( H(X^t_k)+H(X^t_{[3]\setminus \{k\}}| X^t_k) \right)\\
&\geq \frac{1}{3} \left( H(X^t_{[3]})+ \sum_{k\in [3]}  H(X^t_{[3]\setminus \{k\}}| X^t_k) \right),
\end{align}
and thus 
\begin{align}
&2 H(X^t_{[3]}) \geq  \sum_{k\in [3]}  H(X^t_{[3]\setminus \{k\}}| X^t_k) \\
&\geq \sum_{k\in [3]}  H(X^t_{[3]\setminus \{k\}}| Z^{t-1}_k) \label{eq:123 step 1}\\
& = \sum_{k\in [3]}  H(X^t_{[3]\setminus \{k\}}, F_{k}| Z^{t-1}_k) \label{eq:123 step 2}\\
&= \sum_{k\in [3]}  H( F_{k}| Z^{t-1}_k)+ H(X^t_{[3]\setminus \{k\}}| F_{k}, Z^{t-1}_k),  \label{eq:123 step 3} 
\end{align}
where~\eqref{eq:123 step 1} comes from  the fact that  $X^t_k$ is a function of $Z^{t-1}_k$ and conditioning cannot increase entropy, and~\eqref{eq:123 step 2} comes from the decoding constraint for worker $k$ in~\eqref{eq:decoding node k}.
\end{subequations}

Let us focus on worker $1$ and we have 
\begin{align}
H( F_{1}| Z^{t-1}_1)=|F_{1,\{2\}}|+|F_{1,\{2,3\}}|.
\end{align}
In addition, we have $H( X^t_{\{2,3\}}|Z^{t-1}_{1}, F_{1} )\geq |F_{2, \{3\}}|$ from~\eqref{eq:23,1}. Hence,
\begin{align}
&H( F_{1}| Z^{t-1}_1)+H( X^t_{\{2,3\}}|Z^{t-1}_{1}, F_{1} ) \nonumber\\ & \geq |F_{1,\{2\}}|+|F_{1,\{2,3\}}|+|F_{2, \{3\}}|.\label{eq:induction user 1}
\end{align}
Similarly, we have 
\begin{align}
&H( F_{2}| Z^{t-1}_2)+H( X^t_{\{1,3\}}|Z^{t-1}_{2}, F_{2} ) \nonumber\\ & \geq |F_{2,\{3\}}|+|F_{2,\{1,3\}}|+|F_{3, \{1\}}|;\label{eq:induction user 2}\\
&H( F_{3}| Z^{t-1}_3)+H( X^t_{\{1,2\}}|Z^{t-1}_{3}, F_{3} ) \nonumber\\ & \geq |F_{3,\{1\}}|+|F_{3,\{1,2\}}|+|F_{1, \{2\}}|.\label{eq:induction user 3}
\end{align}
By taking~\eqref{eq:induction user 1}-\eqref{eq:induction user 3} to~\eqref{eq:123 step 3}, we have 
\begin{subequations}
\begin{align}
& H(X^t_{[3]}) \geq \frac{1}{2} \sum_{k\in [3]}  H( F_{k}| Z^{t-1}_k)+ H(X^t_{[3]\setminus \{k\}}| F_{k}, Z^{t-1}_k) \\
&= |F_{1, \{2\}}| + |F_{2, \{3\}}|+ |F_{3, \{1\}}| +\frac{|F_{1,\{2,3\}}|}{2}+\frac{|F_{2,\{1,3\}}|}{2}\nonumber\\& +\frac{|F_{3,\{1,2\}}|}{2},
\end{align}
coinciding with~\eqref{eq:K worker nodes lemma}.
\end{subequations}
\hfill$\square$
\end{example}

We now go back to the general proof of Theorem~\ref{thm:converse bound}.
We next consider all the permutations $\mathbf{d}=(d_1,\ldots,d_{\Ksf})$ of $[\Ksf]$ where $d_k\neq k$ for each $k\in[\Ksf]$, and sum together the inequalities in the form of~\eqref{eq:K worker nodes lemma}. 
For an integer $m\in[\Ksf]$, by the symmetry of the problem, the sub-blocks $F_{i,\Wc}$ where $i\in[\Nsf]$, $\usf^{t-1}_{i}\in \Wc$ and $|\Wc|=m$ appear the same number of times in the final sum. In addition, the total number of these sub-blocks in general is $\Nsf\binom{\Ksf-1}{m-1}$ and the total number of such sub-blocks in each inequality in the form of~\eqref{eq:K worker nodes lemma} is $\Nsf\binom{\Ksf-2}{m-1}$. So we obtain
\begin{align}
 \Rsf^{\star}_{\mathrm{u}} 
 &\geq 
 \sum_{m=1}^{\Ksf} \ \
 \sum_{i\in [\Nsf]} \ \ 
 \sum_{\Wc\subseteq [\Ksf]: \ \usf^{t-1}_{i}\in \Wc, \ |\Wc|=m}
 \frac{\binom{\Ksf-2}{m-1}}{m\binom{\Ksf-1}{m-1}}|F_{i,\Wc}|   \frac{\qsf\Ksf}{\Nsf\Bsf}
\\
 &=\sum_{m=1}^{\Ksf}\qsf\Ksf \ x_m \frac{1-(m-1)/(\Ksf-1)}{m}
\label{eq:main constraint}
\\
 &=\sum_{m=1}^{\Ksf}\qsf \ x_m\frac{\Ksf-m}{m}\frac{\Ksf}{\Ksf-1}
\label{eq:main constraint in theorem form},
 \end{align}
where we defined $x_m$ as the total number of bits in the sub-blocks stored by $m$ workers at the end of time slot $t-1$ normalized by the total number of bits $\Nsf\Bsf$, i.e.,
\begin{align}
0\leq x_m:=
\sum_{i\in[\Nsf]} \ \
\sum_{\Wc\subseteq [\Ksf]: \ \usf^{t-1}_{i}\in \Wc, \ |\Wc|=m}
\frac{|F_{i,\Wc}|}{\Nsf\Bsf},
\label{eq:def of xt}
\end{align}
which must satisfy
\begin{align}
\sum_{m\in[\Ksf]}x_m&=1 \ \text{(total size of all data units)},
\label{eq:file size}
\\
\sum_{m\in[\Ksf]}m \, x_m &\leq \frac{\Ksf\Msf}{\Nsf}=\frac{\Msf}{\qsf} \ \text{(total storage size)}.
\label{eq:cache size}
\end{align}

We then use a method based on Fourier-Motzkin elimination~\cite[Appendix D]{networkinformation} for to bound $\Rsf^{\star}_{\mathrm{u}}$ from~\eqref{eq:main constraint} under the constraints in~\eqref{eq:file size} and~\eqref{eq:cache size},  as we did in~\cite{ontheoptimality} for coded caching with uncoded cache placement.
In particular, for each integer $p\in[\Ksf]$,  we multiply~\eqref{eq:file size} by $\frac{-\Nsf(2\Ksf p- p^2+\Ksf-p)}{p(p+1)}$ to obtain 
\begin{align}
\frac{-\Nsf(2\Ksf p- p^2+\Ksf-p)}{p(p+1)}=\sum_{m=1}^{\Ksf}\frac{-\Nsf(2\Ksf p- p^2+\Ksf-p)}{p(p+1)} x_m,\label{eq:from file size}
\end{align}
and we  multiply~\eqref{eq:cache size} by $\frac{\Nsf \Ksf}{(\Ksf-1) p (p+1)}$ to have
\begin{align}
\frac{\Nsf \Ksf}{(\Ksf-1) p (p+1)} \frac{\Ksf\Msf}{\Nsf} \geq \sum_{m=1}^{\Ksf}\frac{\Nsf \Ksf}{(\Ksf-1) p (p+1)} m x_m.\label{eq:from cache size}
\end{align}
We then add~\eqref{eq:from file size},~\eqref{eq:from cache size}, and~\eqref{eq:main constraint} to obtain,
\begin{align}
\Rsf^{\star}_{\mathrm{u}}&\geq \sum_{m=1}^{\Ksf} \frac{\Nsf \Ksf (p-m)(p+1-m)}{m (\Ksf-1)p (p+1)} x_m - \frac{\Nsf \Ksf}{(\Ksf-1) p (p+1)} \frac{\Ksf\Msf}{\Nsf}\nonumber\\&+\frac{\Nsf(2\Ksf p- p^2+\Ksf-p)}{p(p+1)}\\
&\geq - \frac{\Nsf \Ksf}{(\Ksf-1) p (p+1)} \frac{\Ksf\Msf}{\Nsf}+\frac{\Nsf(2\Ksf p- p^2+\Ksf-p)}{p(p+1)}.\label{eq:final converse bound}
\end{align}
Hence, for each integer $p\in[\Ksf]$, the bound in~\eqref{eq:final converse bound} becomes a linear function in $\Msf$. 
When $\Msf=\qsf p$, from~\eqref{eq:final converse bound}  we have $\Rsf^{\star}_{\mathrm{u}}\geq \frac{\Nsf (\Ksf-p)}{(\Ksf-1)p}$. When $\Msf=\qsf (p+1)$,  from~\eqref{eq:final converse bound}  we have $\Rsf^{\star}_{\mathrm{u}}\geq \frac{\Nsf(\Ksf-p-1)}{(\Ksf-1)(p+1)}$. In conclusion, we prove that $\Rsf^{\star}_{\mathrm{u}}$ is lower bounded by the lower convex envelope (also referred to as ``memeory sharing'') of the points $\left(\Msf=\qsf m,\Rsf=\frac{\Nsf (\Ksf-m)}{(\Ksf-1)m}\right)$, where $m\in[\Ksf]$. 

This concludes the proof of Theorem~\ref{thm:converse bound}.

\subsection{Discussion}
\label{subsec:discussionconvese}
 
We conclude this session  the following remarks:
\begin{enumerate}

\item
The corner points from the converse bound are of the form $\left(\Msf/\qsf= m,\Rsf/\qsf=\frac{\Ksf\binom{\Ksf-2}{m-1}}{m\binom{\Ksf-1}{m-1}}\right)$, which may suggest the following placement.

At the end of time slot $t-1$,
each data unit is partitioned into $\binom{\Ksf-1}{m-1}$ equal-length sub-blocks of length $\Bsf/\binom{\Ksf-1}{m-1}$ bits as $F_{i}=(F_{i,\Wc}:\Wc\subseteq [\Ksf], |\Wc|=m, \usf^{t-1}_{i}\in\Wc  )$; by definition $F_{i,\Wc}=\emptyset$ if either $\usf^{t-1}_{i} \notin \Wc$ or $|\Wc|\not=m$.
Each worker $k\in[\Ksf]$ stores all the sub-blocks $F_{i,\Wc}$ if $k\in\Wc$; in other words, worker $k\in [\Ksf]$ stores all the $\binom{\Ksf-1}{m-1}$ sub-blocks of the desired data units, and $\binom{\Ksf-2}{m-2}$ sub-blocks of the  remaining data units. 

In the data shuffling phase of time slot $t$, worker $k\in [\Ksf]$ must decode the missing $\binom{\Ksf-1}{m-1} - \binom{\Ksf-2}{m-2} = \binom{\Ksf-2}{m-1}$ sub-blocks of data unit $F_j$ for all $j\in \Ac^{t}_{k} \backslash  \Ac^{t-1}_{k}$. An interpretation of the converse bound is that, in the worst case, the total number of transmissions is equivalent to at least $\frac{\qsf \Ksf}{m} \binom{\Ksf-2}{m-1}$ sub-blocks.

We will use this interpretation to design the storage update phase 
our proposed Schemes~B and~C.

\item
The converse bound is derived for the objective of minimizing the ``sum load'' $\sum_{k\in[\Ksf]}\Rsf^{\Ac^{t-1} \to \Ac^{t}}_{k}$, see~\eqref{eq:goal}. 

The same derivation would give a converse bound for the ``largest individual load'' $\max_{k\in[\Ksf]}\Rsf^{\Ac^{t-1} \to \Ac^{t}}_{k}$. In the latter case, the corner points from converse bound are of the form $\left(\Msf/\qsf= m,\Rsf/\qsf=\frac{\binom{\Ksf-2}{m-1}}{m\binom{\Ksf-1}{m-1}}\right)$. This view point may suggest that, in the worst case, all the individual loads $\Rsf^{\Ac^{t-1} \to \Ac^{t}}_{k}$ are the same, i.e., the burden of communicating missing data units is equally shared by all the workers.

Our proof technique for Theorem~\ref{thm:converse bound} could also be directly extended to derive a converse bound on the {\it average load} (as opposed to the worst-case load) for all the possible shuffles in the decentralized data shuffling problem when $\Nsf=\Ksf$.

\end{enumerate}

\section{Achievable Schemes for Decentralized Data Shuffling}
\label{sec:achievable}
In this section, we propose three schemes for the decentralized data shuffling problem, and analyze their performances.

\subsection{Scheme~A in Theorem~\ref{thm:Achievable scheme A}}
\label{sub:achievable a}
Scheme~A extends the general centralized data shuffling scheme in Section~\ref{sec:distributed vs shared-link data shuffling} to the distributed model. Scheme~A achieves the load in Theorem~\ref{thm:Achievable scheme A} for each storage size $\Msf=\big(1+g\frac{\Ksf-1}{\Ksf} \big)\qsf$, where $g\in[\Ksf-1]$; the whole storage-load tradeoff piecewise curve is achieved by memory-sharing\footnote{\label{foot:memory sharing}
 Memory-sharing  is an achievability technique originally proposed by Maddah-Ali and Niesen  in~\cite{dvbt2fundamental} for coded caching systems,  which is used to extend achievability in between discrete memory points.
More precisely, focus one storage size $\Msf^{\prime}=\alpha \Msf_1 +(1-\alpha)\Msf_2$, where  $\alpha \in [0,1]$, $\Msf_1=\big(1+g\frac{\Ksf-1}{\Ksf} \big)\qsf$, $\Msf_2= \big(1+(g+1)\frac{\Ksf-1}{\Ksf} \big)\qsf$, and $g\in [\Ksf-1]$. We can divide each data unit into two non-overlapping parts,with $\alpha \Bsf$ and $(1-\alpha)\Bsf$ bits, respectively. The first and second parts of the $\Nsf$ data units are stored and transmitted based on the proposed data shuffling scheme for $\Msf_1 $ and $\Msf_2 $, respectively.} between these points  (given in~\eqref{eq:distributed shuffling achievable bound 1}) and the (trivially achievable) points in~\eqref{eq:distributed shuffling achievable bound 1 trivial M=q}-\eqref{eq:distributed shuffling achievable bound 1 trivial M=Kq}.

\paragraph*{Structural Invariant Data Partitioning and Storage}
This is the same as the one in Section~\ref{sec:distributed vs shared-link data shuffling} for the centralized case. 

\paragraph*{Initialization}
The master directly transmits all data units. 
The storage is as in~\eqref{eq:ravi scheme 1 storageII} given  $\Ac^0$.

\paragraph*{Data Shuffling Phase of time slot $t\in [T]$}
The data shuffling phase is inspired by the delivery in D2D caching~\cite{d2dcaching}.
Recall the definition of sub-block $G^{\prime}_{k,\Wc}$ in~\eqref{eq:ravi scheme 1 storageIII}, where each sub-block is known by $|\Wc|=g$ workers and needed by worker $k$. Partition $G^{\prime}_{k,\Wc}$ into $g$ non-overlapping and equal-length pieces $G^{\prime}_{k,\Wc} =\{ G^{\prime}_{k,\Wc}(j) : j\in\Wc\}$.
Worker $j\in \Jc$ broadcasts 
\begin{align}
W^{t}_{j,\Jc} = \underset{k\in\Jc\setminus\{j\}}{\oplus} G^{\prime}_{k,\Jc\setminus\{k\}}(j), \ \ 
\text{$\forall\Jc\subseteq [\Ksf]$ where $|\Jc|=g+1$},
\label{eq:d2d general multicast}
\end{align}
in other words, one linear combination $W^{t}_{\Jc}$ in~\eqref{eq:ravi scheme 1 X} for the centralized setting becomes $g+1$ linear combinations $W^{t}_{j,\Jc}$ in~\eqref{eq:d2d general multicast} for the decentralized setting, but of size reduced by a factor $g$.
Evidently, each sub-block in $W^{t}_{j,\Jc}$ is stored in the storage of worker  $j$ at the end of time slot $t-1$. In addition, 
each worker  $k\in \Jc\setminus\{j\}$ knows $G^{\prime}_{k_1,\Jc\setminus\{k_1\}}(j)$ where $k_1\in\Jc\setminus \{k,j\}$ such that it can recover its desired block $G^{\prime}_{k,\Jc\setminus\{k\}}(j)$.

Since $|G^{\prime}_{k, \Wc}|\leq \qsf\Bsf/\binom{\Ksf}{g}$, the worst-case  load is 
\begin{align}
\Rsf
\leq 
     \qsf \frac{(g+1)\binom{\Ksf}{g+1}}{g\binom{\Ksf}{g}}
   = \qsf \frac{\Ksf-g}{g} =: \Rsf_\text{\rm Ach.A},
\label{eq:Ach.A}
\end{align}
as claimed in Theorem~\ref{thm:Achievable scheme A}, where the subscript ``$\text{\rm Ach.A}$'' in~\eqref{eq:Ach.A} denotes the worst-case load achieved by   Scheme~A.

\paragraph*{Storage Update Phase of time slot $t\in [T]$}
The storage update phase is the same as the general centralized data shuffling scheme in Section~\ref{sec:distributed vs shared-link data shuffling}, and thus is not repeated here.

\subsection{Scheme~B in Theorem~\ref{thm:Achievable scheme B}}
\label{sub:achievable b}

During the data shuffling phase in time slot $t$ of  Scheme~A, we treat some sub-blocks known by $g+1$ workers as if they were only known by $g$ workers (for example, if $\usf^{t-1}_i \notin \Wc$, $G_{i,\Wc}$ is stored by workers $\{\usf^{t-1}_i\}\cup \Wc$, but Scheme~A treats $G_{i,\Wc}$ as if it was only stored by workers in $\Wc$), which may be suboptimal as more multicasting opportunities may be leveraged. In the following, we propose Scheme~B to remedy for this shortcoming for $\Msf=m\qsf$ for $m\in \{\Ksf-2,\Ksf-1\}$.

\paragraph*{Structural Invariant Data Partitioning and Storage} 
Data units are partitions as inspired by the converse bound (see discussion in Section~\ref{subsec:discussionconvese}), which is as in the improved centralized data shuffling scheme in~\cite{neartoptimalAttia2018}.  
Fix $m\in[\Ksf]$.
Partition each data unit into $\binom{\Ksf-1}{m-1}$ non-overlapping equal-length sub-block of length $\Bsf/\binom{\Ksf-1}{m-1}$ bits.
Write $F_{i}=(F_{i,\Wc}:\Wc\subseteq [\Ksf], |\Wc|=m, \usf^{t}_{i}\in\Wc  )$, and set $F_{i,\Wc}=\emptyset$ if either $\usf^{t}_{i} \notin \Wc$ or $|\Wc|\not=m$.
The storage of worker $k\in[\Ksf]$ at the end of time slot $t$  is as follows,
\begin{align}
Z_k^t
  &=  \Big(
\underbrace{(F_{i,\Wc} : i\in \Ac_k^t, \forall \Wc)}_\text{required data units} \cup 
\underbrace{(F_{i,\Wc} : i\not\in \Ac_k^t, k \in\Wc)}_\text{other data units}
 \Big),
\label{eq:ravi scheme 2 storageI}
\end{align}
that is, worker $k\in [\Ksf]$ stores all the $\binom{\Ksf-1}{m-1}$ sub-blocks of data unit $F_i$ if $i\in \Ac^{t}_{k}$, and $\binom{\Ksf-2}{m-2}$ sub-blocks of data unit $F_j$ if $j\notin \Ac^{t}_{k}$   (the sub-blocks stored are such that $k\in\Wc$), thus the required storage space is
\begin{align}
\Msf = \qsf+(\Nsf-\qsf)\frac{\binom{\Ksf-2}{m-2}}{\binom{\Ksf-1}{m-1}}
     = \qsf+(m-1)\frac{\Nsf-\qsf}{\Ksf-1}
     = m\qsf.
\end{align}
In the following, we shall see that it is possible to maintain the storage structure in~\eqref{eq:ravi scheme 2 storageI} after the shuffling phase.

\paragraph*{Initialization}
The master directly transmits all data units. 
The storage is as in~\eqref{eq:ravi scheme 2 storageI} given  $\Ac^0$.

\paragraph*{Data Shuffling Phase of time slot $t\in [T]$ for $m=\Ksf-1$} 
Each data unit has been partitioned into $\binom{\Ksf-1}{m-1}=\Ksf-1$ sub-blocks and each sub-block is stored by $m=\Ksf-1$ workers.
Similarly to Scheme~A, define the set of sub-blocks needed by worker $k\in[\Ksf]$ at time slot $t$ and not previously stored as
\begin{align}
F^{\prime}_{k,[\Ksf]\setminus\{k\}} 
= \negmedspace\Big( \negmedspace F_{i,\Wc} :  i\in \Ac^{t}_{k}\setminus \Ac^{t-1}_{k}, \Wc = [\Ksf]\setminus\{k\} \negmedspace \Big) \negmedspace, 
  \forall k\in[\Ksf].
\label{eq:ravi scheme 2 storageIII}
\end{align}
Since $F^{\prime}_{k,[\Ksf]\setminus\{k\}}$ (of length $\qsf\Bsf/(\Ksf-1)$ bits in the worst case) is desired by worker $k$ and known by all the remaining   $m=\Ksf-1$ workers, we partition $F^{\prime}_{k,[\Ksf]\setminus\{k\}}$ into $m=\Ksf-1$ pieces (of length $\qsf\Bsf/(\Ksf-1)^2$ bits in the worst case), and write $F^{\prime}_{k,[\Ksf]\setminus\{k\}} =\big( F^{\prime}_{k,[\Ksf]\setminus\{k\}}(j) : j\in[\Ksf]\setminus \{k\} \big)$.
Worker $j\in[\Ksf]$ broadcasts the single linear combination (of length $\qsf\Bsf/(\Ksf-1)^2$ bits in the worst case) given by
\begin{align}
W^{t}_{j} 
= \underset{k\not=j}{\oplus} F^{\prime}_{k,[\Ksf]\setminus\{k\}}(j). 
\label{eq:d2d general multicast 2}
\end{align}

Therefore, the worst-case satisfies
\begin{align}
\Rsf 
\leq \frac{\Ksf}{(\Ksf-1)^2} \qsf 
= \left.\frac{\Ksf-m}{m}\frac{\Ksf}{\Ksf-1}\qsf \right|_{m=\Ksf-1} 
\negmedspace=:\left.\Rsf_\text{\rm Ach.B}\right|_{\Msf=(\Ksf-1)\qsf}
\end{align}
which coincides with the converse bound.

\paragraph*{Storage Upadte Phase of time slot $t\in [T]$ for $m=\Ksf-1$}
In time slot $t-1>0$, we assume that the above storage configuration of each worker $k\in[\Ksf]$ can be done with $Z^{t-2}_k$ and $X^{t-1}_j$ where $j\in [\Ksf]\setminus \{k\}$. We will show  next that at the end of time slot $t$, we can re-create the same configuration of storage, but with permuted data units. Thus by the induction method, we prove the above storage update phase is also structural invariant.

For each worker $k\in [\Ksf]$ and each data unit $F_i$ where $i\in \Ac^{t}_{k}\setminus \Ac^{t-1}_k$, worker  $k$ stores the whole data unit $F_{i}$ in its storage. For each data unit $F_i$ where $i\in \Ac^{t-1}_k\setminus \Ac^{t}_{k}$, instead of storing the whole data unit $F_i$, worker  $k$ only stores the bits of $F_i$ which was stored at the end of time slot $t-1$ by worker  $\usf^{t}_{i}$. For other data units, worker  $k$ does not change the stored bits.
Hence, after the storage  phase in time slot $t$, we can re-create the same configuration of storage as the end of time slot $t-1$ but with permuted data units.

\paragraph*{Data Shuffling Phase of time slot $t\in [T]$ for $m=\Ksf-2$} 
We partition the $\Nsf$ data units into $\qsf$ 
groups as $[\Nsf]= \underset{i\in[\qsf]}{\cup} \Hc_i$, where each group contains $\Ksf$ data units, and such that  
for each group $\Hc_i, i\in[\qsf],$ and  each worker $k\in[\Ksf]$ we have $|\Hc_i\cap \Ac^{t}_{k}|=1$ and $|\Hc_i\cap \Ac^{t-1}_{k}|=1$. 
In other words, the partition is such that, during the data shuffling phase of time slot $t$, among all the $\Ksf$ data units in each group,
each worker requests exactly one data unit  and  knows exactly one data unit. 
Such a partition can always be found~\cite[Lemma 7]{fundamentalshuffling2018}.
The dependance of $\Hc_i$ on $t$ is not specified so not to clutter the notation.
%
For group $\Hc_i, i\in[\qsf],$ we define
\begin{align}
\Uc(\Hc_i) := \{k\in[\Ksf] : \Hc_i\cap \Ac^{t}_{k} \subseteq \Ac^{t-1}_{k}  \}, \quad \forall i\in[\qsf],\label{eq:uhi}
\end{align}
as the set of workers in the group who already have stored the needed data point (i.e., who do not need to shuffle).
Since each worker has to recover at most one data unit in each group, the delivery in each group is as if $\qsf=1$. Hence, to simplify the description, we focus on the case $\qsf=1$, 
in which case there is only one group and thus we simplify the notation $\Uc(\Hc_i)$ to just $\Uc$.
We first use the following example to illustrate the main idea.

\begin{example}
\label{ex:example scheme b}
\rm
Consider the $(\Ksf,\qsf,\Msf)=(5,1,3)$ decentralized data shuffling problem, where $m=\Msf/\qsf=3$.
Let $\Ac^{t-1}=(5,1,2,3,4)$. 
During the storage update phase in time slot $t-1$, we partition each data unit into $6$ equal-length sub-blocks, each of which has $\Bsf/6$ bits, as
\begin{subequations}
\begin{align}
F_1=& (F_{1,\{1,2,3\}}, F_{1,\{1,2,4\}}, F_{1,\{1,2,5\}}, F_{1,\{2,3,4\}},F_{1,\{2,3,5\}}, \nonumber \\& F_{1,\{2,4,5\}}\}, \label{eq:division M=5 m=3  F1}\\
F_2=& (F_{2,\{1,2,3\}}, F_{2,\{1,3,4\}}, F_{2,\{1,3,5\}}, F_{2,\{2,3,4\}},F_{2,\{2,3,5\}},\nonumber \\& F_{2,\{3,4,5\}}\}, \label{eq:division M=5 m=3  F2}\\
F_3=& (F_{3,\{1,2,4\}}, F_{3,\{1,3,4\}}, F_{3,\{1,4,5\}}, F_{3,\{2,3,4\}},F_{3,\{2,4,5\}},\nonumber \\& F_{3,\{3,4,5\}}\}, \label{eq:division M=5 m=3  F3}\\
F_4=& (F_{4,\{1,2,5\}}, F_{4,\{1,3,5\}}, F_{4,\{1,4,5\}}, F_{4,\{2,3,5\}},F_{4,\{2,4,5\}},\nonumber \\& F_{4,\{3,4,5\}}\}, \label{eq:division M=5 m=3  F4}\\
F_5=& (F_{5,\{1,2,3\}}, F_{5,\{1,2,4\}}, F_{5,\{1,2,5\}}, F_{5,\{1,3,4\}},F_{5,\{1,3,5\}},\nonumber \\& F_{5,\{1,4,5\}}\}, \label{eq:division M=5 m=3  F5}
\end{align}
\label{eq:division M=5 m=3}
\end{subequations}
and each worker $k$ stores $F_{i,\Wc}$ if $k\in \Wc$.

In the following, we consider various shuffles in time slot $t$. 
If one sub-block is stored by some worker in $\Uc$, we let this worker transmit it and the transmission is equivalent to   centralized data shuffling; otherwise, we will introduce the  proposed scheme B to transmit it.

We first consider $\Ac^{t}=(1,2,3,4,5)$. 
For each set $\Jc\subseteq [\Ksf]$ of size $|\Jc|=m+1=\Ksf-1=4$, 
we generate  
\begin{align}
V^{t}_{\Jc}=\underset{k\in \Jc}{\oplus}F_{d^t_k,\Jc\setminus\{k\}},
\label{eq:K-3 multicast}
\end{align}
where $d^t_k$ represents the demanded data unit of worker $k$ in time slot $t$ if $\qsf=1$.
The details are illustrated in Table~\ref{tab:example scheme b}. For example, when $\Jc=\{1,2,3,4\}$, we have $  V_{\{1,2,3,4\}}^{t} =          F_{1,\{2,3,4\}} +F_{2,\{1,3,4\}}+F_{3,\{1,2,4\}}+ F_{4,\{1,2,3\}}$ where $F_{4,\{1,2,3\}}$ is empty and we replace $F_{4,\{1,2,3\}}$
by $\Bsf/6$ zero bits. Since $ F_{1,\{2,3,4\}}$, $F_{2,\{1,3,4\}}$, and $F_{3,\{1,2,4\}}$ are all stored by worker $4$, $V_{\{1,2,3,4\}}^{t}$ can be transmitted by worker $4$. Similarly, we let worker $3$ transmit $ V_{\{1,2,3,5\}}^{t}$, worker $2$ transmit $V_{\{1,2,4,5\}}^{t}$, worker $1$ transmit $V_{\{1,3,4,5\}}^{t}$, and worker $5$ transmit $V_{\{2,3,4,5\}}^{t}$. It can be checked that each worker can recover its desired sub-blocks and the achieved load is $5/6$, which coincides with the proposed converse bound.

Let us then focus on $\Ac^{t}=(5,2,3,4,1)$. For this shuffle, $\Ac^{t-1}_1=\Ac^t_1$ such that worker $1$ needs not to decode anything from what the other workers transmit. 
We divide all desired sub-blocks into two sets, stored and not stored by worker $1$ as follows 
\begin{align*}
\Sc_{\{1\}}&=\{F_{1,\{1,2,3\}},F_{1,\{1,2,4\}},F_{2,\{1,3,4\}},F_{2,\{1,3,5\}},F_{3,\{1,2,4\}},\nonumber\\ & F_{3,\{1,4,5\}},F_{4,\{1,2,5\}},F_{4,\{1,3,5\}}\}, 
\\
\Sc_{\emptyset}&=\{F_{1,\{2,3,4\}},F_{2,\{3,4,5\}},F_{3,\{2,4,5\}},F_{4,\{2,3,5\}}\}.
\end{align*}
Since the sub-blocks in $\Sc_{\{1\}}$ are all stored by worker $1$, we can treat worker $1$ as a central server and the transmission of $\Sc_{\{1\}}$ is equivalent to centralized data shuffling with $\Ksf_{\textrm{eq}}=4$, $\Msf_{\textrm{eq}}=2$ and $\qsf_{\textrm{eq}}=1$, where $\Uc_{\textrm{eq}}=\emptyset$. For this centralized problem, the data shuffling schemes in~\cite{neartoptimalAttia2018,fundamentalshuffling2018} are optimal under the constraint of uncoded storage. Alternatively, we can also use the following simplified scheme. By generating $V_{\{1,2,3,4\}}^{t}$ as in~\eqref{eq:K-3 multicast}, and it can be seen that  $V_{\{1,2,3,4\}}^{t}$ is known by workers $1$ and $4$. Similarly,  $V_{\{1,2,3,5\}}^{t}$ is known by workers $1$ and $3$, $V_{\{1,2,4,5\}}^{t}$ is known by workers $1$ and $2$, and $V_{\{1,3,4,5\}}^{t}$ is known by workers $1$ and $5$. Hence, we can let worker $1$ 
transmit $V_{\{1,2,3,4\}}^{t} \oplus V_{\{1,2,3,5\}}^{t}$, $V_{\{1,2,3,4\}}^{t}\oplus  V_{\{1,2,4,5\}}^{t}$, and $V_{\{1,2,3,4\}}^{t}\oplus V_{\{1,3,4,5\}}^{t}$.
Hence, each worker can recover $V_{\{1,2,3,4\}}^{t}$,  $V_{\{1,2,3,5\}}^{t}$, $V_{\{1,2,4,5\}}^{t}$, and $V_{\{1,3,4,5\}}^{t}$.
We then consider the transmission for $\Sc_{\emptyset}=\{F_{1,\{2,3,4\}},F_{2,\{3,4,5\}},F_{3,\{2,4,5\}},F_{4,\{2,3,5\}}\}$, which is equivalent to decentralized data shuffling with $\Ksf_{\textrm{eq}}=4$, $\Msf_{\textrm{eq}}=3$ and $\qsf_{\textrm{eq}}=1$, where $\Uc_{\textrm{eq}}=\emptyset$ defined in~\eqref{eq:uhi}. Hence, we can use the proposed Scheme B for  $m=\Ksf-1$. More precisely, we split each sub-block in $V_{\{2,3,4,5\}}^{t}$ into $3$ non-overlapping and equal-length sub-pieces, e.g., $F_{2,\{3,4,5\}}=\{F_{2,\{3,4,5\}}(3),F_{2,\{3,4,5\}}(4),F_{2,\{3,4,5\}}(5)\}$. We then let 
\begin{align*}
&\textrm{worker $2$ transmit } F_{3,\{2,4,5\}}(2)\oplus F_{4,\{2,3,5\}}(2)\oplus F_{1,\{2,3,4\}}(2);\\
&\textrm{worker $3$ transmit } F_{2,\{3,4,5\}}(3)\oplus F_{4,\{2,3,5\}}(3)\oplus F_{1,\{2,3,4\}}(3);\\
&\textrm{worker $4$ transmit } F_{2,\{3,4,5\}}(4)\oplus F_{3,\{2,4,5\}}(4)\oplus F_{1,\{2,3,4\}}(4);\\ 
&\textrm{worker $5$ transmit } F_{2,\{3,4,5\}}(5)\oplus F_{3,\{2,4,5\}}(5)\oplus F_{4,\{2,3,5\}}(5).  
\end{align*}
In conclusion, the total load for $\Ac^{t}=(5,2,3,4,1)$ is $\frac{3}{6}+\frac{2}{9}=\frac{13}{18}$.

Finally, we consider $\Ac^t=\{5,1,3,4,2\}$. For this shuffle, $\Ac^{t-1}_1=\Ac^t_1$ and  $\Ac^{t-1}_2=\Ac^t_2$  such that workers $1$ and $2$ need not to decode anything from other workers.   
We  divide all desired sub-blocks into three sets
$$
\Sc_{\{1,2\}} =\{F_{2,\{1,2,3\}},F_{3,\{1,2,4\}},F_{4,\{1,2,5\}}\},
$$
stored by workers~1 and~2,
$$
\Sc_{\{1\}}   =\{F_{2,\{1,3,4\}},F_{3,\{1,4,5\}},F_{4,\{1,3,5\}}\},
$$
stored by worker $1$ and not by worker $2$, and
$$
\Sc_{\{2\}}   =\{F_{2,\{2,3,4\}},F_{3,\{2,4,5\}},F_{4,\{2,3,5\}}\} 
$$
stored by worker $2$ and not by worker $1$.

The transmission for $\Sc_{\{1,2\}}$ is equivalent to a centralized data shuffling with $\Ksf_{\textrm{eq}}=3$, $\Msf_{\textrm{eq}}=1$ and $\qsf_{\textrm{eq}}=1$. We use the following simplified scheme to let worker $1$ transmit 
$V_{\{1,2,3,4\}}^{t}\oplus V_{\{1,2,3,5\}}^{t}$ and $V_{\{1,2,3,4\}}^{t} \oplus V_{\{1,2,4,5\}}^{t}$
 such that each worker can recover  $V_{\{1,2,3,4\}}^{t}$, $V_{\{1,2,3,5\}}^{t}$, and $V_{\{1,2,4,5\}}^{t}$ (as illustrated in Table~\ref{tab:example scheme b}, $V_{\{1,2,3,4\}}^{t}=F_{3,\{1,2,4\}}$, $V_{\{1,2,3,5\}}^{t}=F_{2,\{1,2,3\}}$, and $V_{\{1,2,4,5\}}^{t}=F_{4,\{1,2,5\}}$). Similarly, for  $\Sc_{\{1\}}$, we let worker $1$ transmit $V_{\{1,3,4,5\}}^{t}$. For $\Sc_{\{2\}}$, we let worker $2$ transmit $V_{\{2,3,4,5\}}^{t}$. In conclusion the total load for  $\Ac^t=\{5,1,3,4,2\}$ is $\frac{2}{6}+\frac{1}{6}+\frac{1}{6}=\frac{2}{3}$.
\hfill$\square$
\end{example}

\begin{table}
\centering
\protect\caption{Multicast messages for Example~\ref{ex:example scheme b}. Empty sub-blocks are colored in magenta.}
\label{tab:example scheme b}
\begin{align*}
\tiny
\begin{array}{| l |}
\hline
   \text{For $\Ac^{t}=(           1,2,3,4,5)=(F_1,F_2,F_3,F_4,F_5)$}\\
\hline
  \text{Worker $1$ wants } (F_{1,\{2,3,4\}},F_{1,\{2,3,5\}},F_{1,\{2,4,5\}},{\mgt F_{1,\{3,4,5\}}=\emptyset})  
\\
  \text{Worker $2$ wants } (F_{2,\{1,3,4\}},F_{2,\{1,3,5\}},F_{2,\{3,4,5\}},{\mgt F_{2,\{1,4,5\}}=\emptyset})       
\\
  \text{Worker $3$ wants }(F_{3,\{1,2,4\}},F_{3,\{1,4,5\}},F_{3,\{2,4,5\}},{\mgt F_{3,\{1,2,5\}}=\emptyset})                 
\\
  \text{Worker $4$ wants }(F_{4,\{1,2,5\}},F_{4,\{1,3,5\}},F_{4,\{2,3,5\}},{\mgt F_{4,\{1,2,3\}}=\emptyset})                     
\\
\text{Worker $5$ wants }(F_{5,\{1,2,3\}},F_{5,\{1,2,4\}},F_{5,\{1,3,4\}},{\mgt F_{5,\{2,3,4\}}=\emptyset})           
\\
\hline
   V_{\{1,2,3,4\}}^{t} =          F_{1,\{2,3,4\}} +F_{2,\{1,3,4\}}+F_{3,\{1,2,4\}}+{\mgt F_{4,\{1,2,3\}}} 
\\ 
   V_{\{1,2,3,5\}}^{t} =         F_{1,\{2,3,5\}} +F_{2,\{1,3,5\}}+{\mgt F_{3,\{1,2,5\}}}       +F_{5,\{1,2,3\}}     
\\ 
   V_{\{1,2,4,5\}}^{t}  =       F_{1,\{2,4,5\}} +{\mgt F_{2,\{1,4,5\}}}           +F_{4,\{1,2,5\}}+F_{5,\{1,2,4\}}   
\\ 
   V_{\{1,3,4,5\}}^{t}  =     {\mgt F_{1,\{3,4,5\}}}           +F_{3,\{1,4,5\}}+F_{4,\{1,3,5\}}+F_{5,\{1,3,4\}}    
\\ 
   V_{\{2,3,4,5\}}^{t} =          F_{2,\{3,4,5\}}+F_{3,\{2,4,5\}}+F_{4,\{2,3,5\}}+{\mgt F_{5,\{2,3,4\}}}     
\\
\hline\hline
 \text{For $\Ac^{t}=({\magenta 5},2,3,4,1)=(F_5,F_2,F_3,F_4,F_1)$} \\ 
\hline
 \text{Worker $1$ wants }{\mgt (F_{5,\{2,3,4\}},F_{5,\{2,3,5\}},F_{5,\{2,4,5\}},F_{5,\{3,4,5\}})=\emptyset}  
\\
 \text{Worker $2$ wants } (F_{2,\{1,3,4\}},F_{2,\{1,3,5\}},F_{2,\{3,4,5\}},{\mgt F_{2,\{1,4,5\}}=\emptyset})      
\\
 \text{Worker $3$ wants }(F_{3,\{1,2,4\}},F_{3,\{1,4,5\}},F_{3,\{2,4,5\}},{\mgt F_{3,\{1,2,5\}}=\emptyset})                 
\\
 \text{Worker $4$ wants }(F_{4,\{1,2,5\}},F_{4,\{1,3,5\}},F_{4,\{2,3,5\}},{\mgt F_{4,\{1,2,3\}}=\emptyset})                      
\\
\text{Worker $5$ wants }(F_{1,\{1,2,3\}},F_{1,\{1,2,4\}},F_{1,\{2,3,4\}},{\mgt F_{1,\{1,3,4\}}=\emptyset})           
\\
\hline
 V_{\{1,2,3,4\}}^{t} =   {\mgt F_{5,\{2,3,4\}}}+F_{2,\{1,3,4\}}+F_{3,\{1,2,4\}}+{\mgt F_{4,\{1,2,3\}}}        
\\ 
  V_{\{1,2,3,5\}}^{t}= {\mgt F_{5,\{2,3,5\}}}+F_{2,\{1,3,5\}}+{\mgt F_{3,\{1,2,5\}}}         +F_{1,\{1,2,3\}}  
\\ 
  V_{\{1,2,4,5\}}^{t}  = {\mgt F_{5,\{2,4,5\}}}+{\mgt F_{2,\{1,4,5\}}}           +F_{4,\{1,2,5\}}+F_{1,\{1,2,4\}}   
\\ 
 V_{\{1,3,4,5\}}^{t}  = {\mgt F_{5,\{3,4,5\}}}         +F_{3,\{1,4,5\}}+F_{4,\{1,3,5\}}+{\mgt F_{1,\{1,3,4\}}} 
\\ 
  V_{\{2,3,4,5\}}^{t}=       F_{2,\{3,4,5\}}+F_{3,\{2,4,5\}}+F_{4,\{2,3,5\}}+F_{1,\{2,3,4\}} 
\\
\hline\hline
 \text{For $\Ac^{t}=({\magenta 5,1},3,4,2)=(F_5,F_1,F_3,F_4,F_2)$}   \\ 
 \hline
 \text{Worker $1$ wants } {\mgt (F_{5,\{2,3,4\}},F_{5,\{2,3,5\}},F_{5,\{2,4,5\}},F_{5,\{3,4,5\}})=\emptyset}   
\\
 \text{Worker $2$ wants }{\mgt (F_{1,\{1,3,4\}},F_{1,\{1,3,5\}},F_{1,\{3,4,5\}},F_{1,\{1,4,5\}})=\emptyset}  
\\
 \text{Worker $3$ wants }(F_{3,\{1,2,4\}},F_{3,\{1,4,5\}},F_{3,\{2,4,5\}},{\mgt F_{3,\{1,2,5\}}=\emptyset})        
\\
 \text{Worker $4$ wants } (F_{4,\{1,2,5\}},F_{4,\{1,3,5\}},F_{4,\{2,3,5\}},{\mgt F_{4,\{1,2,3\}}=\emptyset})                
\\
 \text{Worker $5$ wants }(F_{2,\{1,2,3\}},F_{2,\{1,3,4\}},F_{2,\{2,3,4\}},{\mgt F_{2,\{1,2,4\}}=\emptyset})    
\\
\hline
  V_{\{1,2,3,4\}}^{t}  =    {\mgt F_{5,\{2,3,4\}}}+{\mgt F_{1,\{1,3,4\}}}+F_{3,\{1,2,4\}}+{\mgt F_{4,\{1,2,3\}}}     
\\ 
 V_{\{1,2,3,5\}}^{t}= {\mgt F_{5,\{2,3,5\}}}+{\mgt F_{1,\{1,3,5\}}}+{\mgt F_{3,\{1,2,5\}}}  +F_{2,\{1,2,3\}} 
\\ 
  V_{\{1,2,4,5\}}^{t} = {\mgt F_{5,\{2,4,5\}}}+{\mgt F_{1,\{1,4,5\}}}           +F_{4,\{1,2,5\}}+{\mgt F_{2,\{1,2,4\}}}  
\\ 
 V_{\{1,3,4,5\}}^{t}  = {\mgt F_{5,\{3,4,5\}}}           +F_{3,\{1,4,5\}}+F_{4,\{1,3,5\}}+F_{2,\{1,3,4\}}   
\\ 
  V_{\{2,3,4,5\}}^{t}=  {\mgt F_{1,\{3,4,5\}}}+F_{3,\{2,4,5\}}+F_{4,\{2,3,5\}}+F_{2,\{2,3,4\}}
\\
\hline
\end{array}
\end{align*}
\end{table}

Now we are ready to introduce Scheme B for $m=\Ksf-2$ as a generalization of Example~\ref{ex:example scheme b}.
Recall that, from our earlier discussion, we can consider without loss of generality $\qsf=1$, and that $\Uc$ represents the set of workers who need  not to recover anything from others. 
We divide all desired sub-blocks by all workers into non-overlapping sets
\begin{align}
\Sc_{\Kc} \negmedspace := \negmedspace \{F_{d_{k},\Wc}\negmedspace :\negmedspace  k\in [\Ksf]\setminus \Uc, \negmedspace |\Wc|\negmedspace = \negmedspace m+1,\negmedspace \Wc\cap \Uc=\Kc, \negmedspace k\notin \Wc\},  
\label{eq:def Sck scheme b}
\end{align} 
where $\Kc\subseteq \Uc$.
We then encode the sub-blocks in each set in $\Sc_{\Kc}$ in~\eqref{eq:def Sck scheme b} as follows: 
\begin{itemize}

\item
For each $\Kc\subseteq \Uc$ where $\Kc \neq \emptyset$, the transmission for $\Sc_{\Kc}$ is equivalent to a centralized data shuffling problem with $\Ksf_{\textrm{eq}}=\Ksf-|\Uc|$, $\qsf_{\textrm{eq}}=1$ and $\Msf_{\textrm{eq}}=m-|\Kc|$, where $\Uc_{\textrm{eq}}=\emptyset$.
It can be seen that $\Ksf_{\textrm{eq}}-\Msf_{\textrm{eq}}\leq 2$. Hence, we can use the optimal centralized data shuffling schemes in~\cite{neartoptimalAttia2018,fundamentalshuffling2018}. 
 
Alternatively, we propose the following simplified scheme.  For each set  $\Jc\subseteq [\Ksf]$ of size $|\Jc|=m+1=\Ksf-1$, where $\Jc\cap \Uc= \Kc$, we generate $V^{t}_{\Jc}$ as in~\eqref{eq:K-3 multicast}. Each sub-block in $\Sc_{\Kc}$ appears in one   $V^{t}_{\Jc}$,  where  $\Jc\subseteq [\Ksf]$, $|\Jc|=\Ksf-1$ and $\Jc\cap \Uc= \Kc$.
It can be seen that for each worker $j \in [\Ksf]\setminus \Uc$, among all  $V^{t}_{\Jc}$  where  $\Jc\subseteq [\Ksf]$, $|\Jc|=\Ksf-1$ and $\Jc\cap \Uc= \Kc$, worker $j$ knows one of them (which is $V^{t}_{[\Ksf]\setminus \{\usf^{t-1}_{d^t_{j}}\}}$). We denote all sets  $\Jc\subseteq [\Ksf]$ where $|\Jc|=\Ksf-1$ and $\Jc\cap \Uc= \Kc$, by $\Jc_1(\Kc)$, $\Jc_2(\Kc)$, \ldots, $\Jc_{\binom{\Ksf-|\Uc|}{\Ksf-1-|\Kc|}}(\Kc)$. For $\Sc_{\Kc}$, we choose one worker in $\Kc$ to transmit  $V^{t}_{\Jc_1(\Kc)} \oplus V^{t}_{\Jc_2(\Kc)}$, \ldots, $V^{t}_{\Jc_1(\Kc)} \oplus V^{t}_{\Jc_{\binom{\Ksf-|\Uc|}{\Ksf-1-|\Kc|}}(\Kc)}$, such that   each worker in $\Ksf\setminus \Uc$ can recover 
all $V^{t}_{\Jc}$  where  $\Jc\subseteq [\Ksf]$, $|\Jc|=\Ksf-1$ and $\Jc\cap \Uc= \Kc$.

\item
For $\Kc=\emptyset$, the transmission for $\Sc_{\Kc}$ is equivalent to  decentralized data shuffling with $\Ksf_{\textrm{eq}}=\Ksf-|\Uc|$, $\qsf_{\textrm{eq}}=1$ and $\Msf_{\textrm{eq}}=m=\Ksf-2$, where $\Uc_{\textrm{eq}}=\emptyset$.
Hence, in this case $\Uc\leq 2$. 

If $|\Uc|=2$, we have $\Msf_{\textrm{eq}}=\Ksf_{\textrm{eq}}$ and thus we  do not transmit anything for $\Sc_{\emptyset}$. 

If $|\Uc|=1$, we have $\Msf_{\textrm{eq}}=\Ksf_{\textrm{eq}}-1$ and thus we can use Scheme B for $m=\Ksf-1$ to transmit $\Sc_{\emptyset}$. 

Finally, we consider $|\Uc|=\emptyset$.  For each set $\Jc\subseteq [\Ksf]$ where $|\Jc|=m+1=\Ksf-1$, among all the workers in $\Jc$, there is exactly one worker  in $\Jc$ where $\usf^{t-1}_{d^t_{k} }\notin \Jc$ (this worker is assumed to be $k$ and we have $\usf^{t-1}_{d^t_{k}}=[\Ksf]\setminus\Jc$ with a slight abuse of notation). We then let worker $k$ transmit $V^{t}_{\Jc} $. 

\end{itemize}
In conclusion,   by comparing the loads for different cases, the worst-cases are when $\Ac^{t-1}_k\cap \Ac^t_k=\emptyset$ for each $k\in [\Ksf]$ and the worst-case load achieved by
 Scheme~B is 
\begin{align}
 \qsf \Ksf/\binom{\Ksf-1}{\Ksf-2}=\left. \qsf  \frac{\Ksf-m}{m}\frac{\Ksf}{\Ksf-1}\right|_{m=\Ksf-1} 
=:\left.\Rsf_\text{\rm Ach.B}\right|_{\Msf=(\Ksf-2)\qsf},
\end{align}
which coincides with the converse bound.

\paragraph*{Storage Update Phase of time slot $t\in [T]$ for $m=\Ksf-2$}
The storage  update phase for  $m=\Ksf-2$ is the same as the one of scheme B for  $m=\Ksf-1$.

\begin{rem}[Scheme~B realizes distributed interference alignment]
\label{rem:dia e scheme B}
In Example~\ref{ex:example scheme b}, from $\Ac^{t-1}=(5,1,2,3,4)$ to $\Ac^{t}=(1,2,3,4,5)$, by the first column of Table~\ref{tab:example scheme b}, we see that each worker  desires $\Ksf-2=3$ of the sub-blocks that need to be shuffled.
Since each worker cannot benefit from its own transmission, we see that the best possible scheme would have each worker recover its $\Ksf-2=3$ desired sub-blocks from the $\Ksf-1=4$ ``useful transmissions,'' that is, the unwanted sub-blocks should ``align'' in a single transmission, e.g., for worker $1$, all of its unwanted sub-blocks are aligned in $V^t_{\{2,3,4,5\}}$.
From the above we see that this is exactly what happens for each worker when $\Ksf-2=m$.  
How to realize distributed interference alignment seems to be a key question in decentralized data shuffling. 
\hfill$\square$
\end{rem}

\begin{rem}[Extension of Scheme~B to other storage sizes]
\label{rem:extension of achievable scheme B}
We can extend Scheme B to any storage size by the following three steps:
\begin{itemize}

\item 
We partition the $\Nsf$ data units into $\qsf$ groups, where each group contains $\Ksf$ data units, and such that  
during the data shuffling phase each worker requests exactly one data unit  and  knows exactly one data unit  among all the $\Ksf$ data units in each group.

\item 
For  each group $\Hc_i$, we partition all desired sub-blocks by all workers into sets depending on which workers in $\Uc(\Hc_i)$ know them. Each set is denoted by $\Sc_{\Kc}(\Hc_i)$, which is known by workers in $\Kc\subseteq \Uc(\Hc_i)$, and is defined similarly to~\eqref{eq:def Sck scheme b}.

\item
For each set $\Sc_{\Kc}(\Hc_i)$, 
\begin{itemize}

\item
if $\Kc\neq \emptyset$, the transmission is equivalent to centralized data shuffling with $\Ksf_{\textrm{eq}}=\Ksf-|\Uc(\Hc_i)|$, $\qsf_{\textrm{eq}}=1$ and $\Msf_{\textrm{eq}}=\Msf-|\Kc|$. We can use the optimal centralized data shuffling scheme in~\cite{fundamentalshuffling2018}; 

\item
if $\Kc=\emptyset$, for each set $\Jc \subseteq ([\Ksf]\setminus \Uc(\Hc_i))$, where $|\Jc|=m+1$, we generate the 
multicast messages $V_{\Jc}^{t}$ as defined in~\eqref{eq:K-3 multicast}.

If there  exists some empty sub-block in $V_{\Jc}^{t}$, we let the worker who demands this sub-block transmit  $V_{\Jc}^{t}$. 

Otherwise, $V_{\Jc}^{t}$ is transmitted as Scheme B for $m=\Ksf-1$.
\end{itemize}

\end{itemize}

 Unfortunately, a general closed-form expression for the load in this general case is not available as it heavily depends on the shuffle. 

Note: in Example~\ref{ex:improved example general scheme b} next we show the transmission for $\Kc=\emptyset$, can be further improved by random linear combinations.
\hfill$\square$
\end{rem}

\begin{example}
\label{ex:improved example general scheme b}
\rm  
Consider the $(\Ksf,\qsf,\Msf)=(5,1,2)$ decentralized data shuffling problem, where $m=\Msf/\qsf=2$.
From the outer bound we have $\Rsf_\text{u}^\star \geq 15/8$; if each sub-block is of size $1/\binom{\Ksf-1}{m-1}=1/4$, 
the outer bound suggests that we need to transmit at least $15/2=7.5$ sub-blocks in total.

Let $\Ac^{t-1}=(5,1,2,3,4)$. 
During the storage update phase in time slot $t-1$, we partition each data unit into $4$ equal-length sub-blocks, each of which has $\Bsf/4$ bits, as
\begin{subequations}
\begin{align}
&F_1= (F_{1,\{1,2\}}, F_{1,\{2,3\}}, F_{1,\{2,4\}}, F_{1,\{2,5\}}\}, \label{eq:division M=5 m=2  F1}\\
&F_2= (F_{2,\{1,3\}}, F_{2,\{2,3\}}, F_{2,\{3,4\}}, F_{2,\{3,5\}}\}, \label{eq:division M=5 m=2  F2}\\
&F_3= (F_{3,\{1,4\}}, F_{3,\{2,4\}}, F_{3,\{3,4\}}, F_{3,\{4,5\}}\}, \label{eq:division M=5 m=2  F3}\\
&F_4= (F_{4,\{1,5\}}, F_{4,\{2,5\}}, F_{4,\{3,5\}}, F_{4,\{4,5\}}\}, \label{eq:division M=5 m=2  F4}\\
&F_5= (F_{5,\{1,2\}}, F_{5,\{1,3\}}, F_{5,\{1,4\}}, F_{5,\{1,5\}}\}, \label{eq:division M=5 m=2  F5}
\end{align}
\label{eq:division M=5 m=2}
\end{subequations}
and each worker $k$ stores $F_{i,\Wc}$ if $k\in \Wc$.

Let $\Ac^{t}=(1,2,3,4,5)$. 
During the data shuffling phase in time slot $t$, each worker must recover $3$ sub-blocks of the desired data unit which it does not store, e.g., 
worker~$1$ must recover $(F_{1,\{2,3\}}, F_{1,\{2,4\}}, F_{1,\{2,5\}})$, 
worker~$2$ must recover $(F_{2,\{1,3\}}, F_{2,\{3,4\}}, F_{2,\{3,5\}})$, etc.

For each set $\Jc\subseteq [\Ksf]$ where $|\Jc|=m+1=3$, we generate $V^{t}_{\Jc}=\underset{k\in\Jc}{\oplus}F_{k,\Jc\setminus\{k\}}$ as in~\eqref{eq:K-3 multicast}. More precisely, we have 
\begin{subequations}
\begin{align}
V^{t}_{\{1,2,3\}}&=F_{1,\{2,3\}} \oplus F_{2,\{1,3\}}, \ \text{(can be sent by worker~3)},\\ 
V^{t}_{\{1,2,4\}}&=F_{1,\{2,4\}}, \\ 
V^{t}_{\{1,2,5\}}&=F_{1,\{2,5\}} \oplus F_{5,\{1,2\}}, \ \text{(can be sent by worker~2)}, \\
V^{t}_{\{1,3,4\}}&=F_{3,\{1,4\}}, \\ 
V^{t}_{\{1,3,5\}}&=F_{5,\{1,3\}}, \\ 
V^{t}_{\{1,4,5\}}&=F_{4,\{1,5\}} \oplus F_{5,\{1,4\}}, \ \text{(can be sent by worker~1)}, \\
V^{t}_{\{2,3,4\}}&=F_{2,\{3,4\}} \oplus F_{3,\{2,4\}}, \ \text{(can be sent by worker~4)}, \\ 
V^{t}_{\{2,3,5\}}&=F_{2,\{3,5\}}, \\ 
V^{t}_{\{2,4,5\}}&=F_{4,\{2,5\}}, \\
V^{t}_{\{3,4,5\}}&=F_{3,\{4,5\}} \oplus F_{4,\{3,5\}}, \ \text{(can be sent by worker~5)}.
\end{align}
\label{eq:MAN M=5 m=2}
\end{subequations}

We deliver these  
multicast messages with a two-phase scheme, as follows.
\begin{itemize}

\item
Phase~1.
It can be seen that 
the multicast messages like $V^{t}_{\{1,2,3\}}$ in~\eqref{eq:MAN M=5 m=2} (which is known by worker $3$ only) 
can be sent by one specific worker.
Similarly, we can let workers $2$, $1$, $4$ and $5$ broadcast $V^{t}_{\{1,2,5\}}$, $V^{t}_{\{1,4,5\}}$,  $V^{t}_{\{2,3,4\}}$ and $V^{t}_{\{3,4,5\}}$, respectively.

\item
Phase~2.  
After the above Phase~1, the remaining messages are known by two workers.
For example, $V^{t}_{\{1,2,4\}}=F_{1,\{2,4\}}$ is known by workers~$2$ and~$4$; we can let worker $2$ transmit  $V^{t}_{\{1,2,4\}}$.
If we do so as Scheme B, since each multicast message in~\eqref{eq:MAN M=5 m=2} has $\Bsf/4$ bits and there are $10$ multicast messages in~\eqref{eq:MAN M=5 m=2}, the total load is $10/4$.
 
  In this example Phase~2 of Scheme B can be improved as follows.
The problem with the above strategy (i.e., assign each 
multicast message to a worker) is that we have not leveraged the fact that, after Phase~1, there are still five sub-blocks to be delivered (one demanded per worker, namely $F_{1,\{2,4\}},F_{2,\{3,5\}},F_{3,\{1,4\}},F_{4,\{2,5\}},F_{5,\{1,3\}}$), each of which is known by two workers. Therefore, we can form random linear combinations so that each worker can recover all three of the unstored sub-blocks. In other words, if a worker receivers from each of the remaining $\Ksf-1=4$ workers $3/4 \times \text{``size of a sub-block''}$ linear equations then it solved for the three missing sub-blocks, that is, each worker broadcasts $\frac{3\Bsf}{16}$ random linear combinations of all bits of the two sub-blocks he stores among $F_{1,\{2,4\}},F_{2,\{3,5\}},F_{3,\{1,4\}},F_{4,\{2,5\}},F_{5,\{1,3\}}$. It can be  checked that for worker $1$, the received $3\Bsf/4$ random linear combinations from other workers are linearly independent known $F_{1,\{2,4\}}$ and $F_{5,\{1,3\}}$ as $\Bsf\to\infty$, such that it can recover all these five sub-blocks. By the symmetry, each other worker can also recover these five sub-blocks.

\end{itemize} 

In conclusion, the total load with this two-phase is $5(1+3/4) \times 1/4 =\frac{35}{16}<10/4$, which is achieved by Scheme B.
By comparison, the load of Scheme~A is $\frac{27}{8}$ and the converse bound under the constraint of uncoded storage   in Theorem~\ref{thm:converse bound} is $\frac{15}{8}$.

As a final remark, note that the five sub-blocks in Phase~2 is symmetric, i.e., the number of sub-blocks stored by each worker is the same and the one known by each worker is also the same. In general case, this symmetry may not hold (thus the conditional linearly independent of the received random linear combinations by each worker may not hold), and the generalization of this scheme is part of ongoing work.
\hfill$\square$
\end{example}

\begin{rem}[Limitation of Scheme~B for small storage size]
\label{rem:limitation of achievable scheme B}
The data shuffling phase with uncoded storage can be represented by a directed graph, where each sub-block demanded by some worker is represented by a node in the graph. A directed edge exists from node $a$ to node $b$ in the graph if the worker demanding the data represented by node $b$ has the data represented by node $a$ in its storage. 
By generating a directed graph for the data shuffling phase as described in Section~\ref{sec:model}, we can see that each 
 multicast message in~\eqref{eq:K-3 multicast} is the 
sum of the sub-blocks contained in a clique, where a clique is a set of nodes where each two of them are connected in two directions.   The sub-blocks in  each 
multicast message in~\eqref{eq:K-3 multicast} also form a distributed clique, where a distributed clique is a clique whose nodes are all known by some worker.

Consider the case where $\Ksf=\Nsf$ is much larger than $\Msf=m=2$ (i.e., small storage size regime). 
Consider a ``cyclic shuffle'' of the form $\Ac^{t-1}_{1}=\{\Ksf\}$ and $\Ac^{t-1}_{k}=\{k-1\}$ for $k\in [2:\Ksf]$, to $\Ac^{t}_{k}=\{k\}$ for $k\in [\Ksf]$. Each data unit is split into $\binom{\Ksf-1}{m-1}=\Ksf-1$ sub-blocks and  each worker needs to recover $\binom{\Ksf-2}{m-1}=\Ksf-2$ sub-blocks. 

If during the data shuffling phase in time slot $t$ we generate the 
multicast messages as above, 
only $2$ of the $\Ksf-2$ demanded sub-blocks are in a distributed clique of size $m=2$. More precisely,
let us focus on worker $1$ who needs to recover $F_1$. Notice that each sub-block of $F_1$ is stored by worker $2$, 
and each of its demanded sub-blocks $F_{1,\{2,j\}}$ where $j\in[\Ksf]\setminus \{1,2\}$, is in the 
 multicast message 
\begin{align}
V^t_{\{1,2,j\}}=F_{1,\{2,j\}}\oplus F_{2,\{1,j\}} \oplus F_{j,\{1,2\}}.
\end{align}
When $j=3$, it can be seen that $F_{3,\{1,2\}}$ is empty because all sub-blocks of $F_3$ are stored by worker $4$,
and thus $V^t_{\{1,2,3\}}=F_{1,\{2,3\}}\oplus F_{2,\{1,3\}}$ could be transmitted by worker $3$. 
However, when $j=4$, both sub-blocks $F_{2,\{1,4\}}$ and $F_{4,\{1,2\}}$ are empty, and thus $V^t_{\{1,2,4\}} = F_{1,\{2,4\}}$. 
Similarly, among all the $\Ksf-2$ demanded sub-blocks by worker $1$, only $F_{1,\{2,3\}}$ and $F_{1,\{2,\Ksf\}}$ are in cliques including two sub-blocks, while the remaining $\Ksf-4$ ones are in cliques including only one sub-block.

If the delivery strategy is to assign each 
multicast message, or distributed clique, to a worker, we see that most of the distributed cliques
have a multicast coding gain of $1$ (where the multicast coding gain is the gain on the transmitted load compared to uncoded/direct transmission, e.g., if we transmit one sub-block in a distributed clique of length $2$, the multicast coding gain to transmit this sub-block is $2$). 
Hence, Scheme~B is generally inefficient for $m\in [2:\Ksf-3]$. 
In this paper we mainly use Scheme~B for $m\in \{\Ksf-2,\Ksf-1,\Ksf\}$ for which  it is   optimal under the constraint of uncoded storage.
\hfill$\square$
\end{rem}

\subsection{Scheme~C in Theorem~\ref{thm:Achievable Scheme C}}
\label{sub:achievable c}
To overcome the limitation of Scheme~B described in Remark~\ref{rem:limitation of achievable scheme B}, in the following we propose Scheme~C for $\Msf/\qsf=m=2$ based on an {\it unconventional   distributed clique-covering} strategy for the two-sender distributed index coding problems  proposed in~\cite{thapa2017chandra}.  

The storage update phase of Scheme~C is the same as Scheme~B, which is structural invariant, and thus we only describe the transition from time slot $t-1$ to time slot $t$. 
The main idea is not to use the conventional distributed clique-covering method which transmits distributed cliques (e.g, 
the multicast messages in Scheme~B),
because most of the distributed 
cliques only contain one sub-block, 
and most sub-blocks are not in any  distributed 
clique including more than one sub-blocks,
as explained in Remark~\ref{rem:limitation of achievable scheme B}. 
Instead, 
we propose a novel decentralized data shuffling scheme
to increase the efficiency (coded multicasting gain) of the transmissions. The main idea is that  we search for cliques with length $m=2$; 
if this clique is not a distributed clique, we add one node to the clique and transmit the three nodes by two binary sums, 
such that each node can be recovered by the corresponding user;
if this clique is a distributed clique we proceed as in Scheme~B.

\begin{figure}
\centerline{\includegraphics[scale=0.4]{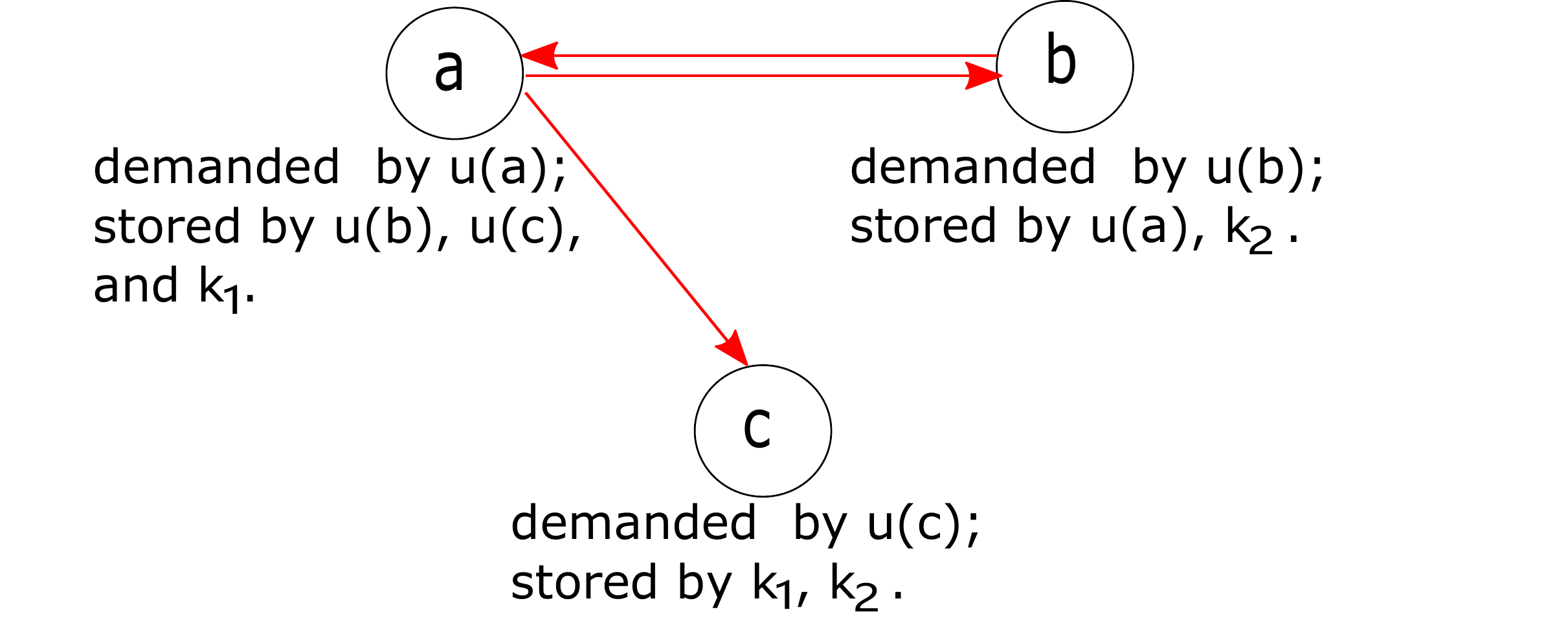}}
\caption{\small Directed graph of the two-sender ($k_1$ and $k_2$) distributed index problem in Proposition~\ref{prop:distributed clique covering}. A direct edge from node $a$ to node $b$, means that receiver $u(b)$ demanding message $b$ stores message $a$ demanded by receiver $u(a)$.} 
\label{fig:propo1}
\end{figure}

Before introducing the details of Scheme C,
we first recall the unconventional clique-covering method for the two-sender distributed index coding problems proposed in~\cite[Theorem 8, Case 33]{thapa2017chandra}. 
\begin{prop}[Unconventional Distributed Clique-covering in~\cite{thapa2017chandra}]
\label{prop:distributed clique covering}
In a distributed index coding problem with two senders (assumed to be $k_1$ and $k_2$), as illustrated Fig.~\ref{fig:propo1}, there are three messages $a$ (demanded by receiver $u(a)$), $b$ (demanded by receiver $u(b)$), and $c$ (demanded by receiver $u(c)$), where  $u(a)$ stores $b$,   $u(b)$ stores $a$, and  $u(c)$ stores $a$. Sender $k_1$ stores $a$ and $c$, and sender $k_2$ stores  $b$ and $c$.
 We can let worker $k_1$ transmit $a\oplus c$ and worker $k_2$ transmit $b\oplus c$, such that workers $u(a)$, $u(b)$ and $u(c)$ can recover node $a$, $b$ and $c$, respectively.

Indeed, receiver $u(a)$ knows  $b$ such that it can recover   $c$, and then recover  $a$. Similarly receiver $u(b)$ can recover  $b$. User $u(c)$ knows  $a$, such that it can recover  $c$   from $a\oplus c$.
\end{prop}


 In the decentralized data shuffling problem, each data unit is divided into sub-blocks depending on which subset of workers stored them before the data shuffling phase; each sub-block desired by a worker is an independent message in the corresponding distributed index coding problem; thus the data shuffling phase is a $\Ksf$-sender distributed index coding problem that contains a number of messages that in general is doubly exponential in the number of workers in the original decentralized data shuffling problem. Hence, it is   non-trivial to use  Proposition~\ref{prop:distributed clique covering} in the decentralized data shuffling problem.


We then illustrate the main idea of Scheme C by means of an example.

\begin{example}
\label{ex:achievable c}
\rm
We consider the same example as Remark~\ref{rem:extension of achievable scheme B}, where $\Ksf=5$, $\qsf=1$ and $\Msf=2$. 
Let $\Ac^{t-1}_{1}=\{5\}$ and $\Ac^{t-1}_{k}=\{k-1\}$ for $k\in [2:5]$, and $\Ac^{t}_{k}=\{k\}$ for $k\in [5]$.
The data units are split as in~\eqref{eq:division M=5 m=2}.

In the data shuffling phase of time slot $t$,
the distributed clique-covering method in Scheme~B has many sub-blocks which are not in any  distributed 
clique including more than one sub-block (e.g., $F_{1,\{2,4\}}$), as explained in Remark~\ref{rem:limitation of achievable scheme B}.
Moreover, the sub-blocks $F_{1,\{2,3\}}$ and $F_{3,\{1,4\}}$ are in a clique in the graph, but none of the workers can transmit $F_{1,\{2,3\}} \oplus F_{3,\{1,4\}}$, and thus it is not a distributed clique.
However, if we add $F_{1,\{2,4\}}$ to the group, and transmit the two sums $F_{1,\{2,3\}} \oplus F_{1,\{2,4\}}$ (by worker~$2$) and $F_{3,\{1,4\}}\oplus F_{1,\{2,4\}}$ (by worker~$4$), we see that worker~$1$ (who knows $F_{3,\{1,4\}}$) can recover $F_{1,\{2,4\}}$ from the second sum, and then recover $F_{1,\{2,3\}}$ from the first sum. 
Similarly, worker~$3$ (who knows $F_{1,\{2,3\}}$) can recover $ F_{1,\{2,4\}}$ from the first sum, and then recover  $F_{3,\{1,4\}}$ from the second sum. 
 It can be seen that $F_{1,\{2,3\},1}$ is message $a$, $F_{3,\{1,4\},1}$ is message $b$, and $F_{1,\{2,4\},1}$ is message $c$ in Proposition~\ref{prop:distributed clique covering}, while $u(a)$ and $u(c)$ are both worker $1$, $u(b)$ is worker $3$. In addition,
workers $2$ and $4$ serve as senders $k_1$ and $k_2$.

Similarly, the sub-blocks $F_{1,\{2,4\}}$ and $F_{4,\{1,5\}}$ are in a clique in the graph, but none of the workers can transmit $F_{1,\{2,4\}} \oplus F_{4,\{1,5\}}$. 
However, if we add $F_{1,\{2,5\}}$ to the group, and transmit $F_{1,\{2,4\}} \oplus F_{1,\{2,5\}}$ (by worker~$2$) and $F_{1,\{2,5\}} \oplus F_{4,\{1,5\}}$ (by worker~$5$), then worker~$1$ (who knows $F_{4,\{1,5\}}$) can recover $F_{1,\{2,5\}}$ from the second sum, and then
recover $F_{1,\{2,4\}}$ from the first sum; also, worker~$4$ (who knows $F_{1,\{2,4\}}$) can recover $F_{1,\{2,5\}}$ from the first sum, and then recover  $F_{4,\{1,5\}}$ from the second sum.

In general, we have the following data shuffling scheme in time slot $t$.
Recall that $\usf^{t-1}_{i}$ denotes the worker who should recover $F_i$ at the end of time slot $t-1$.  
We partition each sub-block into $3$ equal-length sub-pieces as $F_{i,\Wc}=(F_{i,\Wc,1},F_{i,\Wc,2},F_{i,\Wc,3})$ (recall $|\Wc|=m=2$ and $\usf^{t-1}_{i}\in \Wc$).
In general, we consider a pair $(a,b)$, where $a\in [5]$ and $b\in [5] \setminus \{a,\usf^{t-1}_{a}\}$, i.e., $a$ is a worker with the request $F_a$ in time slot $t$, while $b$  is another worker which is not the one who requests $F_a$ in time slot $t-1$.
We have two cases:
\begin{enumerate}

\item
If $a\neq \usf^{t-1}_{b}$, we find the group of sub-blocks $F_{a,\{\usf^{t-1}_{a},b\}}$, $F_{a,\{\usf^{t-1}_{a},\usf^{t-1}_{b}\}}$, and $F_{b,\{a,\usf^{t-1}_{b}\}}$. We pick one of untransmitted sub-pieces for each of these three sub-blocks. Assume the picked sub-pieces are $n_1$, $n_2$, and $n_3$, respectively. We let worker $\usf^{t-1}_{a}$ transmit $F_{a,\{\usf^{t-1}_{a},b\},n_1} \oplus F_{a,\{\usf^{t-1}_{a},\usf^{t-1}_{b}\},n_2}$, and let worker $\usf^{t-1}_{b}$ transmit $F_{a,\{\usf^{t-1}_{a},\usf^{t-1}_{b}\},n_2}\oplus F_{b,\{a,\usf^{t-1}_{b}\},n_3}$. It can be seen that worker $a$ can recover $F_{a,\{\usf^{t-1}_{a},b\},n_1}$ and $F_{a,\{\usf^{t-1}_{a},\usf^{t-1}_{b}\},n_2}$ while worker $b$ can recover $F_{b,\{a,\usf^{t-1}_{b}\},n_3}$.

\item
If  $a= \usf^{t-1}_{b}$, we assume $c=\usf^{t-1}_{a}$ and $d=\usf^{t-1}_{c}$ (e.g, if $a=1$ and $b=5$, we have $c=\usf^{t-1}_{1}=2$ and $d=\usf^{t-1}_{c}=3$), i.e.,  worker $a$ requests $F_a$ in time slot $t$, worker $c$ requests $F_a$ in time slot $t-1$ and requests $F_c$ in time slot $t$, worker $d$ requests $F_c$ in time slot $t-1$ and requests $ F_d$ in time slot $t$.
We find the group of sub-blocks $F_{a,\{c,b\}}$, $F_{a,\{c,d\}}$, and $F_{c,\{a,d\}}$. Notice that  $F_{a,\{c,d\}}$ and $F_{c,\{a,d\}}$ form  a distributed clique.
We pick one of untransmitted sub-pieces for each of these three sub-blocks (assumed to be $n_1$, $n_2$, and $n_3$, respectively). We let worker $c$  transmit $F_{a,\{c,b\},n_1} $, and let worker $d$ transmit $F_{a,\{c,d\},n_2} \oplus F_{c,\{a,d\},n_3}$. It can be seen that worker $a$ can recover $F_{a,\{c,b\},n_1}$ and $F_{a,\{c,d\},n_2}$ while worker $c$ can recover $F_{c,\{a,d\},n_3}$.

\end{enumerate}
By this construction, see also Table~\ref{tab:example}, each sub-block appears in three groups such that each of its sub-pieces is transmitted.
We use two binary sums to encode each group containing $3$ sub-pieces, such that the coding gain of this scheme is $2/3$. The achieved worst-case load is $5/2$, while the achieved loads by Schemes~A and~B  are $\frac{27}{8}$ and $\frac{10}{4}$, respectively. 

The introduced scheme in this example has the same load as Scheme~B. However, in general when $\Msf=2\qsf$, the coding gain of the new scheme is $2/3$, and this is independent of $\Ksf$. For the other schemes, the coding gain of Scheme~B is close to $1$ if $\Ksf$ is large, and the same holds for Scheme~A (as it will be shown in Section~\ref{sub:optimality}).  Therefore, Scheme~C is preferable to the other schemes  when $\Msf=2\qsf$ and $\Ksf$ is large. 
\hfill$\square$
\end{example}

\begin{table*}
\protect\caption{Transmission of Scheme C for Example~\ref{ex:achievable c}}\label{tab:example}
\centering{}
\begin{tabular}{|c|c|c|c|}
\hline 
Considered vectors & Groups of sub-pieces & First sum & Second sum \tabularnewline
\hline 
\hline 
$(1,3)$ & $F_{1,\{2,3\},1},F_{1,\{2,4\},1},F_{3,\{1,4\},1}$ & $F_{1,\{2,3\},1}\oplus F_{1,\{2,4\},1}$ & $F_{1,\{2,4\},1}\oplus F_{3,\{1,4\},1}$ \tabularnewline
\hline 
$(1,4)$ & $F_{1,\{2,4\},2},F_{1,\{2,5\},1},F_{4,\{1,5\},1}$ & $F_{1,\{2,4\},2}\oplus F_{1,\{2,5\},1}$ & $F_{1,\{2,5\},1}\oplus F_{4,\{1,5\},1}$ \tabularnewline
\hline 
$(1,5)$ & $F_{1,\{2,5\},2},F_{1,\{2,3\},2},F_{2,\{1,3\},1}$ & $F_{1,\{2,5\},2}$ & $F_{1,\{2,3\},2}\oplus F_{2,\{1,3\},1}$ \tabularnewline
\hline 
$(2,4)$ & $F_{2,\{3,4\},1},F_{2,\{3,5\},1},F_{4,\{2,5\},1}$ & $F_{2,\{3,4\},1}\oplus F_{2,\{3,5\},1}$ & $F_{2,\{3,5\},1}\oplus F_{4,\{2,5\},1}$ \tabularnewline
\hline
$(2,5)$ & $F_{2,\{3,5\},2},F_{2,\{1,3\},2},F_{5,\{1,2\},1}$ & $F_{2,\{3,5\},2}\oplus F_{2,\{1,3\},2}$ & $F_{2,\{1,3\},2}\oplus F_{5,\{1,2\},1}$ \tabularnewline
\hline
$(2,1)$& $F_{2,\{1,3\},3},F_{2,\{3,4\},2},F_{3,\{2,4\},1}$ & $F_{2,\{1,3\},3}$ & $F_{2,\{3,4\},2}\oplus F_{3,\{2,4\},1}$ \tabularnewline
\hline
$(3,5)$&$F_{3,\{4,5\},1},F_{3,\{1,4\},2},F_{5,\{1,3\},1}$ & $F_{3,\{4,5\},1}\oplus F_{3,\{1,4\},2}$ & $F_{3,\{1,4\},2}\oplus F_{5,\{1,3\},1}$ \tabularnewline
\hline
$(3,1)$&$F_{3,\{1,4\},3},F_{3,\{2,4\},2},F_{1,\{2,3\},3}$ & $F_{3,\{1,4\},3}\oplus F_{3,\{2,4\},2}$ & $F_{3,\{2,4\},2}\oplus F_{1,\{2,3\},3}$ \tabularnewline
\hline
$(3,2)$&$F_{3,\{2,4\},3},F_{3,\{4,5\},2},F_{4,\{3,5\},1}$ & $F_{3,\{2,4\},3}$ & $F_{3,\{4,5\},2}\oplus F_{4,\{3,5\},1}$ \tabularnewline
\hline
$(4,1)$&$F_{4,\{1,5\},2},F_{4,\{2,5\},2},F_{1,\{2,4\},3}$ & $F_{4,\{1,5\},2}\oplus F_{4,\{2,5\},2}$ & $F_{4,\{2,5\},2}\oplus F_{1,\{2,4\},3}$ \tabularnewline
\hline
$(4,2)$&$F_{4,\{2,5\},3},F_{4,\{3,5\},2},F_{2,\{3,4\},3}$ & $F_{4,\{2,5\},3}\oplus F_{4,\{3,5\},2}$ & $F_{4,\{3,5\},2}\oplus F_{2,\{3,4\},3}$ \tabularnewline
\hline
$(4,3)$&$F_{4,\{3,5\},3},F_{4,\{1,5\},3},F_{5,\{1,4\},1}$ & $F_{4,\{3,5\},3}$ & $F_{4,\{1,5\},3}\oplus F_{5,\{1,4\},1}$ \tabularnewline
\hline
$(5,2)$&$F_{5,\{1,2\},2},F_{5,\{1,3\},2},F_{2,\{3,5\},3}$ & $F_{5,\{1,2\},2}\oplus F_{5,\{1,3\},2}$ & $F_{5,\{1,3\},2}\oplus F_{2,\{3,5\},3}$ \tabularnewline
\hline
$(5,3)$&$F_{5,\{1,3\},3},F_{5,\{1,4\},2},F_{3,\{4,5\},3}$ & $F_{5,\{1,3\},3}\oplus F_{5,\{1,4\},2}$ & $F_{5,\{1,4\},2}\oplus F_{3,\{4,5\},3}$ \tabularnewline
\hline
$(5,4)$&$F_{5,\{1,4\},3},F_{5,\{1,2\},3},F_{1,\{2,5\},3}$ & $F_{5,\{1,4\},3}$ & $F_{5,\{1,2\},3}\oplus F_{1,\{2,5\},3}$ \tabularnewline
\hline
\end{tabular}
\end{table*}

We are now ready to generalize the scheme in Example~\ref{ex:achievable c} to the case $\Msf/\qsf=m=2$. 

\paragraph*{Structural  Invariant Data Partitioning and Storage}
This is the same as in Scheme~B (described in Section~\ref{sub:achievable b}), i.e., 
each sub-block of $F_{i}, i\in [\Nsf],$ is stored by worker $\usf^{t}_{i}$ and by another worker in $[\Ksf]\setminus \{\usf^{t}_{i}\}$. 
The length of each sub-block is $\frac{\Bsf}{\binom{\Ksf-1}{m-1}}=\frac{\Bsf}{\Ksf-1}$.

\paragraph*{Data Shuffling Phase of time slot $t\in [T]$}
As in Scheme~B, we partition the $\Nsf$ data units into $\qsf$ equal-length groups such that, during the data shuffling phase of time slot $t$, among all the $\Ksf$ data units in each group,
each worker requests exactly one data unit  and  knows exactly one data unit. 
To simplify the description, as in Scheme~B, we focus on one group and remove the $\Hc_i$ in the notation. In addition, we can  restrict attention to $\Uc=\emptyset$, because if $\Uc\neq \emptyset$ we can divide all desired sub-blocks by all workers into sets as we did in Scheme B. For each set which is known by some worker in $\Uc$, the transmission for this set is equivalent to centralized data shuffling. Thus we only need to consider the set which is not known by any worker in $\Uc$, and the transmission for this set is equivalent to decentralized data shuffling with $\Ksf_{\textrm{eq}}=\Ksf-|\Uc|$, $\qsf_{\textrm{eq}}=1$ and $\Msf_{\textrm{eq}}=m$.
Hence, for the simplicity, we only consider $\Uc=\emptyset$ for Scheme C.

We define a set of pair of workers  as
\begin{align}
\Yc:=\Big\{(a,b): a \in [\Ksf], b\in \big([\Ksf]\setminus \{\usf^{t-1}_{d^t_{a} }, a\} \big) \Big\}.\label{eq:def of V_H}
\end{align}
For each vector $(a,b) \in \Yc$, we divide $F_{d^t_{a},\{b,\usf^{t-1}_{d^t_{a} }\}}$ into $3$ non-overlapping and equal-length sub-pieces,  $F_{d^t_{a},\{b,\usf^{t-1}_{d^t_{a}}\},1}$, $F_{d^t_{a},\{b,\usf^{t-1}_{d^t_{a}}\},2}$ and $F_{d^t_{a},\{b,\usf^{t-1}_{d^t_{a}}\},3}$.
For each vector $(a,b) \in \Yc$, we consider two cases:
\begin{itemize}

\item 
Case $\usf^{t-1}_{d^t_{b}} \neq a$: 
For each one of  
$F_{d^t_{a}, \{\usf^{t-1}_{d^t_{a}}, b \}}$, 
$F_{d^t_{b}, \{\usf^{t-1}_{d^t_{b}}, a\}}$, and 
$F_{d^t_{a}, \{\usf^{t-1}_{d^t_{a}}, \usf^{t-1}_{d^t_{b}} \}}$,
we select one of its non-transmitted sub-pieces and assume they are 
$F_{d^t_{a}, \{\usf^{t-1}_{d^t_{a}}, b \},n_1}$,
$F_{d^t_{b}, \{\usf^{t-1}_{d^t_{b}}, a\},n_2}$, and
$F_{d^t_{a}, \{\usf^{t-1}_{d^t_{a}}, \usf^{t-1}_{d^t_{b}} \},n_3}$.
By Proposition~\ref{prop:distributed clique covering},  
\begin{align*}
&\textrm{worker }\usf^{t-1}_{d^t_{a}} \textrm{ transmits } F_{d^t_{a}, \{\usf^{t-1}_{d^t_{a}}, b \},n_1} \oplus F_{d^t_{a}, \{\usf^{t-1}_{d^t_{a}}, \usf^{t-1}_{d^t_{b}} \},n_3};
\\
&\textrm{worker }\usf^{t-1}_{d^t_{b}} \textrm{ transmits } F_{d^t_{b}, \{\usf^{t-1}_{d^t_{b}}, a\},n_2} \oplus F_{d^t_{a}, \{\usf^{t-1}_{d^t_{a}}, \usf^{t-1}_{d^t_{b}} \},n_3},
\end{align*}
such that each of the above linear combinations can be decoded by its requesting worker. 
For example in Table~\ref{tab:example}, for pair $(1,3)$, we let worker~$2$ transmit $F_{1,\{2,3\},1}\oplus F_{1,\{2,4\},1}$,
and worker~$4$ transmit $F_{1,\{2,4\},1}\oplus F_{3,\{1,4\},1}$. 

\item 
Case $\usf^{t-1}_{d^t_{b}} = a$:
Let $\usf^{t-1}_{d^t_{a}}=c$ and $\usf^{t-1}_{d^t_{c}}=d$. 
For each one of  
$F_{d^t_{a}, \{c, b \}}$, 
$F_{d^t_{a}, \{c, d\}}$, and
$F_{d^t_{c},  \{a, d \}}$,
we select one of its non-transmitted sub-pieces and assume they are 
$F_{d^t_{a}, \{c, b \},n_1}$,
$F_{d^t_{a}, \{c, d\},n_2}$, and 
$F_{d^t_{c},  \{a, d \},n_3}$.
By Proposition~\ref{prop:distributed clique covering},  
\begin{align*}
&\textrm{worker }c \textrm{ transmits } F_{d^t_{a}, \{c, b \},n_1};
\\
&\textrm{worker }d \textrm{ transmit } F_{d^t_{a}, \{c, d\},n_2} \oplus F_{d^t_{c},  \{a, d \},n_3},
\end{align*}
such that each of the above sub-pieces can be decoded by its requesting worker. 
For example in Table~\ref{tab:example}, for vector $(1,5)$, we let worker $2$ transmit $F_{1,\{2,5\},2}$ and worker $3$ transmit $F_{1,\{2,3\},2}\oplus F_{2,\{1,3\},1}$.
Notice that in the case if  $d=b$, we have $F_{d^t_{a}, \{c, b \}}=F_{d^t_{a}, \{c, d\}}$, and thus we transmit two different sub-pieces of $F_{d^t_{a}, \{c, b \}}$ for the vector $(a,b)$.

\end{itemize}

Next we prove that after considering all the pairs in $\Yc$, each sub-piece of sub-block $F_{d^t_{a_1},\{\usf^{t-1}_{d^t_{a_1}},b_1\}}$ has been transmitted for $(a_1,b_1) \in \Yc$. 
For each $(a_1,b_1)\in\Yc$,  if there exists one worker $c\not\in\{ a_1,b_1 \}$ such that $c=\usf^{t-1}_{d^t_{a_1}}$, $b_1=\usf^{t-1}_{d^t_{c}}$, and $a_1=\usf^{t-1}_{d^t_{b_1}}$ (in Example~\ref{ex:achievable c}, this pair $(a_1,b_1)$ does not exist), we transmit two sub-pieces of $F_{d^t_{a_1},\{\usf^{t-1}_{d^t_{a_1}},b_1\}}$ in the transmission for the pair $(a_1,b_1)$ and one sub-piece in the transmission for the pair $(b_1,c)$.
Otherwise, we transmit the three sub-pieces of $F_{d^t_{a_1},\{\usf^{t-1}_{d^t_{a_1}},b_1\}}$ in the following three transmissions for three different pairs:
\begin{enumerate}
\item
The transmission for the pair $(a_1,b_1)$.

\item
The transmission for the pair $(a_1,b_2)$ where $\usf^{t-1}_{d^t_{b_2}}=b_1$ if the requested data unit by worker $\usf^{t-1}_{d^t_{a_1}}$    in time slot $t$   
  was not stored by worker $b_1$ at the end of time slot $t-1$, 
(e.g., in Table~\ref{tab:example},  let $(a_1,b_1)=(1,4)$, one sub-piece of  $F_{1,\{2,4\}}$ appears in the transmission for vector $(a_1,b_2)=(1,3)$).

Otherwise, the transmission for the pair $(a_1,b_3)$ where $\usf^{t-1}_{d^t_{b_3}}=a_1$ (e.g., in Table~\ref{tab:example},  let $(a_1,b_1)=(1,3)$, one sub-piece of  $F_{1,\{2,3\}}$ appears in the transmission for vector $(a_1,b_3)=(1,5)$).

\item
The transmission for the pair $(b_1,a_1)$ if $\usf^{t-1}_{d^t_{b_1}} \neq a_1$ (e.g., in Table~\ref{tab:example},  let $(a_1,b_1)=(1,3)$, one sub-piece of  $F_{1,\{2,3\}}$ appears in the transmission for vector $(b_1,a_1)=(3,1)$).

Otherwise, the transmission for the pair $(b_1,b_4)$ where $\usf^{t-1}_{d^t_{b_4}}=b_1$ (e.g., in Table~\ref{tab:example},  let $(a_1,b_1)=(1,5)$, one sub-piece of  $F_{1,\{2,5\}}$ appears in the transmission for vector $(b_1,b_4)=(5,4)$).

\end{enumerate}
This shows that   $F_{d^t_{a_1},\{\usf^{t-1}_{d^t_{a_1}},b_1\}}$ is transmitted.
In Scheme C, we transmit each three sub-pieces in two sums, and thus the multicast coding gain is $2/3$.

Finally, by comparing the loads for different cases, the worst-cases are when $\Ac^{t-1}_k\cap \Ac^t_k=\emptyset$ for each $k\in [\Ksf]$ and the worst-case load achieved by
 Scheme~C is 
\begin{align}
 \qsf \frac{2 \Ksf (\Ksf-2)}{3(\Ksf-1)}
=:\left.\Rsf_\text{\rm Ach.C}\right|_{\Msf=2\qsf},
\label{eq:worst-case load achieved by Scheme C}
\end{align}
which is optimal within a factor  $4/3$ compared to the converse bound 
$ \frac{\Ksf (\Ksf-2)}{2(\Ksf-1)}$ under the constraint of uncoded storage for $\Msf/\qsf=2$.

\paragraph*{Storage Update Phase of time slot $t\in [T]$}
The storage update phase of Scheme~C is the same as the one of Scheme~B.

\subsection{Optimality Results of the Combined Achievable Scheme}
\label{sub:optimality}

Since the proposed converse bound is a piecewise linear curve with corner points $\big(m \qsf, \qsf \frac{\Ksf-m}{m}\frac{\Ksf}{\Ksf-1} \big)$ for $m\in [\Ksf]$ and these corner points are  successively convex, it follows immediately that the combined scheme in Corollary~\ref{cor:combined scheme} is optimal under the constraint of uncoded storage when $\Msf/\qsf=1$ or $\Msf/\qsf\in[\Ksf-2,\Ksf]$, thus proving Theorem~\ref{thm:exact optimality scheme 2} (and also Corollary~\ref{cor:optimality until 4}).

In order to prove Theorem~\ref{thm:order optimality} (i.e., an order optimality result for the cases not covered by  Theorem~\ref{thm:exact optimality scheme 2} or Corollary~\ref{cor:optimality until 4}) we proceed as follows.
Recall that the proposed converse bound is a piecewise linear curve with successively convex  corner points, and that the straight line in the storage-load tradeoff between two achievable points is also achievable by memory-sharing. Hence, in the following, we focus on each corner point  of the converse bound $\big(m \qsf, \qsf \frac{\Ksf-m}{m}\frac{\Ksf}{\Ksf-1} \big)$ where $m\in [\Ksf]$, and characterize the multiplicative gap between the combined scheme and the converse when $\Msf=m \qsf$. Thus the   multiplicative gap between  the achievability and the  converse  curves is upper bounded by the maximum of the obtained  $\Ksf$ gaps.

We do not consider the corner points where $m\in \{1,\Ksf-2,\Ksf-1,\Ksf\}$ because the optimality of the combined scheme has been proved. We have:
\begin{itemize}

\item \ $\Msf=2\qsf$:
It was proved in Section~\ref{sub:achievable c} that the multiplicative gap between the Scheme C and the converse bound is $4/3$.

\item \ Interpolate the achievable bound for Scheme~A in~\eqref{eq:distributed shuffling achievable bound 1} between $\Msf_1=\big(1+g\frac{\Ksf-1}{\Ksf} \big)\qsf$ and $\Msf_2=\big(1+(g+1)\frac{\Ksf-1}{\Ksf} \big)\qsf$ to match the converse bound in~\eqref{eq:distributed shuffling converse bound} at $\Msf_3=(g+1)\qsf$ where $g\in[2:\Ksf-4]$:
%
For each $g\in[2:\Ksf-4]$, we have
\begin{align}
&\Msf_1=\left(1+g\frac{\Ksf-1}{\Ksf}\right)\qsf,
  \  \Rsf_\text{\rm Ach.A}(\Msf_1)=\qsf\frac{\Ksf-g}{g},
\\
&\Msf_2 \negmedspace = \negmedspace \left(\negmedspace 1\negmedspace+\negmedspace(g+1)\frac{\Ksf-1}{\Ksf} \negmedspace \right)\negmedspace\qsf,
  \ \Rsf_\text{\rm Ach.A}(\Msf_2)\negmedspace= \negmedspace \qsf\frac{\Ksf-g-1}{g+1}.
\end{align}
 By memory-sharing between $(\Msf_1,\Rsf_\text{\rm Ach.A}(\Msf_1))$ and $(\Msf_2,\Rsf_\text{\rm Ach.A}(\Msf_2))$ with coefficient $\alpha=(\Ksf-1-g)/(\Ksf-1)$, we get 
\begin{align}
\Msf_3 & = \left. \alpha \Msf_1 + (1 - \alpha) \Msf_2 \right|_{\alpha=1-g/(\Ksf-1)}
\nonumber
\\&= (1+g)\qsf,
\end{align}
 as in the converse bound for $m=g+1\in[3:\Ksf-3]$, and
 \begin{align}
\Rsf_\text{\rm Ach.A}(\Msf_3)
  &=\alpha\Rsf_\text{\rm Ach.A}(\Msf_1) + (1 - \alpha) \Rsf_\text{\rm Ach.A}(\Msf_2)
\nonumber
\\&= \frac{\qsf\Ksf-g}{g}\frac{\Ksf-1-g}{\Ksf-1}+\frac{\qsf g}{\Ksf-1}\frac{\Ksf-g-1}{g+1} \nonumber
\\&=\frac{\qsf (\Ksf-g-1)(\Ksf g+\Ksf-g)}{(\Ksf-1)g(g+1)}.
\label{eq:scheme 1 for M=1+g}
\end{align}
From (the proof of) Theorem~\ref{thm:converse bound}, we know that 
\begin{align}
\Rsf_\text{\rm Out}(\Msf_3)\geq \Nsf\frac{1-g/(\Ksf-1)}{g+1} = \qsf\frac{\Ksf}{\Ksf-1}\frac{\Ksf-g-1}{g+1}.\label{eq:converse bound in this case}
\end{align}
Hence, from~\eqref{eq:scheme 1 for M=1+g} and~\eqref{eq:converse bound in this case}, we have
\begin{align}
\frac{\Rsf_\text{\rm Ach.A}(\Msf_3)}{\Rsf_\text{\rm Out}(\Msf_3)}
  &\leq  \frac{\Ksf g+\Ksf-g}{g\Ksf}=1-\frac{1}{\Ksf}+\frac{1}{g} \nonumber\\&
\leq  1-0+\frac{1}{2} = \frac{3}{2} \ \text{(since $g\geq 2$)}.
\label{eq:final order optimality first case}
\end{align}
\end{itemize}

We then focus on $m \qsf \leq \Msf \leq (m+1)\qsf$, where $m\in [\Ksf-1]$. The converse bound in Theorem~\ref{thm:converse bound} for $m\qsf \leq \Msf \leq (m+1)\qsf$ is a straight line between $(\Msf^{\prime},\Rsf^{\prime})=(m\qsf, \frac{(\Ksf-m)\Ksf}{m(\Ksf-1)})$ and $(\Msf^{\prime\prime},\Rsf^{\prime\prime})=\left((m+1)\qsf, \frac{(\Ksf-m-1)\Ksf}{(m+1)(\Ksf-1)} \right)$.
\begin{itemize}
\item $m=1$.  The combined scheme can achieve $(\Msf^{\prime},\Rsf^{\prime})$ and $(\Msf^{\prime\prime},4\Rsf^{\prime\prime}/3)$. Hence, by memory-sharing, the multiplicative gap between the combined scheme and the converse bound is less than $4/3$.
\item $m=2$. The combined scheme can achieve $\left(\Msf^{\prime},4 \Rsf^{\prime}/3 \right)$ and $\left(\Msf^{\prime\prime},\left(1-\frac{1}{\Ksf}+\frac{1}{2} \right)\Rsf^{\prime\prime} \right)$. Hence, by memory-sharing, the multiplicative gap between the combined scheme and the converse bound is less than $1-\frac{1}{\Ksf}+\frac{1}{2}$.
\item $m \in [3: \Ksf-4]$. The combined scheme can achieve $\left(\Msf^{\prime},\left(1-\frac{1}{\Ksf}+\frac{1}{m-1} \right)\Rsf^{\prime} \right)$ and $\left(\Msf^{\prime\prime},\left(1-\frac{1}{\Ksf}+\frac{1}{m} \right)\Rsf^{\prime\prime} \right)$. Hence, by memory-sharing, the multiplicative gap between the combined scheme and the converse bound is less than $1-\frac{1}{\Ksf}+\frac{1}{m-1}$.
\item $m = \Ksf-3$. The combined scheme can achieve $\left(\Msf^{\prime},\left(1-\frac{1}{\Ksf}+\frac{1}{m-1} \right)\Rsf^{\prime} \right)$ and $\left(\Msf^{\prime\prime}, \Rsf^{\prime\prime} \right)$. Hence, by memory-sharing, the multiplicative gap between the combined scheme and the converse bound is less than $1-\frac{1}{\Ksf}+\frac{1}{m-1}$.
\item $m  \in \{ \Ksf-2,\Ksf-1\}$. The combined scheme can achieve $\left(\Msf^{\prime}, \Rsf^{\prime} \right)$ and $\left(\Msf^{\prime\prime}, \Rsf^{\prime\prime} \right)$.  Hence, the combined scheme coincides with the converse bound.
\end{itemize}
This concludes the proof of Theorem~\ref{thm:order optimality}.

\paragraph*{On  communication cost of peer-to-peer operations}
By comparing the decentralized data shuffling converse bound and the optimal centralized data shuffling load (denoted by $\Rsf_{\textrm{Opt.Cen}}(\Msf)$), we have $\Rsf_\text{\rm Out}(\Msf)/\Rsf_{\textrm{Opt.Cen}}(\Msf)=\Ksf/(\Ksf-1)$ for any $\qsf\leq \Msf\leq \Ksf\qsf$. In addition, the maximum multiplicative gap between the achieved load by the combined scheme and $\Rsf_\text{\rm Out}(\Msf)$, is $\max\left\{ 1-\frac{1}{\Ksf}+\frac{1}{m-1},  \frac{4}{3} \right\}$, where $m \geq 3$.  Hence, the maximum multiplicative gap between the achieved load by the combined scheme and $\Rsf_{\textrm{Opt.Cen}}(\Msf)$ is 
\begin{align}
&\frac{\Ksf}{\Ksf-1}\max\left\{ 1-\frac{1}{\Ksf}+\frac{1}{m-1} ,\frac{4}{3} \right\}\nonumber\\&=\max\left\{1+\frac{\Ksf}{(m-1)(\Ksf-1)},  \frac{4\Ksf}{3(\Ksf-1)}  \right\}, \label{eq:cost of p2p}
\end{align}
which is no more than $5/3$ if $\Ksf\geq 5$.
In addition, when $\Ksf\leq 4$, by Corollary~\ref{cor:optimality until 4}, the combined scheme is optimal such that the communication cost of peer-to-peer operations is $\Ksf/(\Ksf-1)\leq 2$.
In general, the communication cost of peer-to-peer operations is no more than a factor of $2$ as stated in Corollary~\ref{rem:cost}.

In addition, with a similar proof as above to analyse each storage size regime  $m \qsf \leq \Msf \leq (m+1)\qsf$ where $m\in [\Ksf-1]$, we prove   Corollary~\ref{rem:cost}.

\section{Conclusions}
\label{sec:conclusion}
In this paper, we introduced the decentralized data shuffling problem and studied its fundamental limits.
We proposed a converse bound under the constraint of uncoded storage and three achievable schemes.
In general, under the constraint of uncoded storage, our schemes are order optimal to within a factor of $3/2$, and exactly optimal for
small and large storage sizes, 
or the systems with no more than four workers.

\appendix

\section{Proof of a Key Lemma}
\label{sec:Proof of Lemma}
We   define 
\begin{align}
\Vc_{\Sc}:=\{k\in \Sc : d_k\in \Sc\} , \ \forall \Sc\subseteq [\Ksf],  
\label{eq:def Vs_Sc}
\end{align} 
where $\Vc_{\Sc}$ represents the subset of workers in $\Sc $ whose demanded data units in time slot $t$, indexed by $\Ac^{t}_k=\Ac^{t-1}_{d_k}$ in~\eqref{eq:something:Actk}, were stored by some workers in $\Sc$ at the end of time slot $t-1$\footnote{For example, if $\Ksf=4$ and $(d_{1},\ldots,d_{4})=(2,3,4,1)$, we have $\Vc_{\{2,3\}}=\{2\}$ because $d_{2}=3$  and thus  the requested data unit by worker $2$ in time slot $t$ was stored by worker $3$ at the end of time slot $t-1$; similarly, we have $\Vc_{\{2,4\}}=\emptyset$ and $\Vc_{\{1,2,4\}}=\{1,4\}$.}.

In addition, we also define
 $Y^{t}_{\Sc}$ as the sub-blocks that any worker in $\Sc$ either needs to store at the end of time slot $t$ or has stored at the end of time slot $t-1$, that is,
\begin{align}
Y^{t}_{\Sc}&:= \Big\{ F_{i}:i\in \cup_{k\in \Sc}\Ac^{t}_{k} \Big\} \cup \Big\{ Z^{t-1}_{k}:k\in \Sc \Big\} 
\notag\\
&= \Big\{ F_{i}:i\in \cup_{k\in \Sc}(\Ac^{t}_{k}\cup  \Ac^{t-1}_{k}) \Big\} \nonumber\\& \cup
   \Big\{ F_{i,\Wc}:  i\notin \cup_{k\in \Sc}(\Ac^{t}_{k}\cup  \Ac^{t-1}_{k}), \Wc\cap \Sc\neq \emptyset  \Big\}.
\label{eq:def Y^{t+1}_{Sc}}
\end{align}

With the above definitions and recalling  $X^{t}_{\Sc}$ defined in~\eqref{eq:def X^{t+1}_{Sc}} represents the messages sent by the workers in $\Sc$ during time slot $t$,
we have the following lemma:
\begin{lem}[Induction Lemma]
\label{lem:induction lemma}
For each  non-empty set $\Sc\subseteq  [\Ksf]  $, we have 
\begin{align}
&H\Big(  X^{t}_{\Sc}|Y^{t}_{[\Ksf]\setminus \Sc } \Big)
 \nonumber\\&
 \geq 
\sum_{m=1}^{|\Sc|} \ \
\sum_{k\in \Vc_{\Sc}} \ \
\sum_{i\in \Ac^{t}_{k}} \ \
\sum_{\substack{\Wc\subseteq \Sc \setminus \{k\} : \\ \usf^{t-1}_{i}\in\Wc,   |\Wc|=m}}
\frac{|F_{i,\Wc}|}{m}.\label{eq:induction lemma}
\end{align}
\end{lem}

Lemma~\ref{lem:induction lemma} is the key novel contribution of our proof. The bound in~\eqref{eq:induction lemma} can be intuitively explained as follows: $H(  X^{t}_{\Sc}|Y^{t}_{[\Ksf]\setminus \Sc } )$ is lower bounded  by the size of the requested sub-blocks by the workers in $\Vc_{\Sc}$ (instead of in $\Sc$ as in the distributed computing problem~\cite{distributedcomputing}) because each requested data unit by the workers in $\Sc\setminus \Vc_{\Sc}$ was requested in the previous time slot  by  some workers in $[\Ksf]\setminus  \Sc  $ because of the storage constraint in~\eqref{eq:additional storage constraint node k} and the definition of $\Vc_{\Sc}$ in~\eqref{eq:def Vs_Sc}.

This lemma is proved by induction, inspired by~\cite{distributedcomputing}.

\begin{IEEEproof}
Case $|\Sc|=1$:
If $\Sc=\{k\}$ where $k\in[\Ksf]$, we have that $\Vc_{\{k\}}=\emptyset$  and thus the RHS of~\eqref{eq:induction lemma} is $0$; 
thus~\eqref{eq:induction lemma} holds for $|\Sc|=1$ because entropy is non-negative.

Case $|\Sc| \leq s$:
Assume that~\eqref{eq:induction lemma} holds for all  non-empty $\Sc\subseteq [\Ksf]$ where $|\Sc|\leq s$ for some integer $s\in[\Ksf-1]$. 

Case $|\Sc| = s+1$:
Having assumed that the lemma holds for all $\Sc\subseteq [\Ksf]$ where $|\Sc|\leq s$, we aim to show that 
for any set $\Jc\subseteq [\Ksf]$ where $|\Jc|=s+1$, we have 
\begin{align}
&H(X^{t}_{\Jc}|Y^{t}_{[\Ksf]\setminus \Jc})\geq\nonumber\\& \sum_{m=1}^{|\Jc|}\sum_{k\in \Vc_{\Jc}} \sum_{i\in \Ac^{t}_{k}} \ \ \sum_{\substack{\Wc\subseteq (\Jc\setminus \{k\}):\\ |\Wc|=m,\usf^{t-1}_{i}\in \Wc}}\frac{|F_{i,\Wc}|}{m}.
\label{eq:induction equation}
\end{align}
From the independence bound on entropy we have
\begin{subequations} 
\begin{align}
&H(X^{t}_{\Jc}|Y^{t}_{[\Ksf]\setminus \Jc})\nonumber\\
&=\frac{1}{|\Jc|}\sum_{k\in\Jc}\Big( H(X^{t}_{\Jc\setminus\{k\}}|X^{t}_{k},Y^{t}_{[\Ksf]\setminus \Jc})+H(X^{t}_{k}|Y^{t}_{[\Ksf]\setminus \Jc})\Big)\\
&\geq \frac{1}{|\Jc|}\Big(\sum_{k\in\Jc} H(X^{t}_{\Jc\setminus\{k\}}|X^{t}_{k},Y^{t}_{[\Ksf]\setminus \Jc})+ H(X^{t}_{\Jc} |Y^{t}_{[\Ksf]\setminus \Jc})\Big),
\label{eq:here no tight}
\end{align}
and thus 
\begin{align}
&(|\Jc|-1)H(X^{t}_{\Jc}|Y^{t}_{[\Ksf]\setminus \Jc})
 \geq \sum_{k\in \Jc}H(X^{t}_{\Jc\setminus\{k\}}|X^{t}_{k},Y^{t}_{[\Ksf]\setminus \Jc})
\label{eq:here no tight again}
\\
&\geq \sum_{k\in \Jc}H(X^{t}_{\Jc\setminus\{k\}}|X^{t}_{k},Y^{t}_{[\Ksf]\setminus \Jc},Z^{t-1}_{k})
\label{eq:from adding Z_k}
\\
&= \sum_{k\in \Jc}H(X^{t}_{\Jc\setminus\{k\}},  \{F_{i}:i\in \Ac^{t}_{k}\} |X^{t}_{k},Y^{t}_{[\Ksf]\setminus \Jc},Z^{t-1}_{k})
\label{eq:from decoding condition}
\\
&=\sum_{k\in \Jc} H(\{F_{i}:i\in \Ac^{t}_{k}\}|Z^{t-1}_{k},Y^{t}_{[\Ksf]\setminus \Jc})  
\nonumber\\& +\sum_{k\in \Jc} H(X^{t}_{\Jc\setminus\{k\}}|\{F_{i}:i\in \Ac^{t}_{k}\},Z^{t-1}_{k},Y^{t}_{[\Ksf]\setminus \Jc})
\label{eq: X^{t+1}_{Jc-k} contains F_{Ac^{t+1}_{k}}}
\\
&=\sum_{k\in \Jc} H(\{F_{i}:i\in \Ac^{t}_{k}\}|Z^{t-1}_{k},Y^{t}_{[\Ksf]\setminus \Jc})
 \nonumber\\& +\sum_{k\in \Jc} H\big(X^{t}_{\Jc\setminus \{k\}}|Y^{t}_{([\Ksf]\setminus \Jc)\cup\{k\})}\big)
\label{eq:sum of two terms, t2},
\end{align} 
\end{subequations}
where~\eqref{eq:from adding Z_k} follows because we added $Z^{t-1}_{k}$ in the conditioning, and conditioning cannot increase entropy,
where~\eqref{eq:from decoding condition} follows because $\{F_{i}:i\in \Ac^{t}_{k}\}$ is a function of $(Z^{t-1}_{k},X^{t})$ by the decoding constraint in~\eqref{eq:decoding node k} (note that the knowledge of $(Y^{t}_{[\Ksf]\setminus \Jc},Z^{t-1}_{k})$ implies the knowledge of $Z^{t-1}_{\{k\} \cup [\Ksf]\setminus \Jc}$ and thus of $X^{t}_{\{k\} \cup [\Ksf]\setminus \Jc}$ by the encoding constraint in~\eqref{eq:encoding node k}), 
where~\eqref{eq: X^{t+1}_{Jc-k} contains F_{Ac^{t+1}_{k}}} follow because $X^{t}_{k}$ is a function of $Z^{t-1}_{k}$ (see the encoding constraint in~\eqref{eq:encoding node k}), 
and~\eqref{eq:sum of two terms, t2} from the definition in~\eqref{eq:def Y^{t+1}_{Sc}}.

Next we bound the first term of~\eqref{eq:sum of two terms, t2} by using the independence of the sub-blocks, and the second term of~\eqref{eq:sum of two terms, t2} by the induction assumption. More precisely, 
\begin{itemize}

\item
First term of~\eqref{eq:sum of two terms, t2}. 
For each $k\in \Jc$,  if $k\notin \Vc_{\Jc}$ , we have $\{F_{i}:i\in \Ac^{t}_{k}\} \subseteq  Y^{t}_{[\Ksf]\setminus \Jc} $. So for each $k\in \Jc$, by independence of sub-blocks, we have~\eqref{eq: first term original} 
\begin{figure*}
\begin{align}
 &H(\{F_{i}:i\in \Ac^{t}_{k}\}|Z^{t-1}_{k},Y^{t}_{[\Ksf]\setminus \Jc})\nonumber\\
 &=\begin{cases}
 \sum_{m=1}^{|\Jc|}\sum_{i\in \Ac^{t}_{k}} \ \ \sum_{\Wc\subseteq (\Jc\setminus \{k\}):|\Wc|=m,\usf^{t-1}_i\in \Wc} |F_{i,\Wc}|,
 & k\in \Vc_{\Jc} \\
 0 & \textrm{otherwise} \\
 \end{cases}
  \label{eq: first term original}
\end{align}
\end{figure*}
and thus we rewrite the first term of~\eqref{eq:sum of two terms, t2} as
\begin{align}
  &\sum_{k\in \Jc}H(\{F_{i}:i\in \Ac^{t}_{k}\}|Z^{t-1}_{k},Y^{t}_{[\Ksf]\setminus \Jc})
  \notag 
\\&=\sum_{k\in \Vc_{\Jc}} \sum_{m\in[|\Jc|]} \sum_{i\in \Ac^{t}_{k}} \ \ \sum_{\substack{\Wc\subseteq (\Jc\setminus \{k\}):\\ |\Wc|=m,\usf^{t-1}_i\in \Wc}} |F_{i,\Wc}|.\label{eq: first term}
\end{align}

\item
Second term of~\eqref{eq:sum of two terms, t2}. 
By the induction assumption,  
\begin{align}
  &\sum_{k\in \Jc} H\big(X^{t}_{\Jc\setminus \{k\}}|Y^{t}_{([\Ksf]\setminus \Jc)\cup \{k\}}\big) \nonumber\\
&  =\sum_{k\in \Jc} H\big(X^{t}_{\Jc\setminus \{k\}}|Y^{t}_{[\Ksf]\setminus (\Jc\setminus \{k\})}\big) 
\notag
\\&\geq \sum_{k\in \Jc} \sum_{u\in \Vc_{\Jc\setminus \{k\}}} \sum_{m=1}^{|\Jc|-1}\sum_{i\in \Ac^{t}_{u}} \ \ \sum_{ \substack{\Wc\subseteq (\Jc\setminus  \{k, u\}):\\ |\Wc|=m,\usf^{t-1}_i\in \Wc}}\frac{|F_{i,\Wc}|}{m}.\label{eq:second term original}
\end{align}

\end{itemize}

In order to combine~\eqref{eq: first term} with~\eqref{eq:second term original}, both terms need to have the summations in the same form.
Let us focus on one worker $u^{\prime}\in \Vc_{\Jc}$ and one sub-block $F_{i^{\prime},\Wc^{\prime}}$, where $i^{\prime}\in \Ac^{t}_{u^{\prime}}$, $\Wc^{\prime}\subseteq \Jc\setminus \{u^{\prime}\}$, $|\Wc^{\prime}|=m$, and $\usf^t_{i^{\prime}}\in \Wc^{\prime}$. 
On the RHS of~\eqref{eq:second term original}, for each $k\in \Jc \setminus (\Wc^{\prime}\cup \{u^{\prime}\})$,  it can be seen that $F_{i^{\prime},\Wc^{\prime}}$ appears once in the sum 
\begin{align}
\sum_{m\in[|\Jc|-1]}\sum_{u\in \Vc_{\Jc\setminus \{k\}}}\sum_{i\in \Ac^{t}_{u}} \ \ \sum_{\substack{\Wc\subseteq (\Jc\setminus  \{k, u\}):\\ |\Wc|=m,\usf^{t-1}_i\in \Wc}}\frac{|F_{i,\Wc}|}{m},
\end{align}
hence, the coefficient of $F_{i^{\prime},\Wc^{\prime}}$ on the RHS of~\eqref{eq:second term original} is $(|\Jc|-m-1)/m$.
Thus, from~\eqref{eq:second term original}, we have 
\begin{subequations} 
\begin{align}
&\sum_{k\in \Jc}H\big(X^{t}_{\Jc\setminus \{k\}}|Y^{t}_{[\Ksf]\setminus \Jc)\cup \{k\}}\big) \\
&\geq\sum_{u^{\prime}\in \Vc_{\Jc}} \sum_{m\in[|\Jc|-1]}\sum_{i^{\prime}\in \Ac^{t}_{u^{\prime}}}  \sum_{\substack{\Wc^{\prime}\subseteq (\Jc\setminus \{u^{\prime}\}):\\ |\Wc^{\prime}|=m,\usf^{t-1}_{i^{\prime}}\in \Wc^{\prime}}} \negmedspace\negmedspace\negmedspace\negmedspace\negmedspace \frac{|F_{i^{\prime},\Wc^{\prime}}|(|\Jc|-m-1)}{m}\\
&=\sum_{u^{\prime}\in \Vc_{\Jc}} \sum_{m\in[|\Jc|]}\sum_{i^{\prime}\in \Ac^{t}_{u^{\prime}}}  \sum_{\substack{\Wc^{\prime}\subseteq (\Jc\setminus \{u^{\prime}\}):\\ |\Wc^{\prime}|=m,\usf^{t-1}_{i^{\prime}}\in \Wc^{\prime}}} \negmedspace\negmedspace\negmedspace\negmedspace\negmedspace \frac{|F_{i^{\prime},\Wc^{\prime}}|(|\Jc|-m-1)}{m}.
\label{eq:second term}
\end{align}
\end{subequations} 

We next take~\eqref{eq: first term} and~\eqref{eq:second term} into~\eqref{eq:sum of two terms, t2} to obtain,
\begin{subequations} 
\begin{align}
&H(X^{t}_{\Jc}|Y^{t}_{[\Ksf]\setminus \Jc})\nonumber\\
&\geq \frac{1}{|\Jc|-1}\sum_{k\in \Vc_{\Jc}} \sum_{m=1}^{|\Jc|}\sum_{i\in \Ac^{t}_{k}} \sum_{\substack{\Wc\subseteq (\Jc\setminus \{k\}):\\ |\Wc|=m,\usf^{t-1}_i\in \Wc}} |F_{i,\Wc}| \nonumber\\
&+\frac{1}{|\Jc|-1}\sum_{k\in \Vc_{\Jc}} \sum_{m=1}^{|\Jc|}\sum_{i\in \Ac^{t}_{k}} \sum_{\substack{\Wc\subseteq (\Jc\setminus \{k\}):\\ |\Wc|=m,\usf^{t-1}_i\in \Wc}} \frac{|F_{i,\Wc}|(|\Jc|-m-1)}{m}\\
&=\sum_{k\in \Vc_{\Jc}} \sum_{m\in[|\Jc|]}\sum_{i\in \Ac^{t}_{k}} \ \ \sum_{\substack{\Wc\subseteq (\Jc\setminus \{k\}):\\ |\Wc|=m,\usf^{t-1}_i \in \Wc}} \frac{|F_{i,\Wc}|}{m},
\end{align}
\end{subequations}
which proves Lemma~\ref{lem:induction lemma}.
\end{IEEEproof}

\bibliographystyle{IEEEtran}
\bibliography{IEEEabrv,IEEEexample}

					\begin{IEEEbiographynophoto}
						{Kai Wan} (S '15 -- M '18)
						received  the M.Sc. and Ph.D. degrees in Communications from Universit{\'e} Paris Sud-CentraleSup{\'e}lec, France, in 2014 and 2018.  He is currently a post-doctoral researcher with the Communications and Information Theory Chair (CommIT) at Technische Universit\"at Berlin, Berlin, Germany. His research interests include coded caching,  index coding, distributed storage, wireless communications, and distributed computing.
					\end{IEEEbiographynophoto}
					
										\begin{IEEEbiographynophoto}{Daniela Tuninetti}  (M '98 -- SM '13)
 is currently a Professor within the Department of Electrical
and Computer Engineering at the University of Illinois at Chicago (UIC),
which she joined in 2005. Dr. Tuninetti got her Ph.D. in Electrical Engineering
in 2002 from ENST/T{\'e}l{\'e}com ParisTech (Paris, France, with work done at the
Eurecom Institute in Sophia Antipolis, France), and she was a postdoctoral
research associate at the School of Communication and Computer Science
at the Swiss Federal Institute of Technology in Lausanne (EPFL, Lausanne,
Switzerland) from 2002 to 2004. Dr. Tuninetti is a recipient of a best paper
award at the European Wireless Conference in 2002, of an NSF CAREER
award in 2007, and named University of Illinois Scholar in 2015. Dr. Tuninetti
was the editor-in-chief of the IEEE Information Theory Society Newsletter
from 2006 to 2008, an editor for IEEE COMMUNICATION LETTERS from
2006 to 2009, for IEEE TRANSACTIONS ON WIRELESS COMMUNICATIONS
from 2011 to 2014; and for IEEE TRANSACTIONS ON INFORMATION
THEORY from 2014 to 2017. She is currently a distinguished lecturer for the
Information Theory society. Dr. Tuninetti's research interests are in the
ultimate performance limits of wireless interference networks (with special
emphasis on cognition and user cooperation), coexistence between radar and
communication systems, multi-relay networks, content-type coding, cache-aided
systems and distributed private coded computing.

					\end{IEEEbiographynophoto}
					
					\begin{IEEEbiographynophoto}{Mingyue Ji}
(S '09 -- M '15) received the B.E. in Communication Engineering from Beijing University of Posts and Telecommunications (China), in 2006, the M.Sc. degrees in Electrical Engineering from Royal Institute of Technology (Sweden) and from University of California, Santa Cruz, in 2008 and 2010, respectively, and the PhD from the Ming Hsieh Department of Electrical Engineering at University of Southern California in 2015. He subsequently was a Staff II System Design Scientist with Broadcom Corporation (Broadcom Limited) in 2015-2016. He is now an Assistant Professor of Electrical and Computer Engineering Department and an Adjunct Assistant Professor of School of Computing at the University of Utah. He received the IEEE Communications Society Leonard G. Abraham Prize for the best IEEE JSAC paper in 2019, the best paper award in IEEE ICC 2015 conference, the best student paper award in IEEE European Wireless 2010 Conference and USC Annenberg Fellowship from 2010 to 2014. He is interested the broad area of information theory, coding theory, concentration of measure and statistics with the applications of caching networks, wireless communications, distributed computing and storage, security and privacy and (statistical) signal processing.
					\end{IEEEbiographynophoto}
					
					\begin{IEEEbiographynophoto}{Giuseppe Caire}
 (S '92 -- M '94 -- SM '03 -- F '05) 
was born in Torino in 1965. He received the B.Sc. in Electrical Engineering  from Politecnico di Torino in 1990, 
the M.Sc. in Electrical Engineering from Princeton University in 1992, and the Ph.D. from Politecnico di Torino in 1994. 
He has been a post-doctoral research fellow with the European Space Agency (ESTEC, Noordwijk, The Netherlands) in 1994-1995,
Assistant Professor in Telecommunications at the Politecnico di Torino, Associate Professor at the University of Parma, Italy, 
Professor with the Department of Mobile Communications at the Eurecom Institute,  Sophia-Antipolis, France,
a Professor of Electrical Engineering with the Viterbi School of Engineering, University of Southern California, Los Angeles,
and he is currently an Alexander von Humboldt Professor with the Faculty of Electrical Engineering and Computer Science at the
Technical University of Berlin, Germany.

He received the Jack Neubauer Best System Paper Award from the IEEE Vehicular Technology Society in 2003,  the
IEEE Communications Society \& Information Theory Society Joint Paper Award in 2004 and in 2011, the 
Leonard G. Abraham Prize for best IEEE JSAC paper in 2019,  the Okawa Research Award in 2006,
the Alexander von Humboldt Professorship in 2014, the Vodafone Innovation Prize in 2015, and an ERC Advanced Grant in 2018.
Giuseppe Caire is a Fellow of IEEE since 2005.  He has served in the Board of Governors of the IEEE Information Theory Society from 2004 to 2007,
and as officer from 2008 to 2013. He was President of the IEEE Information Theory Society in 2011. 
His main research interests are in the field of communications theory, information theory, channel and source coding
with particular focus on wireless communications.   
					\end{IEEEbiographynophoto}
						
											\begin{IEEEbiographynophoto}{Pablo Piantanida}
(SM '16) received both B.Sc. in Electrical Engineering and the M.Sc (with honors) from the University of Buenos Aires (Argentina) in 2003, and the Ph.D. from Universit{\'e} Paris-Sud (Orsay, France) in 2007. Since October 2007 he has joined the Laboratoire des Signaux et Syst{\`e}mes (L2S), at CentraleSup{\'e}lec together with CNRS (UMR 8506) and Universit{\'e} Paris-Sud, as an Associate Professor of Network Information Theory. He is currently associated with Montreal Institute for Learning Algorithms (Mila) at Universit{\'e} de Montr{\'e}al, Quebec, Canada.  He is an IEEE Senior Member and serves as Associate Editor for IEEE Transactions on Information Forensics and Security. He served as General Co-Chair of the 2019 IEEE International Symposium on Information Theory (ISIT). His research interests lie broadly in information theory and its interactions with other fields, including multi-terminal information theory, Shannon theory, machine learning and representation learning, statistical inference, cooperative communications, communication mechanisms for security and privacy.
					\end{IEEEbiographynophoto}


\end{document}